\documentclass[preprint]{aastex61}

\newcommand\aastex{AAS\TeX}

\accepted{February 4, 2020}

\shorttitle{\aastex\ Subaru infrared AO imaging of nearby ULIRGs}
\shortauthors{Imanishi et al.}

\begin{document}

\title{Subaru Infrared Adaptive Optics-assisted High-spatial-resolution 
Imaging Search for Luminous Dual Active Galactic Nuclei 
in Nearby Ultraluminous Infrared Galaxies}

\correspondingauthor{Masatoshi Imanishi}
\email{masa.imanishi@nao.ac.jp}

\author[0000-0001-6186-8792]{Masatoshi Imanishi}
\affil{National Astronomical Observatory of Japan, National Institutes 
of Natural Sciences (NINS), 2-21-1 Osawa, Mitaka, Tokyo 181-8588, Japan}
\affil{Department of Astronomy, School of Science, The Graduate University 
for Advanced Studies, SOKENDAI, Mitaka, Tokyo 181-8588}

\author{Taiki Kawamuro}
\affil{National Astronomical Observatory of Japan, National Institutes 
of Natural Sciences (NINS), 2-21-1 Osawa, Mitaka, Tokyo 181-8588, Japan}

\author{Satoshi Kikuta}
\affil{Department of Astronomy, School of Science, The Graduate University 
for Advanced Studies, SOKENDAI, Mitaka, Tokyo 181-8588}
\affil{National Astronomical Observatory of Japan, National Institutes 
of Natural Sciences (NINS), 2-21-1 Osawa, Mitaka, Tokyo 181-8588, Japan}

\author{Suzuka Nakano}
\affil{Department of Astronomy, School of Science, The Graduate University 
for Advanced Studies, SOKENDAI, Mitaka, Tokyo 181-8588}
\affil{National Astronomical Observatory of Japan, National Institutes 
of Natural Sciences (NINS), 2-21-1 Osawa, Mitaka, Tokyo 181-8588, Japan}

\author{Yuriko Saito}
\affil{Department of Astronomy, School of Science, The Graduate University 
for Advanced Studies, SOKENDAI, Mitaka, Tokyo 181-8588}
\affil{National Astronomical Observatory of Japan, National Institutes 
of Natural Sciences (NINS), 2-21-1 Osawa, Mitaka, Tokyo 181-8588, Japan}



\begin{abstract}

We present infrared $K'$- (2.1 $\mu$m) and $L'$-band (3.8 $\mu$m) 
high-spatial-resolution ($<$0$\farcs$3) imaging observations of 17 nearby 
(z $<$ 0.17) ultraluminous infrared galaxies (ULIRGs) assisted with 
the adaptive optics of the Subaru Telescope.
We search for compact red $K'-L'$ color emission as the indicator of 
luminous active galactic nuclei (AGNs) due to AGN-heated hot dust emission.
Two luminous dual AGN candidates are revealed.
Combining these results with those of our previous study, we can state that 
the detected fraction of luminous dual AGNs in nearby ULIRGs is much less 
than unity ($<$20\%), even when infrared wavelengths $>$2 $\mu$m are used 
that should be sensitive to buried AGNs due to small dust extinction effects. 
For ULIRGs with resolved multiple nuclear $K'$-band emission, we estimate 
the activation of supermassive black holes (SMBHs) at individual galaxy 
nuclei in the form of AGN luminosity normalized by SMBH mass 
inferred from host galaxy stellar luminosity.
We confirm a trend that more massive SMBHs in $K'$-band brighter primary 
galaxy nuclei are generally more active, with higher SMBH-mass-normalized 
AGN luminosity than less massive SMBHs in $K'$-band fainter secondary 
galaxy nuclei, as predicted by numerical simulations of gas-rich major 
galaxy mergers.
In two sources, the presence of even infrared elusive 
extremely deeply buried AGNs is indicated by comparisons with 
available (sub)millimeter data.
Non-synchronous SMBH activation (i.e., less activation of 
less massive SMBHs) and the possible presence of such infrared elusive 
AGNs may be responsible for the small fraction of 
infrared-detected luminous dual AGNs in nearby merging ULIRGs.

\end{abstract}




\section{Introduction} \label{sec:intro}

Current standard galaxy formation scenarios postulate that 
small gas-rich galaxies collide, merge, and then evolve into more 
massive galaxies \citep{whi78}.
If a supermassive black hole (SMBH) is ubiquitously present at the center 
of each merging galaxy \citep{mag98,kor13}, 
then many galaxy mergers are expected to contain multiple SMBHs. 
In gas-rich galaxy mergers, a sufficient amount of gas can accrete onto 
the existing SMBHs, and such SMBHs can become active galactic nuclei (AGNs) 
by emitting strong radiation from their surrounding accretion 
disks \citep{hop06}.
If more than one SMBH become luminous AGNs in gas-rich galaxy mergers, 
many dual AGNs will exist in the universe.
Dual AGNs with kiloparsec-scale physical separation in the local universe 
(z $\lesssim$ 0.2) can be resolved spatially with sub-arcsecond-resolution 
observations. Attempts to find such dual AGNs have recently been 
conducted using various methods.

One of the standard ways of the dual AGN search is to look for AGNs 
with double-peaked narrow emission lines in large optical spectroscopic 
databases (e.g., Sloan Digital Sky Survey; SDSS) \citep{yor00}, as 
kiloparsec-scale-separation dual AGNs with two independent narrow 
line-emitting regions (NLRs) are one scenario that can explain the observed 
double-peaked optical narrow emission lines (e.g.,) 
\citep{wan09,liu10,smi10,pil12,ge12,bar13}.
However, given that other factors, such as outflows, rotating disks, and 
the complex kinematics of NLRs, can also account for the observed 
double-peaked optical narrow emission line properties, 
high-spatial-resolution follow-up observations at other wavelengths 
(e.g., infrared, radio, and X-ray) are needed to confirm or disprove 
the dual AGN scenario (e.g.,) 
\citep{fu11,ros11,she11,tin11,com12,fu12,liu13,mcg15,mul15,liu18,rub19}.
The results from these follow-up studies show that the majority 
of double-peaked optical narrow emission line AGNs can be 
better explained by scenarios other than dual AGNs.
In particular, \citet{rub19} argued that mergers are better indicators 
of dual AGNs than double-peaked optical narrow emission line AGNs.
Numerical simulations predict that strong dual AGN activity occurs 
during the late phases of gas-rich galaxy mergers at small nuclear 
separations \citep{hop06,van12,cap17,ble18}. Under these conditions, 
a large fraction of AGNs can be buried deeply in nuclear gas and dust in 
virtually all lines of sight without well-developed classic NLRs; 
thus, they tend to be optically elusive \citep{mai03,ima06}. 
Optical dual AGN searches have this limitation and so observations 
at wavelengths of stronger gas/dust penetrating power are required 
to properly detect luminous, but buried, dual AGNs.

An AGN usually emits much stronger X-ray emission than a star-forming 
region with the same bolometric luminosity \citep{sha11}.
Hard X-ray observations at $>$2 keV can be particularly powerful for 
investigating optically elusive luminous buried AGNs, due to reduced 
extinction effects, compared to optical and soft X-ray observations 
at $<$2 keV.
In fact, luminous dual AGNs were found through hard X-ray observations 
of several infrared luminous gas/dust-rich major galaxy mergers 
\footnote
{In this paper, we denote mergers of comparable galaxy mass 
(mass ratio $< $4) as major galaxy mergers and mergers of 
largely different galaxy mass (mass ratio with $\gtrsim$4) 
as minor galaxy mergers, following \citet{cap15}.}
with no obvious 
optical AGN signatures (e.g.,) 
\citep{kom03,bal04,bia08,pic10,kos11,kos12,fab11,kos16}, 
providing strong evidence that some fraction of such galaxy mergers 
actually contain optically elusive luminous dual AGNs.
However, in many infrared luminous gas/dust-rich galaxy mergers, X-ray 
emission from putative luminous buried AGNs is not clearly detected 
or is only marginally detected when it is difficult to securely estimate 
intrinsic AGN luminosity from model fitting of limited X-ray 
photon number spectra (e.g.,) \citep{ten09,iwa11,liu13,ric17}. 
A large number of dual AGNs in gas/dust-rich galaxy mergers could be 
missed, even with hard X-ray observations.

Infrared $>$2 $\mu$m observations can be another effective tool for
investigating optically elusive, luminous dual buried AGNs in 
gas/dust-rich galaxy mergers, thanks to much smaller dust extinction 
effects than those of optical observations \citep{nis08,nis09}.
More importantly, AGN activity can be differentiated from star-forming 
activity based on infrared observations.
Because the radiative energy generation efficiency of an AGN 
(mass accreting SMBH; 6--42\% of Mc$^{2}$) \citep{bar70,tho74} is much 
higher than that of star formation (nuclear fusion inside stars; 
$\sim$0.7\% of Mc$^{2}$), in an AGN, large luminosity can 
come from a compact area and an emission surface brightness can be very high.
An AGN can create a larger amount of hot ($>$100 K) dust than 
star-forming activity, and so the infrared spectral shape and color 
can be largely different between these two kinds of activity, 
in particular in the near- to mid-infrared (2--8 $\mu$m) range (e.g.,) 
\citep{ris06,alo06,ima08,ima10b,ris10,jar11,ste12,lee12,mat12}.
In fact, infrared observations have revealed optically elusive, but 
luminous buried AGNs in many late-stage infrared-luminous gas/dust-rich 
galaxy mergers, by distinguishing from star-formation, based on 
infrared colors at 3--5 $\mu$m \citep{sat14}.
In such mergers, it is often the case that infrared observations have 
successfully provided the signatures of luminous buried AGNs that 
are elusive, even in the hard X-ray regime (e.g.,) 
\citep{alo06,arm07,ima07,ima08,vei09,ten09,nar10,nar11,ten15}.
This is possibly due to the presence of a large column density of 
dust-free, X-ray-absorbing gas around luminous AGNs inside the dust 
sublimation radius, which can make X-ray absorption substantially 
greater than that expected from dust extinction in the infrared and 
Galactic extinction curve (e.g.,) \citep{alo97,gra97,geo11,bur15,ich19}.

Using the WISE 3.4-$\mu$m and 4.6-$\mu$m infrared colors, 
\citet{sat17} have attempted to detect luminous dual AGNs in infrared 
luminous gas/dust-rich galaxy mergers.
However, given the limited spatial resolution of WISE ($\sim$6$''$) 
\citep{wri10}, these WISE data only indicate that luminous buried AGNs 
are present at least in one galaxy nucleus, and infrared colors of small 
separation galaxy nuclei cannot be constrained individually. 
Follow-up ground-based near-infrared $<$2.5 $\mu$m spectroscopy and/or 
hard X-ray $>$2 keV observations with higher spatial resolution are 
needed to confirm luminous dual AGNs \citep{sat17,pfe19}.
For near-infrared spectroscopic AGN confirmation, because high excitation 
coronal emission lines are used, deeply buried AGNs without well 
developed NLRs can still be missed.
It is highly desirable to conduct high-spatial-resolution infrared 
observations to identify directly AGN-like infrared colors in individual 
merging galaxy nuclei separately, to obtain a better census of luminous 
buried AGNs in infrared luminous gas/dust-rich galaxy mergers.   

\citet{ima14} conducted ground-based adaptive-optics (AO)-assisted, 
high-spatial-resolution ($<$0$\farcs$3) infrared 
$K'$- (2.1 $\mu$m) and $L'$-band (3.8 $\mu$m) observations of nearby 
infrared luminous gas/dust-rich galaxy mergers at $z <$ 0.22 using 
the Subaru 8.2 m telescope, as much higher spatial resolution is 
achievable, particularly in the $L'$-band (3.8 $\mu$m), than with 
any infrared satellite so far launched. 
\citet{ima14} have argued that AGNs, including deeply buried ones, and 
star-forming regions are distinguishable from the infrared $K'-L'$ 
colors, as an AGN should show much redder $K'-L'$ color 
\citep{iva00,alo03,vid13}. 
For stellar emission in a normal star-forming region, the infrared 
$L'$-band flux is usually much smaller than the $K'$-band flux; however, 
$L'$-band emission is strong in a luminous AGN because of AGN-heated 
hot ($>$100 K) dust emission (Figure 1).
This excess $L'$-band emission by AGN-heated hot dust is so strong that 
the infrared $L'$-band to bolometric luminosity ratio in a pure luminous 
AGN is nearly two orders of magnitude higher than in pure 
star formation \citep{ris10}.
An AGN with a modest bolometric contribution (e.g., $\sim$20\%) is 
detectable from a significantly redder $K'-L'$ color than star formation 
\citep{ima14}.
Most importantly, this red $K'-L'$ color method should be sensitive 
to an optically elusive luminous buried AGN without well-developed NLRs, 
as we probe the excess $L'$-band emission coming from AGN-heated hot dust 
at the inner part of the obscuring material 
\footnote{
This excess $L'$-band emission should be strong not only for optically 
elusive luminous {\it buried} AGNs surrounded by dust with a large 
covering factor, but also for {\it unobscured} AGNs as long as some 
amount of nuclear dust is present in directions perpendicular to our 
line of sight.  
}. 
Unlike rest frame optical wavelengths, there are no extremely strong 
emission and absorption features at infrared 2--5 $\mu$m 
(the rest frame).
The modestly strong 3.3 $\mu$m polycyclic aromatic hydrocarbons 
(PAH) emission feature is usually seen in star-formation 
(Figure 1, left) (rest equivalent width $\lesssim$ 0.15 $\mu$m; 
\citet{ima08,ima10b}).
The 3.4 $\mu$m carbonaceous dust absorption feature is detected in 
some fraction of obscured AGNs with $\sim$50\% dip in the strongest case 
(e.g.,) \citep{imd00,ris03,ima06,ima08}.
However, the effects of these features to the observed $L'$-band flux 
are limited ($<$0.2 mag).
Thus, as long as we observe nearby ($z \lesssim$ 0.2) galaxies, 
it is very unlikely that $K'-L'$ color can change drastically 
only over a particular redshift range, due to emission 
or absorption features.
This method was applied to 29 nearby infrared luminous gas/dust-rich  
galaxy mergers and successfully detected four dual AGN candidates 
\citep{ima14}. 
However, because of the limited sample size, our observational 
constraints on the ubiquity of luminous dual AGNs in 
gas/dust-rich galaxy mergers still have room for improvement. 
In this paper, we extend our successful approach to a larger number 
of objects to better understand optically elusive deeply 
buried luminous AGNs in infrared luminous gas/dust-rich galaxy mergers.

Throughout this paper, quoted magnitudes are based on the Vega system, 
and we adopt H$_{0}$ $=$ 71 km s$^{-1}$ Mpc$^{-1}$, $\Omega_{\rm M}$ = 0.27, 
and $\Omega_{\rm \Lambda}$ = 0.73.
The luminosity distance (in Mpc) and the relationship between the physical 
and apparent scales (in kpc arcsec$^{-1}$) under these cosmological 
parameters are obtained using the calculator provided by \citet{wri06}. 

\section{Targets} 

In the nearby universe at $z \lesssim$ 0.2, ultraluminous infrared 
galaxies (ULIRGs) with infrared luminosity 
L$_{\rm IR}$ $\gtrsim$ 10$^{12}$L$_{\odot}$ are mostly gas/dust-rich galaxy 
mergers (e.g.,) \citep{san88,cle96,mur96,duc97}.
\citet{ima14} selectively observed 23 well-studied nearby 
ULIRGs in the IRAS 1 Jy sample \citep{kim98}, which are relatively 
bright at the observed infrared $K'$- and $L'$-bands, and additional six 
interesting galaxies with slightly lower infrared luminosity 
at L$_{\rm IR}$ $<$ 10$^{12}$L$_{\odot}$.
As a next step, we extend our observations to less studied ULIRGs 
in the IRAS 1 Jy sample at larger distance and with fainter flux 
at the same observed bands.
Even using the world's largest ground-based 8--10 m telescopes, 
it is very difficult to significantly detect $L'$-band 
emission fainter than $L'$ $>$ 14 mag in Vega because of large 
atmospheric background noise from Earth. 
We thus exclude ULIRGs whose WISE 3.4-$\mu$m magnitudes are fainter 
than 14 mag in Vega \citep{wri10}.
This criterion poses some bias against ULIRGs without luminous buried 
AGNs, because $>$3 $\mu$m emission is much brighter for luminous AGNs 
than for star-forming regions when normalized at the bolometric 
luminosity.
Thus, our sample is not statistically unbiased in any sense. 
Table 1 summarizes our observed ULIRG sample (17 sources).
In addition to ULIRGs with no obvious optical AGN signatures 
(i.e., LINER, HII-region, and unclassified types in Table 1), 
those with optically identified AGNs (Seyfert 1 and 2 types) are 
included. 
Verifying red $K'-L'$ colors at the primary galaxy nuclei of these 
optically identified AGNs will further strengthen our proposed $K'-L'$ 
color method as a tool for detecting luminous AGNs, including optically 
elusive buried ones.
For both optically AGN-type and non-AGN-type ULIRGs, our primary 
scientific goal is to investigate whether luminous dual AGNs are 
common using our proposed potentially powerful infrared approach.
In particular, achieved high spatial resolution with $<$0$\farcs$3 
assisted by the adaptive optics (AO) of ground-based 8--10 m telescopes 
will allow the discovery of small-separation luminous dual AGNs that 
may not be resolvable, even with X-ray data of the highest spatial 
resolution ($\sim$0$\farcs$5) provided by the Chandra satellite.

\section{Observations and Data Analyses} 

We used the $K'$- (2.1$\pm$0.2 $\mu$m) and $L'$-band 
(3.8$\pm$0.4 $\mu$m) filters of the infrared camera and spectrograph 
(IRCS) \citep{kob00} of the Subaru 8.2 m telescope atop Mauna Kea, 
Hawaii \citep{iye04}, to conduct our infrared observations.
The 188-element adaptive optics (AO) system, which employs laser-guide 
stars (LGS) or natural-guide stars (NGS) \citep{hay08,hay10}, was used 
to achieve higher spatial resolution ($<$0$\farcs$3) than natural seeing 
(0$\farcs$4--1$\farcs$0 in the $K'$- and $L'$-bands).
We chose the LGS-AO mode whenever possible; however, NGS-AO was used 
for observations in 2015 September and 2018 May (Table 2) because of 
technical issues with LGS-AO.
For LGS-AO, a star or compact object brighter than 18--19 mag in the 
optical $R$-band (0.6 $\mu$m) within $\sim$90$''$ from the target was 
needed as a guide star for tip-tilt correction. 
The AO correction itself was made with a laser spot created by the LGS-AO 
system with an optical $R$-band magnitude of $\sim$11--14.5 mag, 
depending on the target elevation, Earth's atmospheric conditions, and 
LGS-AO system performance.
The laser spot magnitude was generally fainter than those of our previous 
observations before 2013 \citep{ima14} because of degraded 
LGS-AO performance.
For NGS-AO, a star or compact source brighter than $R$ = 16.5 mag 
within $\sim$30$''$ from the target was necessary as a guide star for 
reasonable AO correction.
Table 2 summarizes the details of our observations, including 
standard stars for photometric calibration and guide stars for tip-tilt 
or AO correction. 

During the observations of the ULIRGs in Table 2, the sky was clear.
For our $K'$-band observations, the 52 mas (52.77 mas pixel$^{-1}$) imaging 
mode was employed. 
The field of view was 54$\farcs$04 $\times$ 54$\farcs$04 for 
the full-array mode (1024 $\times$ 1024 pixels$^{2}$). 
For our $L'$-band observations, we used the 20 mas (20.57 mas pixel$^{-1}$) 
imaging mode to avoid saturation by large background emission 
from Earth's atmosphere.
The field of view was 21$\farcs$06 $\times$ 21$\farcs$06 in the 
full-array mode.  
Even with the 20 mas mode, a subarray mode (768 $\times$ 768 pixels$^{2}$ 
or 896 $\times$ 896 pixels$^{2}$) was necessary to avoid 
saturation for some ULIRGs, depending on conditions (e.g., Earth's 
atmospheric temperature and/or precipitable water vapor above the 
observation site). 
For the $L'$-band observations, because object signals are more difficult 
to recognize in short exposure images due to large Earth's 
background noise compared to the $K'$-band, we first took a $K'$-band 
image using the 20 mas mode, moved the ULIRG's nuclear emission to the 
center of the array, and then inserted the $L'$-band filter.
 
For the $K'$-band observations of ULIRGs, the exposure times were 
1--30 sec, and 2--60 coadds were applied.
For the $L'$-band observations, the exposure times were 0.07--0.12 sec, 
and there were 250--400 coadds. 
The individual exposure times were determined to set signal levels at 
the object positions well below the linearity level of the IRCS imaging  
array ($\lesssim$4000 ADU).
We used nine-point dithering patterns for ULIRGs in both the $K'$- and 
$L'$-bands to correct for the effects of bad pixels by placing object 
signals at nine positions on the array.
This nine-point dithering sequence was repeated multiple times for faint 
ULIRGs when necessary.
For $L'$-band fainter ULIRGs, we generally integrated for a longer time. 
In every observation run, photometric $K'$- and $L'$-band standard stars 
(Table 2) were observed for flux calibration.

We adopted standard data analysis procedures using IRAF
\footnote{IRAF is distributed by the National Optical Astronomy
Observatories, which are operated by the Association of Universities
for Research in Astronomy, Inc. (AURA), under cooperative agreement
with the National Science Foundation.}.
First, we created median-combined sky frames from the nine-point dithered 
data set to make a sky flat image after masking the positions of 
bright objects and bad pixels.
Second, we subtracted sky emission from individual frames and divided 
the resulting images with the sky flat frames for flat fielding.
Third, we shifted the sky-subtracted, flat-fielded images to make the
peak position of the target on the same pixel in the array using 
the emission of compact bright sources within the field of view whenever 
available.
However, no appropriate compact emission required for accurate 
estimate of this shift was found in the $L'$-band data of some 
fraction of ULIRGs, for which offset values were calculated from 
the input values of the dithering amplitude and pixel scale of the 
instrument.
We then average-combined the shifted frames and obtained final images.
For ULIRGs without bright compact objects within the field of view, 
the final image size of compact emission could have been affected by 
possible mechanical pointing errors of the Subaru Telescope at each 
dithering position, in addition to Earth's atmospheric seeing, and thus 
could have been slightly worse than ULIRGs with bright compact objects 
within the field of view.

For standard star photometry, we adopt a 2$\farcs$5 radius aperture 
to recover almost all emission flux.
For ULIRGs, however, the aperture size for flux measurements has 
to be carefully considered.
Our science goal is to search for AGN-originated compact, red $K'-L'$ 
(= large $K'-L'$ value) emission at galaxy nuclei by minimizing 
contamination from spatially extended ($>$kpc) star formation 
emission in the host galaxy.
The smallest possible aperture size is preferred, as long as the bulk 
of the compact nuclear emission is recovered.
However, this is not a trivial task for ground-based AO data, 
because, in addition to a spatially compact ($<$0$\farcs$3) AO-corrected 
core component, a seeing-sized spatially extended AO halo component 
is also present. 
If we set the aperture to a seeing size, almost all compact nuclear 
emission can be recovered, but at the same time, significant contamination 
from spatially extended ($>$kpc) star formation emission in the host 
galaxy will be included, in particular in the $K'$-band, which can make 
the observed $K'-L'$ color bluer than the intrinsic color of the 
compact nuclear emission.
This systematic uncertainty coming from the choice of aperture size has  
to be considered more carefully than statistical photometric uncertainty, 
which is usually much smaller in the $K'$-band and is also small in 
the $L'$-band (except for $L'$-band very faint ULIRGs).  

We investigated the growth curve of signals as a function of aperture 
size using AO-corrected data of standard stars and confirmed that 
$\gtrsim$75\% of point source emission was recovered with 0$\farcs$5 radius 
aperture measurements, as in the case of \citet{ima14}.
For standard stars, which are very bright in the optical, their AO 
correction is generally good. 
However, it is not obvious whether similarly good AO correction is achieved 
for ULIRGs, because AO guide stars are generally fainter than 
bright standard stars in the optical and there is some separation 
between AO guide stars and target ULIRGs.
We created the growth curve of signals using bright, compact objects 
(compact ULIRGs themselves and/or compact sources other than the target 
ULIRGs whenever available) inside the field of view of ULIRG data 
taken from 2015 to 2019; these are shown in Appendix A.
We confirmed that $\gtrsim$75\% of the compact emission was usually 
recovered with a 0$\farcs$5 radius aperture. 
We thus make photometry of the ULIRG's compact nuclear emission with 
a 0$\farcs$5 radius aperture, where possible photometric uncertainty 
is estimated to be $\lesssim$0.3 mag (i.e., the difference between 
$\gtrsim$75\% and 100\% recovery).
For $L'$-band undetected faint ULIRG nuclei, we assume that their radial 
emission profile is similar to that of $L'$-band detected bright ULIRG 
nuclei within the same field of view or in a different image taken 
on the same night (for ULIRGs with no $L'$-band detection in any nuclei).
This assumption is reasonable, because ULIRG's nuclear $L'$-band 
emission is usually dominated by a compact component, with minimum 
contamination from spatially extended star formation emission 
(Figure 2). 
Because the same guide stars were used for the $K'$- and $L'$-band 
observations of each ULIRG, the possible photometric uncertainty of 
the $K'-L'$ colors of the ULIRG's compact nuclear emission 
is expected to be smaller than the uncertainty in the individual 
$K'$- and $L'$-band photometry of the emission, say $<$0.2 mag 
(i.e., the difference in compact emission signal recovery within 
the 0$\farcs$5 radius aperture between $K'$ and $L'$).

The laser spot magnitudes during ULIRG observation (Table 1) were 
fainter than those in 2011--2013; thus, the AO correction 
may have been slightly worse than in our previous study \citep{ima14}.
We may have lost a larger fraction of the ULIRG's compact nuclear emission 
than in \citet{ima14} with 0$\farcs$5 radius aperture photometry.
We thus use a slighter larger 0$\farcs$75 radius aperture as well.
If compact red $K'-L'$ emission is detected in both the 
0$\farcs$5 and 0$\farcs$75 radius aperture photometry, we will be able 
to argue strongly about the presence of a $L'$-band continuum emitting, 
luminous AGN. 
Remaining possible uncertainties of $K'-L'$ color in our AO data 
measurements of ULIRG nuclei are explained in Appendix B.

In summary, the derived $K'-L'$ color of the compact nuclear 
emission of a ULIRG is expected to have systematic uncertainty 
of up to 0.2 mag or so.
This level of uncertainty will not significantly affect our discussion 
of the presence of a luminous AGN, because the difference in 
$K'-L'$ color between pure star formation ($K'-L'$ $<$ 1 mag) 
\citep{hun02} and a pure luminous AGN ($K'-L'$ $\sim$ 2 mag) 
\citep{iva00,alo03,vid13} is much larger than $\sim$0.2 mag.

\section{Results} 

Figure 2 displays infrared $K'$- (2.1 $\mu$m) and  
$L'$-band (3.8 $\mu$m) images of observed ULIRGs. 
Although all ULIRGs are clearly detected in the $K'$-band, the 
detection rate in the $L'$-band is lower because of the lower sensitivity 
due to the larger Earth's background noise in ground-based observations.

In the $K'$-band, multiple emission components are evident in a 
large fraction (12 out of 17) of observed sources, which suggests 
the presence of multiple merging galaxy nuclei and supports the widely 
accepted merger origin scenario of nearby ULIRGs \citep{san88,tan98}.
In particular, for IRAS 09039$+$0503, IRAS 10035$+$2740, 
IRAS 12072$-$0444, and IRAS 15206$+$3342, although multiple nuclear 
emission with small separation was not identified in previously taken 
seeing-limited imaging data \citep{kim02}, our AO-assisted 
high-spatial-resolution imaging data clearly resolve multiple 
nuclear emission (Figure 2).
For IRAS 09039$+$0503, ALMA high-spatial-resolution ($<$0$\farcs$2) 
continuum emission data at $\sim$1.2 mm also reveal two nuclear 
components with $\sim$0$\farcs$5 separation along the southwest (SW) 
to northeast (NE) direction, with the SW nucleus significantly 
brighter than the NE nucleus, in a similar way to our infrared 
$K'$-band data \citep{ima19}.

In the $L'$-band, multiple nuclear emission is clearly seen only in 
IRAS 10190$+$1322, IRAS 12072$-$0444, and IRAS 12112$+$0305 (Figure 2).
Table 3 summarizes nuclear $K'$- and $L'$-band magnitudes and 
$K'-L'$ color in 0$\farcs$5 and 0$\farcs$75-radius aperture 
measurements. 

\section{Discussion} 

\subsection{Luminous AGNs in individual ULIRG nuclei}

We proceed our discussion of AGN contribution to the observed 
compact nuclear $L'$-band flux, following to that presented 
in \citet{ima14}.
Although the intrinsic $K'-L'$ colors are slightly different among  
individual star formation and AGNs, we adopt $K'-L'$ = 0.5 mag 
for pure star formation \citep{hun02} and $K'-L'$ = 2.0 mag for 
a pure luminous AGN \citep{iva00,alo03,vid13}.
The observed nuclear $K'-L'$ color is expected to increase 
(become redder) with increasing AGN contribution.
As discussed in \citet{ima14}, a red $K'-L'$ color is 
primarily caused by hot dust emission heated by a luminous AGN 
rather than dust reddening of star formation, whose possible effect 
is estimated to be limited ($<$0.3 mag for A$_{\rm V}$ = 10 mag dust 
extinction of star formation).
We calculate the contribution of star formation ($K'-L'$ = 0.5 mag) 
and an AGN ($K'-L'$ = 2.0 mag) to the observed compact nuclear 
$L'$-band emission, to reproduce the observed $K'-L'$ colors measured 
with the 0$\farcs$5 aperture.
For example, when the observed $K'-L'$ color is 
$\lesssim$0.5, 0.8, 1.0, 1.5, $\gtrsim$2.0 mag, the AGN contribution 
to the observed compact nuclear $L'$-band flux is estimated to be 
0, 32, 50, 80, 100\%, respectively \citep{ima14}.
The estimated AGN contribution in individual ULIRG nuclei is 
summarized in Table 3 (column 8). 
We classify ULIRG nuclei with $K'-L'$ $>$ 1.0 mag as those 
containing luminous AGNs, because $>$50\% AGN contribution 
is indicated.

In Table 3, for a large fraction of ULIRGs, the observed 
$K'-L'$ colors are $>$1.0 mag at the $K'$-band brightest 
primary nuclei, which suggests the presence of luminous AGNs there.
In particular, red $K'-L'$ colors with $>$1.0 mag are confirmed at the 
primary nuclei of all four ULIRGs with optically identified AGNs 
(i.e., optical Seyferts in Table 1; IRAS 08559$+$1053, 
IRAS 12072$-$0444, IRAS 14394$+$5332, and IRAS 21219$-$1757), 
which supports the validity of our method.
For ULIRGs whose 2--5 $\mu$m emission is dominated by single 
nuclei, WISE $W1$ (3.4 $\mu$m) and $W2$ (4.6 $\mu$m) photometry with 
larger beam sizes ($\sim$6$''$) can also be used to identify luminous 
AGNs in the primary galaxy nuclei based on $W1-W2$ $>$ 0.8 mag 
\citep{ste12}.
Table 4 summarizes the WISE $W1-W2$ colors of observed ULIRGs.
For all ULIRGs with WISE-classified luminous AGNs, 
our higher-spatial-resolution AO imaging data support the presence of 
luminous AGNs from red $K'-L'$ colors ($>$1.0 mag), which reinforces 
the argument that our $K'-L'$ color-based AGN selection is effective and 
highly complete.
For two ULIRGs (IRAS 09039$+$0503 and IRAS 13539$+$2920), 
luminous AGN signatures are not clearly seen in WISE data but are found 
in our AO data (Table 4), possibly because our higher-spatial-resolution 
AO data are more sensitive to luminous AGNs, by probing only compact 
nuclear regions with reduced contaminations from spatially extended 
star formation emission in the host galaxies.

Of three ULIRGs with multiple detected $L'$-band emission 
(IRAS 10190$+$1322, IRAS 12072$-$0444, and IRAS 12112$+$0305) ($\S$4), 
only IRAS 12072$-$0444 has two galaxy nuclei with $K'-L'$ color 
significantly redder than 1.0 mag (Table 3); thus, it is classified 
as a dual AGN by our definition.
Since the two red $K'-L'$ nuclei of IRAS 12072$-$0444 are found 
in two independent data using different AO guide stars with largely 
different properties (Table 2), the dual AGN classification should 
be solid.  
IRAS 12112$+$0305 also shows two galaxy nuclei with $K'-L'$ $=$ 1.1 mag 
(0$\farcs$5 radius aperture) (Table 3); given the possible 
systematic uncertainty of $\lesssim$0.2 mag in the $K'-L'$ color ($\S$3), 
the dual AGN classification of IRAS 12112$+$0305 is only 
provisional.
However, because the AO guide star of IRAS 12112$+$0305 is relatively 
bright among the observed ULIRGs (Table 2), it is expected that 
a large fraction of compact emission resides in the AO core component 
and that the uncertainty in the measured $K'-L'$ color is small. 
We thus classify IRAS 12112$+$0305 as a possible dual AGN candidate.
For IRAS 10190$+$1322, the $K'-L'$ colors of both galaxy nuclei are 
$<$1.0 mag; thus, there is no clear indication of a luminous AGN 
in any galaxy nucleus.

In other ULIRGs in Table 3, although many primary  
galaxy nuclei show $K'-L'$ $>$ 1.0 mag, $K'$-band fainter secondary 
galaxy nuclei are mostly undetected in the $L'$-band, which precludes 
meaningful constraints on the $K'-L'$ colors and identification of 
dual AGNs. 

In summary, our infrared $K'$- and $L'$-band AO-assisted 
high-spatial-resolution imaging observations have revealed a strong 
dual AGN candidate in IRAS 12072$-$0444 and another possible candidate 
in IRAS 12112$+$0305.
These ULIRGs have not previously been recognized as dual AGNs and so 
are infrared-identified dual AGNs by our observations.
The infrared-detected dual AGN fraction in our new nearby ULIRG 
sample is only $\sim$12\% (= 2/17). 
\citet{ima14} previously used the same infrared method and 
found a low dual AGN fraction in nearby merging 
ULIRGs ($\sim$13\% = 3/23) and less infrared luminous galaxies 
($\sim$17\% = 1/6). 
Our new and previous results together suggest that the 
infrared-detected dual AGN fraction in nearby merging ULIRGs is much 
less than unity ($<$20\%).

In \citet{ima14} and this paper, 19 out of 23 and 
12 out of 17 ULIRGs show spatially-resolved $K'$-band multiple nuclear 
components, respectively.
We also estimate the infrared-detected dual AGN fraction relative to 
ULIRGs with $K'$-band resolved multiple nuclei, because there may exist 
luminous dual AGNs which are not spatially resolved simply due to 
limited spatial resolution of our imaging data.
The fraction is $<$20\% (= 5/31), still much smaller than unity.
Given the small dust extinction effects in the infrared $>$2 $\mu$m, 
we regard it as unlikely that the dominant fraction of putative 
luminous AGNs is missed because of dust obscuration.
It may be that not all SMBHs in merging ULIRGs become sufficiently 
luminous AGNs to be detectable through our infrared search. 
In the next subsection, we investigate the activation of SMBHs 
in multiple galaxy nuclei separately.

\subsection{Activation of SMBHs in the nuclei of individual ULIRGs}

For ULIRGs in which $K'$-band emission is detected in multiple nuclei 
and $L'$-band emission is detected in at least one nucleus, 
we compare in Figure 3(a) the ratio of $K'$-band luminosity within 
central 4 kpc in diameter \citep{kim02} and that of nuclear $L'$-band 
luminosity between the primary and secondary 
galaxy nucleus, following \citet{ima14}.
These ratios are shown in Table 5.
 
The $K'$-band luminosity ratio is taken as the $K'$-band stellar 
emission luminosity ratio, which can be converted into the central SMBH 
mass ratio between the primary and secondary galaxy nuclei given the 
correlation between $K'$-band stellar emission luminosity and the 
central SMBH mass in galaxies \citep{mar03,vik12}.
Nearby ULIRGs usually consist of (1) energetically-dominant,  
compact ($<$500 pc), highly obscured nuclear regions and (2) 
energetically-insignificant spatially extended stellar emission 
in the host galaxies \citep{soi00,dia10,ima11}.
While the compact nuclei of nearby ULIRGs can be extreme, the properties 
of the spatially extended stellar emission in the host galaxies are 
not so different from those of normal galaxies.
Additionally, excluding two unobscured luminous AGNs, IRAS 21219$-$1757 
and Mrk 231 in our new sample and \citet{ima14}, respectively, 
the central 4 kpc diameter $K'$-band luminosity is expected to largely 
come from the host galaxy stellar emission, because AGN-origin $K'$-band 
emission from the highly obscured compact nuclei is flux attenuated.
In fact, the 4 kpc diameter $K'$-band luminosity \citep{kim02} is much 
brighter than our nuclear 0$\farcs$5 radius aperture $K'$-band luminosity 
in a large fraction of observed ULIRG nuclei (Table 3 and \citet{ima14}).
We thus use the 4 kpc diameter $K'$-band luminosity to roughly 
estimate SMBH masses at ULIRG nuclei, because this is currently only 
one practical way to do so in a large number of sources.
If AGN-origin $K'$-band emission were important for the 4 kpc 
diameter $K'$-band luminosity particularly in ULIRG nuclei with 
luminous AGNs (= active massive SMBHs), SMBH masses would be 
overestimated and AGN luminosity normalized by SMBH masses would be 
underestimated in such nuclei. 
This will not alter our main discussion in the subsequent paragraphs of 
this subsection.

The nuclear $L'$-band luminosity ratio traces the AGN luminosity ratio 
between the primary and secondary galaxy nuclei, particularly for 
active SMBHs because their $L'$-band fluxes predominantly come from 
AGN-heated hot dust emission ($\S$1).
In both the $K'$- and $L'$-band luminosity ratios, luminosity 
at the primary galaxy nucleus is 
divided by that at the secondary galaxy nucleus. 
If SMBHs in multiple galaxy nuclei are simultaneously activated with 
comparable SMBH-mass-normalized AGN luminosity, 
such ULIRGs are expected to be located around the solid straight line 
in Figure 3(a).
If an SMBH at the primary nucleus of a merging ULIRG is more active 
with higher SMBH-mass-normalized AGN luminosity than that at the 
secondary nucleus, such a ULIRG should show a larger $L'$-band 
luminosity ratio than the $K'$-band luminosity ratio between two galaxy 
nuclei and so should be plotted above the solid line.
In Figure 3(a), although the $L'$-band luminosity ratios are only lower 
limits in some multiple nuclei ULIRGs below the solid line, the overall 
trend is consistent with the scenario that more massive SMBHs at the 
primary nuclei are generally more active with higher 
SMBH-mass-normalized AGN luminosity than less massive SMBHs at 
the secondary nuclei in nearby merging ULIRGs.
Namely, SMBH activation in nearby merging ULIRGs is asynchronous, 
as suggested from previous infrared observations by \citet{ima14}.

In Figure 3(b), we use the AGN-origin nuclear $L'$-band luminosity ratio, 
after excluding possible stellar contaminations to the observed nuclear 
$L'$-band emission, in the ordinate. 
This correction can have significant effects, compared to the distribution 
in Figure 3(a), particularly for less active SMBHs, because their 
$L'$-band emission can be significantly contaminated 
by non-AGN components.
The overall trend of higher activity for more massive SMBHs in the 
primary galaxy nuclei is still evident in a similar way as in Figure 3(a).

In Figure 4, we investigate the ratio of SMBH activation between the 
primary and secondary galaxy nuclei as a function of projected nuclear  
separation in kiloparsecs (Table 5), because numerical simulations 
\citep{hop06,van12,cap17,ble18} and observations \citep{kos12,kos18} 
suggest that AGN activity can be particularly strong in late-stage, 
gas-rich galaxy mergers. 
As the ordinate is only the lower limit for a large fraction of the 
observed sources, it is not easy to assign a strong constraint.
However, in merging ULIRGs with small projected nuclear separation 
($<$14 kpc) (possibly biased to a late merging stage), our results are 
consistent with the scenario that SMBHs in primary galaxy nuclei tend 
to be more active with higher SMBH-mass-normalized AGN luminosity than 
SMBHs in secondary galaxy nuclei.
It is not clear from our data whether this trend is true even for ULIRGs 
with large projected nuclear separation ($>$14 kpc), because all such 
sources show only lower limits in the ordinate.

In summary, our infrared $K'$- and $L'$-band high-spatial-resolution 
($<$0$\farcs$3) imaging observations suggest that activation of multiple 
SMBHs in merging ULIRGs is not synchronous, which supports the main features 
of numerical simulations of gas-rich major galaxy mergers 
\citep{van12,cap17}.
We also find that SMBH activation (SMBH-mass-normalized AGN 
luminosity) is generally higher in primary galaxy nuclei 
(hosting more massive SMBHs) than secondary galaxy 
nuclei (hosting less massive SMBHs).
This trend will be even strengthened if possible overestimations of 
SMBH masses in the primary nuclei hosting active massive SMBHs 
(paragraph 2 of this subsection) are corrected.
The low activation of a less massive SMBH makes $L'$-band detection 
of AGN emission in the secondary galaxy nuclei difficult and 
lowers the detection rate of dual AGN.
Numerical simulations of major galaxy mergers also predict that 
a sufficient amount of gas can be fed, by angular momentum removal, 
toward more massive SMBHs of primary galaxy nuclei, which can 
preferentially trigger luminous AGN (and possibly starburst) activity 
there compared to less massive SMBHs of secondary galaxy 
nuclei \citep{cap15,cap17}.
Our observational results of SMBH activation in nearby merging ULIRGs 
are reproduced by these numerical simulations.
We note that higher SMBH activation for less massive SMBHs was argued 
for optically selected dual AGNs, many of which are minor galaxy mergers 
\citep{com15}.
This trend is different from ours for ULIRGs (i.e., gas-rich major galaxy 
mergers including many optically elusive, buried AGNs).  
\citet{cap15} predicts that in minor galaxy mergers, less massive SMBHs 
can be more active, because a companion galaxy is not massive enough to 
significantly affect the gas dynamics around a primary SMBH.
These various observational trends of SMBH activation for dual AGNs 
can be explained under different galaxy merger properties.

\subsection{Infrared elusive dual AGNs?}

For IRAS 12112$+$0305, which is classified optically as a non-AGN 
(LINER) (Table 1), our infrared $K'$- and $L'$-band imaging observations 
detect double nuclear emission, with the SW nucleus being 
brighter than the NE nucleus in both infrared bands 
(Figure 2 and Table 3).
Both the infrared brighter SW and fainter NE nuclei have $K'-L'$ colors 
of $\sim$1.1 mag, which tentatively indicates the possible presence of 
a luminous buried AGN in both galaxy nuclei (see $\S$5.1).
The infrared estimated AGN luminosity is higher in the SW nucleus 
than in the NE nucleus (Table 3, column 9).
However, ALMA (sub)millimeter observations reveal that the 
NE nucleus of IRAS 12112$+$0305 is brighter than the SW nucleus 
in the $\sim$0.9 mm and $\sim$1.2 mm continuum as well as dense 
molecular rotational J-transition lines (i.e., HCN and HCO$^{+}$) 
\citep{ima16b,ima18,ima19}.
The infrared fainter, but (sub)millimeter brighter, NE nucleus shows 
(1) (sub)millimeter dense molecular rotational J-transition line flux 
ratios (i.e., HCN and HCO$^{+}$) often seen in luminous AGNs and 
(2) signatures of fairly strong vibrationally excited HCN J=3--2 and 
J=4--3 emission lines (HCN-VIB) at $\sim$1.2 mm and $\sim$0.9 mm, 
respectively \citep{ima16b,ima18,ima19}, which are naturally explained 
by mid-infrared ($\sim$14 $\mu$m) radiative pumping by AGN-heated 
hot dust emission \citep{sak10,aal15,ima16a,ima17}.
The stronger (sub)millimeter continuum and brighter molecular emission  
lines of plausible AGN origin suggest that a (sub)millimeter-detectable  
buried AGN with higher intrinsic luminosity may be present in the 
infrared fainter NE nucleus of IRAS 12112$+$0305. 

Our infrared $K'$-band imaging observations of IRAS 22491$-$1808 
(optical non-AGN; Table 1) reveal double nuclear emission along 
the east-west direction with the western (W) nucleus brighter than 
the eastern (E) nucleus (Figure 2 and Table 3).
As no significant $L'$-band emission is detected in either galaxy 
nuclei, we cannot meaningfully constrain the presence of a luminous AGN 
from a red $K'-L'$ color.  
However, ALMA (sub)millimeter observations in the continuum and 
dense molecular rotational J-transition lines at $\sim$0.9 mm 
and $\sim$1.2 mm show that the E nucleus is brighter than 
the W nucleus \citep{ima16b,ima18,ima19}.
This E nucleus of IRAS 22491$-$1808 displays similar dense molecular 
rotational J-transition and HCN-VIB line properties to IRAS 12112$+$0305 NE, 
which indicates that 
a (sub)millimeter-detectable, intrinsically luminous buried AGN may be 
present in IRAS 22491$-$1808 E \citep{ima16b,ima18,ima19}.

These two examples suggest that some fraction of dual AGNs could be 
missed or their intrinsic luminosities are underestimated, even in 
infrared $K'$- and $L'$-band observations, 
possibly because of extremely high dust obscuration of the very 
compact nuclei of nearby merging ULIRGs \citep{soi00,dia10,ima11}.
No strong signatures of luminous buried AGNs have been detected in 
2--10 keV hard X-ray observations of these two ULIRGs 
\citep{fra03,ten05,iwa11}.
Such extremely deeply buried AGNs may be detectable only in the 
(sub)millimeter wavelength range because of even smaller dust 
extinction effects than in the infrared and X-ray regimes \citep{hil83}.
In fact, numerical simulations by \citet{roe16} and \citet{ble18} 
suggest that luminous buried AGNs are likely missed in some fraction 
of merging ULIRGs, if not the dominant fraction, through 
infrared-based AGN searches.
The possible presence of such infrared-elusive extremely 
deeply buried AGNs may be an additional factor that lowers the 
detection rate of infrared-identified dual AGNs in the nearby 
merging ULIRG population.

\section{Summary} 

We conducted infrared $K'$- (2.1 $\mu$m) and $L'$-band (3.8 $\mu$m), 
AO-assisted, high-spatial-resolution ($<$0$\farcs$3) imaging 
observations of 17 nearby merging ULIRGs at $z <$ 0.17.   
We searched for compact nuclear red $K'-L'$ emission as the 
signature of luminous AGNs by distinguishing from starbursts, 
as such emission is naturally explained by hot dust radiation 
heated by a luminous AGN.
Given the small dust extinction effects at these infrared wavelengths, 
our method should be sensitive to buried AGNs without 
well-developed classic NLRs photoionized by 
AGN radiation.
We found the following main results.

\begin{enumerate}

\item We detected $K'$-band emission in all observed nearby ULIRGs.
Multiple $K'$-band emission was clearly identified in 
a large fraction ($\sim$71\% = 12/17) of sources, including 
those that had been classified as single nucleus sources in previously 
obtained seeing-limited ($\gtrsim$1$\farcs$0) $K'$-band images.
It was confirmed that multiple merging nuclei are common in 
nearby ULIRGs.

\item $L'$-band emission was also detected in the bulk 
($\sim$88\% = 15/17) of observed 
ULIRGs at the $K'$-band brightest primary nuclei.
$L'$-band emission was clearly detected from $K'$-band fainter  
secondary nuclei in IRAS 10190$+$1322, IRAS 12072$-$0444, and 
IRAS 12112$+$0305, totaling three sources with detected multiple 
emission components in both the $K'$- and $L'$-bands.

\item Of the above mentioned three ULIRGs, IRAS 12072$-$0444 showed 
two emission components whose $K'-L'$ colors were significantly redder 
than those naturally explained by star formation processes, 
which makes this ULIRG a strong dual AGN candidate.
Another ULIRG, IRAS 12112$+$0305, also showed two emission components, 
both of which had red $K'-L'$ colors possibly indicative of luminous 
AGNs.

\item When we combined our new results with those from a previous 
study using the same method \citep{ima14} (totaling 40 sources), 
we found that the fraction of infrared-detected dual AGNs in 
nearby merging ULIRGs is much less than unity ($<$20\%). 

\item For ULIRGs with clearly resolved multiple nuclei, we estimated 
the activation of SMBHs at individual galaxy nuclei using the
SMBH-mass-normalized AGN luminosity.
Our results showed that activation of multiple SMBHs in nearby merging 
ULIRGs is not synchronous, in that more massive SMBHs at primary 
galaxy nuclei are generally more active with higher 
SMBH-mass-normalized AGN luminosity than less massive SMBHs at 
secondary galaxy nuclei.
This is predicted by numerical simulations of gas-rich major galaxy 
mergers in which more efficient gas fueling can happen toward the 
central SMBHs of the primary galaxy nuclei.
The low activation of less massive SMBHs in secondary 
galaxy nuclei makes the intrinsic AGN-origin $L'$-band emission weak and 
makes its detection difficult, which may be partly responsible for 
the low detected dual AGN fraction in nearby merging ULIRGs in our 
ground-based infrared observations.

\item Two ULIRGs, IRAS 12112$+$0305 and IRAS 22491$-$1808, displayed 
multiple $K'$-band emission; however, the longer (sub)millimeter 
wavelength continuum and dense molecular line emission detected with 
ALMA were brighter in the infrared $K'$-band fainter galaxy nuclei 
in both sources.
Dense molecular line observations using ALMA had suggested that 
the $K'$-band fainter nuclei of both ULIRGs may contain infrared 
elusive, but (sub)millimeter-detectable, extremely deeply buried 
luminous AGNs, given even smaller dust extinction effects 
in the (sub)millimeter regime.
The presence of infrared elusive extremely 
deeply buried luminous AGNs in the very dusty nuclei of some nearby 
ULIRGs may also lower the dual AGN detection rate in the infrared. 

\end{enumerate}

Measuring the masses of SMBHs at the extremely highly obscured 
ULIRG nuclei is very difficult.
We used the $K'$-band host galaxy stellar emission luminosity and the 
well-established relation between the luminosity and the central SMBH 
mass in normal galaxies. 
Although this is applicable to many nearby ULIRG nuclei, some ambiguities 
admittedly remain. 
Recently, ALMA very-high-spatial-resolution molecular line observations 
at the almost-dust-extinction-free millimeter wavelength have been 
applied to measure the SMBH mass in the very nearby infrared luminous 
merging galaxy NGC 6240 ($z=$ 0.024) through gas dynamics \citep{med19}.
Our discussion could be verified or improved if such more direct SMBH 
mass estimates are available in a large number of ULIRG nuclei in the near 
future.

\acknowledgments

We thank Drs. Ji Hoon Kim, Yuhei Takagi, and Etsuko Mieda for their 
observing support at the Subaru Telescope, and the referee, 
Dr. Emanuele Nardini, for his valuable comments which helped 
improve the clarity of this manuscript.
M.I. is supported by the Japan Society for the 
Promotion of Science (JSPS) KAKENHI Grant Number 15K05030.
T.K. and S.K. acknowledge supports from JSPS KAKENHI Grant 
Number 17J09016 and 18J11477, respectively.
This publication makes use of data products from the Wide-field 
Infrared Survey Explorer, which is a joint project of the 
University of California, Los Angeles, and the Jet Propulsion 
Laboratory/California Institute of Technology, funded by the 
National Aeronautics and Space Administration, and 
NASA's Astrophysics Data System and the
NASA/IPAC Extragalactic Database (NED) which is operated by the Jet
Propulsion Laboratory, California Institute of Technology, under
contract with the National Aeronautics and Space Administration. 

%

\vspace{5mm}
\facilities{Subaru}






\appendix

\section{Signal growth curve of compact objects within the field of 
view of ULIRG data}

We investigated the signal growth curve of modestly bright 
compact objects (stars, possibly very compact galaxies, and 
point-source-like very compact ULIRGs), 
whenever available, inside the field of view of several ULIRG's data.
This can be used to estimate what fraction of the point source signal 
is recovered with the 0$\farcs$5 and 0$\farcs$75 radius 
apertures under the AO correction at the time of ULIRG observations.
This fraction can be applied to the compact nuclear emission components 
of ULIRGs.
In practice, this estimate is possible only for selected ULIRGs that 
have appropriate compact objects inside the field of view, not for 
all ULIRGs. 
Also, the fraction can vary among ULIRGs depending on AO correction.
However, we can roughly estimate what fraction of compact nuclear 
emission components of ULIRGs are recovered consistently 
with 0$\farcs$5 and 0$\farcs$75 radius aperture photometry in our data.
Figures 5(a), (b), and (c) display the signal growth curve 
for LGS-AO data in the $K'$-band, NGS-AO data in the $K'$-band, and 
NGS-AO data in the $L'$-band, respectively.
We can confirm that $\gtrsim$75\% and $\gtrsim$85\% of compact 
emission is usually recovered with the 0$\farcs$5 and 0$\farcs$75 
radius aperture photometry, respectively, for our AO data of the 
observed ULIRGs.

\section{Remaining possible photometric uncertainty of K'-L' color 
in our AO data measurements of compact emission at ULIRG nuclei}

We have a few caveats about the derived $K'-L'$ colors of the compact 
nuclear emission components of ULIRGs.
First, as mentioned in $\S$3, small aperture photometry can better 
probe the $K'-L'$ color of compact nuclear emission affected by 
luminous AGNs, with reduced contamination from spatially extended 
star formation emission, but may lose a larger fraction of signals 
in the AO halo component of compact emission.  
When we increase the aperture size, the possible flux loss of the AO 
halo component will be smaller. However, the star formation contamination  
will increase, in particular in the $K'$-band, which makes the 
observed $K'-L'$ color bluer. Thus, the signature of a luminous AGN 
could be diluted.  
The probed physical scale and thereby possible 
contaminations from spatially extended star formation emission can 
be larger for more distant sources.
We, however, do not see a trend of systematically bluer observed 
$K'-L'$ colors with increasing luminosity distance (Figure 6a), 
suggesting that luminous buried AGNs are properly detected in our method.
At higher redshift, the fraction of ULIRGs with higher infrared 
luminosities becomes higher (Figure 6b).
This may be partly responsible for the high detection rate of 
red $K'-L'$ sources ($>$1.0 mag) at large distance, because it is known 
that AGN's bolometric contributions tend to increase with 
increasing infrared luminosity for nearby ULIRGs \citep{nar10}.

Second, although the same AO guide star was used for the $K'$- and 
$L'$-band observations of each ULIRG (Table 2), AO correction is likely 
to be better in the $L'$-band, because of the longer wavelength, than the 
$K'$-band.
A larger fraction of compact emission should reside in the AO core 
component in the $L'$-band; thus, the possible flux loss of compact 
emission measured with employed apertures should be smaller in 
the $L'$-band. 
This could slightly redden the $K'-L'$ color to a larger value; thus, 
the possible AGN contribution to the compact nuclear emission could be 
overestimated.
 
Finally, for ULIRGs with faint $L'$-band emission, we need to take 
into account additional possible uncertainty.
$K'$-band emission of all ULIRGs and $L'$-band emission of some 
fraction of ULIRGs were clearly detected in individual frames at each 
dithering position, for which we could combine data after confirming the 
peak position of the ULIRG's signals.
However, some fraction of ULIRGs were too faint in the $L'$-band to 
be clearly detected at individual dithering positions.
For such $L'$-band faint ULIRGs, we blindly summed nine-point dithered 
images based on the input dithering amplitude and pixel scale of the 
IRCS instrument.
When this nine-point dithering sequence was repeated multiple times, 
we simply summed these data in the same manner.
We could barely see some signals only after these data combinations.
Because the telescope pointing accuracy at each dithering position may 
not have been perfect, compact emission from these $L'$-band faint ULIRGs 
could have been blurred at a sub-arcsecond level, and the resulting final 
image size could have been larger than that determined by actual Earth 
atmospheric seeing.
For bright objects, we confirmed that this blind summation worked well 
in one sequence (i.e., nine-point dithering), with a resulting 
image size comparable to that obtained with the summation after peak 
position confirmation.
However, for $L'$-band faint ULIRGs, this blind summation was applied 
to data of multiple sequences with longer observation durations; thus, 
some caution is required in that $L'$-band photometry with the apertures 
used may be fainter and $K'-L'$ colors may be slightly bluer (smaller) 
than the actual values of compact nuclear emission.
In summary, possible AGN contribution to the compact nuclear emission 
could be underestimated for $L'$-band faint ULIRGs because of this 
uncertainty.

Comparing the $K'-L'$ colors measured with the 0$\farcs$5 and 0$\farcs$75 
radius aperture (Table 3) can provide some indication that the derived 
$K'-L'$ colors were not strongly affected by these uncertainties.

\clearpage


\begin{deluxetable}{lcccrrrrccc}
\tabletypesize{\scriptsize}
\tablecaption{Basic Properties of the Observed Ultraluminous Infrared Galaxies 
\label{tbl-1}} 
\tablewidth{0pt}
\tablehead{
\colhead{Object} & \colhead{Redshift} & 
\colhead{d$_{\rm L}$} & \colhead{Scale} & 
\colhead{f$_{\rm 12}$} & 
\colhead{f$_{\rm 25}$} & 
\colhead{f$_{\rm 60}$} & 
\colhead{f$_{\rm 100}$} & 
\colhead{log L$_{\rm IR}$} & 
\colhead{log L$_{\rm FIR}$} &
\colhead{Optical} 
\\
\colhead{} & \colhead{} & \colhead{[Mpc]} & \colhead{[kpc/$''$]}  
& \colhead{[Jy]}
& \colhead{[Jy]} & \colhead{[Jy]} & \colhead{[Jy]}  &
\colhead{[L$_{\odot}$]} & \colhead{[L$_{\odot}$]} & \colhead{Class} \\  
\colhead{(1)} & \colhead{(2)} & \colhead{(3)} & \colhead{(4)} & 
\colhead{(5)} & \colhead{(6)} & \colhead{(7)} & \colhead{(8)} & 
\colhead{(9)} & \colhead{(10)} & \colhead{(11)}  
}
\startdata
IRAS 00456$-$2904 & 0.110 & 504 & 2.0 & $<$0.08 & 0.14 & 2.60 & 3.38 
& 12.2 & 12.2 & HII \\ 
IRAS 04103$-$2838 & 0.118 & 543 & 2.1 & 0.08 & 0.54 & 1.82 & 1.71 & 12.2 &
12.0 & LINER \\ 
IRAS 08559$+$1053 & 0.148 & 695 & 2.6 & $<$0.10 & 0.19 & 1.12 & 1.95 & 12.2 
& 12.1 & Sy2 \\
IRAS 09039$+$0503 & 0.125 & 578 & 2.2 & 0.07 & 0.12 
& 1.48 & 2.06 & 12.1 & 12.0 & LINER \\  
IRAS 09116$+$0334 & 0.146 & 685 & 2.5 & $<$0.09 & $<$0.14 & 1.09 & 1.82 
& 12.2 & 12.1 & LINER \\  
IRAS 10035$+$2740 & 0.165 & 784 & 2.8 & $<$0.14 & $<$0.17 & 1.14 & 1.63 
& 12.3 & 12.2 & unclass \\
IRAS 10190$+$1322 & 0.077 & 345 & 1.4 & $<$0.07 & 0.38 & 3.33 
& 5.57 & 12.0 & 12.0 & HII \\   
IRAS 11506$+$1331 & 0.127 & 588 & 2.2 & $<$0.10 & 0.19 & 2.58 & 3.32 
& 12.4 & 12.3 & HII \\  
IRAS 12072$-$0444 & 0.129 & 598 & 2.3 & $<$0.12 & 0.54 & 2.46 & 2.47 
& 12.4 & 12.3 & Sy2 \\
IRAS 12112$+$0305 & 0.073 & 326 & 1.4 & 0.12 & 0.51 
& 8.50 & 9.98 & 12.3 & 12.3 & LINER  \\
IRAS 13539$+$2920 & 0.108 & 494 & 2.0 & $<$0.09 & 0.12 & 1.83 & 2.73 
& 12.1 & 12.0 & HII \\
IRAS 14202$+$2615 & 0.159 & 752 & 2.7 & 0.18 & 0.15 & 1.49 & 1.99 & 
12.5 & 12.3 & HII \\
IRAS 14394$+$5332 & 0.105 & 479 & 1.9 & 0.03 & 0.35 & 1.95 & 2.39 
& 12.1 & 12.0 & Sy2 \\
IRAS 15206$+$3342 & 0.125 & 578 & 2.2 & 0.08 & 0.35 & 1.77 & 1.89 & 12.3 
& 12.1 & HII \\
IRAS 20414$-$1651 & 0.086 & 387 & 1.6 & $<$0.65 & 0.35 
& 4.36 & 5.25 & 12.3 & 12.1 & HII \\
IRAS 21219$-$1757 & 0.112 & 514 & 2.0 & 0.21 & 0.45 & 1.07 & 1.18 & 
12.1 & 11.8 & Sy1 \\
IRAS 22491$-$1808 & 0.076 & 340 & 1.4 & 0.05 
& 0.55 & 5.44 & 4.45 & 12.2 & 12.1 & HII  \\ \hline
\enddata

\tablecomments{
Col.(1): Object name. 
Col.(2): Redshift by \citet{kim02}. 
Col.(3): Luminosity distance (in Mpc). 
Col.(4): Physical scale (in kpc arcsec$^{-1}$). 
Col.(5)--(8): f$_{12}$, f$_{25}$, f$_{60}$, and f$_{100}$ are 
{\it IRAS} fluxes at 12 $\mu$m, 25 $\mu$m, 60 $\mu$m, and 100 $\mu$m (in Jy),
respectively, taken from \citet{kim98}.
Col.(9): Decimal logarithm of infrared (8$-$1000 $\mu$m) luminosity
in units of solar luminosity (L$_{\odot}$), calculated with
$L_{\rm IR} = 2.1 \times 10^{39} \times$ $d_{\rm L}$(Mpc)$^{2}$
$\times$ (13.48 $\times$ $f_{12}$ + 5.16 $\times$ $f_{25}$ +
$2.58 \times f_{60} + f_{100}$) (ergs s$^{-1}$) \citep{sam96}.
Col.(10): Decimal logarithm of far-infrared (40$-$500 $\mu$m) luminosity
in units of solar luminosity (L$_{\odot}$), calculated with
$L_{\rm FIR} = 2.1 \times 10^{39} \times$ $d_{\rm L}$(Mpc)$^{2}$
$\times$ ($2.58 \times f_{60} + f_{100}$) (ergs s$^{-1}$) \citep{sam96}.
Col.(11): Optical spectroscopic classification by \citet{vei99}.
``LINER'', ``HII'', ``unclass'', ``Sy2'', ``Sy1'' mean LINER, HII-region, 
unclassified, Seyfert 2, and Seyfert 1, respectively. 
Seyfert 1 and 2 are usually regarded as optically identified AGNs.
}

\end{deluxetable}

\begin{deluxetable}{lclcccccc}
\tabletypesize{\scriptsize}
\tablecaption{Observation Log\label{tbl-2}} 
\tablewidth{0pt}
\tablehead{
\colhead{Object} & \colhead{Band} & 
\colhead{Date} & \colhead{Exposure} & \multicolumn{2}{c}{Standard Star} &
\multicolumn{3}{c}{LGS-AO or NGS-AO Guide Star}
\\
\colhead{} & \colhead{} & \colhead{(UT)} & \colhead{(min)} & 
\colhead{Name} & \colhead{(mag)} & \colhead{Name} & \colhead{R-band}  &
\colhead{Separation} \\
\colhead{} & \colhead{} & \colhead{} & \colhead{} & \colhead{} & 
\colhead{} & \colhead{USNO} & \colhead{(mag)} & \colhead{(arcsec)} \\
\colhead{(1)} & \colhead{(2)} & \colhead{(3)} & \colhead{(4)} & 
\colhead{(5)} & \colhead{(6)} & \colhead{(7)} & \colhead{(8)} & 
\colhead{(9)} 
}
\startdata
IRAS 00456$-$2904 & $K'$ & 2015 August 21 & 18 & FS1 & 12.98 & Nucleus & 15 & 0 \\
                  & $L'$ & 2015 August 21 & 25.2 & G158-27 & 6.99 & Nucleus & 15 & 0 \\
IRAS 04103$-$2838 & $K'$ & 2015 February 1 & 9 & FS112 & 10.86 & USNO 0614-0044366 & 15 & 35 \\ 
                  & $L'$ & 2015 February 1 & 9 & HD22686 & 7.20 & USNO 0614-0044366 & 15 & 35 \\
IRAS 08559$+$1053 & $K'$ & 2015 February 1 & 9 & FS126 & 11.64 & USNO 1006-0166801 & 18 & 45 \\
                  & $L'$ & 2015 February 1 & 9 & HD77281 & 7.04 & USNO 1006-0166801 & 18 & 45 \\
IRAS 09039$+$0503 & $K'$ & 2019 April 20 & 13.5 & FS126 & 11.64 & USNO 0948-0171026 & 15 & 43 \\
                  & $L'$ & 2019 April 21 & 22.5 & HD77281 & 7.04 & USNO 0948-0171026 & 15 & 43 \\ 
IRAS 09116$+$0334 & $K'$ & 2015 February 1 & 9 & FS126 & 11.64 & USNO 0933-0204204 & 14 & 30 \\
                  & $L'$ & 2015 February 1 & 18 & HD77281 & 7.04 & USNO 0933-0204204 & 14 & 30 \\
IRAS 10035$+$2740 & $K'$ & 2019 April 20 & 13.5 & FS126 & 11.64 & USNO 1174-0222148 & 18 & 32 \\
                  & $L'$ & 2019 April 20 & 13.5 & HD105601 & 6.67 & USNO 1174-0222148 & 18 & 32 \\
IRAS 10190$+$1322 & $K'$ & 2015 February 1 & 9 & FS126 & 11.64 & USNO 1031-0209138 & 17 & 19 \\
                  & $L'$ & 2015 February 1 & 18 & HD77281 & 7.04 & USNO 1031-0209138 & 17 & 19 \\
IRAS 11506$+$1331 & $K'$ & 2019 April 21 & 9 & FS129 & 10.64 & USNO 1032-0218808 & 15 & 36 \\
                  & $L'$ & 2019 April 21 & 13.5 & HD106965 & 7.31 & USNO 1032-0218808 & 15 & 36 \\
IRAS 12072$-$0444 & $K'$ & 2019 April 21 & 9 & FS129 & 10.64 & USNO 0849-0231140 & 18 & 38 \\
                  & $L'$ & 2019 April 21 & 13.5 & HD77281 & 7.04 & USNO 0849-0231140 & 18 & 38 \\
 & $K'$ & 2018 May 20 & 4.5 & FS19 & 13.79 & Nucleus \tablenotemark{a} & 15 & 0 \\
                  & $L'$ & 2018 May 20 & 13.5 & GL347A & 7.37 & Nucleus \tablenotemark{a} & 15 & 0 \\
IRAS 12112$+$0305 & $K'$ & 2015 February 1 & 9 & FS132 & 11.84 & USNO 0927-0276322 & 13 & 65 \\
                  & $L'$ & 2015 February 1 & 18 & HD106965 & 7.31 & USNO 0927-0276322 & 13 & 65 \\
IRAS 13539$+$2920 & $K'$ & 2016 April 17 & 13.5 & p138-c & 11.10 & USNO 1190-0215111 & 16 & 29 \\
                  & $L'$ & 2016 April 17 & 11.3 & HD105601 & 6.67 & USNO 1190-0215111 & 16 & 29 \\
IRAS 14202$+$2615 & $K'$ & 2016 April 17 & 4.5 & p138-c & 11.10 & Nucleus & 15 & 0 \\
                  & $L'$ & 2016 April 17 & 11.3 & HD105601 & 6.67 & Nucleus & 15 & 0 \\
IRAS 14394$+$5332 & $K'$ & 2019 April 20 & 9 & FS131 & 11.32 & USNO 1433-0256687 & 15 & 66 \\
                  & $L'$ & 2019 April 20 & 11.3 & HD105601 & 6.67 & USNO 1433-0256687 & 15 & 66 \\
IRAS 15206$+$3342 & $K'$ & 2016 April 17 & 9 & p138-c & 11.10 & USNO 1235-0241068 & 16 & 40 \\
                  & $L'$ & 2016 April 17 & 15 & HD105601 & 6.67 & USNO 1235-0241068 & 16 & 40 \\
IRAS 20414$-$1651 & $K'$ & 2015 September 19 & 27 & FS153 & 10.89 & USNO 0733-0845610 \tablenotemark{a} & 13 & 30 \\
                  & $L'$ & 2015 September 19 & 37.8 & G158-27 & 6.99 & USNO 0733-0845610 \tablenotemark{a} & 13 & 30 \\
IRAS 21219$-$1757 & $K'$ & 2015 September 19 & 9 & FS34 & 13.00 & Nucleus \tablenotemark{a} & 14 & 0 \\ 
                  & $L'$ & 2015 September 19 & 12.6 & GL811.1 & 6.69 & Nucleus \tablenotemark{a} & 14 & 0 \\
IRAS 22491$-$1808 & $K'$ & 2015 August 21 & 18 & S667-D & 11.54 & USNO 0721-1176564 & 15 & 58 \\ 
                  & $L'$ & 2015 August 21 & 18.8 & G158-27 & 6.99 & USNO 0721-1176564 & 15 & 58 \\
\hline
\enddata

\tablenotetext{a}{NGS-AO guide star}

\tablecomments{
Col.(1): Object name.  
Col.(2): Observed band. $K'$- (2.1 $\mu$m) or $L'$-band (3.8 $\mu$m).
Col.(3): Observation date (in UT).
Col.(4): Net on-source exposure time (in min).
Col.(5): Standard star's name.
Col.(6): Standard star's magnitude in the $K'$- or $L'$-band.
Col.(7): Guide star name (USNO number) used for LGS-AO tip-tilt
correction or NGS-AO correction.
Col.(8): Guide star's optical $R$-band magnitude.
Col.(9): Separation between the target object and guide star (in arcsec).
}

\end{deluxetable}

\begin{deluxetable}{l|crr|ccc|rc}
\tabletypesize{\scriptsize}
\tablecaption{Nuclear Photometry and Estimated AGN Contribution 
\label{tbl-3}} 
\tablewidth{0pt}
\tablehead{
\colhead{} & \multicolumn{3}{|c|}{0$\farcs$5} & \multicolumn{3}{c|}{0$\farcs$75} & 
\colhead{f$_{\rm AGN}$} & \colhead{L$_{\rm AGN}$} \\ 
\colhead{Object} & \colhead{$K'$} & \colhead{$L'$} & 
\colhead{$K'-L'$} & \colhead{$K'$} & \colhead{$L'$} & 
\colhead{$K'-L'$} & \colhead{[\%]} & \colhead{[10$^{44}$ ergs s$^{-1}$]} \\
\colhead{(1)} & \colhead{(2)} & \colhead{(3)} & \colhead{(4)} &
\colhead{(5)} & \colhead{(6)} & \colhead{(7)}  & \colhead{(8)} & 
\colhead{(9)} 
}
\startdata 
IRAS 00456$-$2904 SW & 14.6 & 14.0 & 0.7 & 14.1 & 13.5 & 0.6 & 22 & 0.03 \\
IRAS 00456$-$2904 NE & 18.1 & $>$14.4 & $<$3.7 & 17.6 & $>$13.9 & $<$3.6 
& $<$100 & $<$0.2 \\
IRAS 04103$-$2838 & 14.1 & 12.5 & 1.6 & 13.8 & 12.3 & 1.5 & 85 & 0.6 \\
IRAS 08559$+$1053 & 13.1 & 11.3 & 1.9 & 12.9 & 11.1 & 1.8 & 97 & 3 \\
IRAS 09039$+$0503 $\tablenotemark{A}$ & 15.1 & 13.2 & 1.8 & 14.6 & 12.7 & 1.9 & 93 & 0.4 \\
IRAS 09116$+$0334 W & 14.1 & 13.3 & 0.7 & 13.7 & 13.0 & 0.7 & 22 & 0.1\\
IRAS 09116$+$0334 SE & 17.0 & $>$14.4 & $<$2.7 & 16.7 & $>$14.0 
& $<$2.7 & $<$100 & $<$0.2 \\
IRAS 09116$+$0334 NE & 17.9 & $>$14.4 & $<$3.5 & 17.3 & $>$14.0 
& $<$3.3 & $<$100 & $<$0.2 \\
IRAS 10035$+$2740 N & 15.8 & $>$14.9 & $<$0.8 & 15.5 (0$\farcs$6) & 
$>$14.9 (0$\farcs$6) & $<$0.6 (0$\farcs$6) & $<$33 & $<$0.05\\
IRAS 10035$+$2740 S & 15.9 & $>$14.9 & $<$0.9 & 15.6 (0$\farcs$6) & 
$>$14.9 (0$\farcs$6) & $<$0.7 (0$\farcs$6) & $<$42 & $<$0.07 \\
IRAS 10190$+$1322 E & 14.1 & 13.2 & 0.8 & 13.6 & 12.9 & 0.7 & 32 & 0.05 \\
IRAS 10190$+$1322 W & 14.6 & 14.2 & 0.4 & 14.2 & 14.3 & $-$0.1 & 0 & 0 \\
IRAS 11506$+$1331 & 14.3 & 13.0 & 1.3 & 14.0 & 12.7 & 1.3 & 70 & 0.4 \\
IRAS 12072$-$0444 N (2019 Apr) & 14.4 & 12.3 & 2.1 & 14.3 (0$\farcs$6) 
& 12.3 (0$\farcs$6) & 2.0 (0$\farcs$6) & 100 & 1 \\
IRAS 12072$-$0444 S (2019 Apr) & 15.3 & 13.9 & 1.4 & --- $\tablenotemark{B}$ 
& --- $\tablenotemark{B}$ & --- $\tablenotemark{B}$ & 75 & 0.2 \\            
IRAS 12072$-$0444 N (2018 May) & 14.7 & 12.3 & 2.5 & 14.5 (0$\farcs$6) & 
12.1 (0$\farcs$6) & 2.4 (0$\farcs$6) & 100 & 1 \\
IRAS 12072$-$0444 S (2018 May) & 15.6 & 14.0 & 1.5 & --- $\tablenotemark{B}$ 
& --- $\tablenotemark{B}$ & --- $\tablenotemark{B}$ & 80 & 0.2 \\               
IRAS 12112$+$0305 SW & 14.6 & 13.5 & 1.1 & 14.4 & 13.3 & 1.0 & 57 & 0.06 \\
IRAS 12112$+$0305 NE & 15.4 & 14.3 & 1.1 & 14.9 & 13.9 & 1.0 & 57 & 0.03 \\
IRAS 13539$+$2920 NW & 15.1 & 13.4 & 1.6 & 14.5 & 13.0 & 1.6 & 85 & 0.2 \\
IRAS 13539$+$2920 SE & 18.0 & $>$13.9 & $<$4.1 & 17.4 & $>$13.5 & $<$4.0 
& $<$100 & $<$0.2 \\     
IRAS 14202$+$2615 SE & 14.6 & 12.6 & 2.0 & 14.2 & 12.4 & 1.8 & 100 & 1 \\
IRAS 14202$+$2615 NW & 16.6 & $>$13.8 & $<$2.9 & 16.1 & $>$13.6 & $<$2.5 
& $<$100 & $<$0.5 \\
IRAS 14394$+$5332 SW $\tablenotemark{C}$ & 14.3 & 12.8 & 1.5 & 13.9 
& 12.7 & 1.3 & 80 & 0.3 \\
IRAS 14394$+$5332 NE $\tablenotemark{C}$ & 15.8 & $>$15.3 & $<$0.5 
& 15.6 (0$\farcs$6) & $>$15.2 (0$\farcs$6) & $<$0.4 (0$\farcs$6) & 0 & 0 \\
IRAS 15206$+$3342 NE & 14.7 & 12.8 & 2.0 & 14.5 (0$\farcs$6) 
& 12.6 (0$\farcs$6) & 1.9 (0$\farcs$6) & 100 & 0.6 \\
IRAS 15206$+$3342 SW & 15.4 & $>$14.5 & $<$0.9 & 15.1 (0$\farcs$6) & 
$>$14.4 (0$\farcs$6) & $<$0.7 (0$\farcs$6) & $<$42 & $<$0.06 \\
IRAS 20414$-$1651 & 14.4 & 13.8 & 0.6 & 14.1 & 13.5 & 0.6 & 12 & 0.01 \\
IRAS 21219$-$1757 & 11.7 & 9.8 & 2.0 & 11.6 & 9.7 & 1.9 & 100 & 8 \\ 
IRAS 22491$-$1808 E & 15.8 & $>$14.5 & $<$1.3 & 15.1 & $>$14.1 & $<$1.1 
& $<$70 & $<$0.04 \\ 
IRAS 22491$-$1808 W & 15.6 & $>$14.5 & $<$1.0 & 15.1 & $>$14.1 & $<$1.0 
& $<$50 & $<$0.03 \\ 
\hline
\enddata

\tablenotetext{A}{We are unable to achieve reliable photometry of the 
northeastern (NE) faint secondary nucleus of IRAS 09039$+$0503 
detected in Figure 2 because it is too close to the much brighter 
southwestern (SW) primary nucleus.}

\tablenotetext{B}{Not derived because of significant contamination from the 
northern (N) brighter primary galaxy nucleus with $\sim$0$\farcs$9 
separation.}

\tablenotetext{C}{For IRAS 14394$+$5332, the SW and NE nuclei correspond 
to the E and EE nuclei defined by \citet{kim02}, respectively.}

\tablecomments{ 
Col.(1): Object name. 
For IRAS 12072$-$0444, data from both 2019 April and 2018 May are 
presented, denoted as ``2019 Apr'' and ``2018 May'', respectively.
Cols.(2)--(4): $K'$-band (2.1 $\mu$m) magnitude, $L'$-band (3.8 $\mu$m) 
magnitude, and $K'-L'$ color in mag, respectively, measured with 
a 0$\farcs$5 radius aperture.
Cols.(5)--(7): $K'$-band magnitude, $L'$-band magnitude, and 
$K'-L'$ color in mag, respectively, measured with a 0$\farcs$75 
radius aperture.
For a few ULIRGs with small nuclear separation, a 0$\farcs$6 radius aperture 
is used instead of the 0$\farcs$75 radius aperture to minimize 
contamination from another galaxy nucleus. 
These sources are denoted as ``(0$\farcs$6)'' in columns (5)--(7).
Uncertainty is up to $\sim$0.3 mag for the $K'$- and $L'$-band 
photometry of compact nuclear emission and up to $\sim$0.2 mag 
for its $K'-L'$ color. 
Col.(8): AGN fraction in the $L'$-band (in \%), derived from the $K'-L'$ color 
measurements with the 0$\farcs$5 radius aperture.
See $\S$5.1.
Col.(9): AGN luminosity in 10$^{44}$ ergs s$^{-1}$ estimated from 
AGN-origin $\nu$F$_{\nu}$ value at $L'$ after the removal of stellar 
emission contribution, based on the 0$\farcs$5 radius aperture photometry.
The possible uncertainty is $<$0.2 mag ($<$20\%) in the sense that 
the AGN luminosity may be underestimated. 
We assume an AGN surrounded by dust in all directions, where the 
surrounding dust has a strong temperature gradient (inner dust has 
higher temperature) and luminosity is transferred from hotter inside 
close to the innermost dust sublimation radius (emitting $\sim$3 $\mu$m 
infrared light) to cooler outside (emitting longer infrared wavelength)
\citep{ima07,ima09,ima10,ima10b}.
}

\end{deluxetable}

\clearpage

\begin{deluxetable}{lcccc}
\tabletypesize{\scriptsize}
\tablecaption{WISE Data and Infrared AGN Selection \label{tbl-4}}
\tablewidth{0pt}
\tablehead{
\colhead{Object} & \colhead{$W1$ (3.4 $\mu$m)}  
& \colhead{$W2$ (4.6 $\mu$m)} & \colhead{$W1-W2$} & \colhead{$K'-L'$}  \\ 
\colhead{} & \colhead{(mag)} & \colhead{(mag)} & \colhead{(mag)} 
& \colhead{(mag)} \\
\colhead{(1)} & \colhead{(2)} & \colhead{(3)} & \colhead{(4)} 
& \colhead{(5)} 
}
\startdata
IRAS 00456$-$2904 & 12.7 & 12.0 & 0.7 & 0.7 \\
IRAS 04103$-$2838 & 12.3 & 11.0 & 1.3 (AGN) & 1.6 (AGN) \\ 
IRAS 08559$+$1053 & 11.6 & 10.5 & 1.1 (AGN) & 1.9 (AGN) \\
IRAS 09039$+$0503 & 13.5 & 12.8 & 0.7 & 1.8 (AGN) \\
IRAS 09116$+$0334 & 12.7 & 12.2 & 0.5 & 0.7 \\
IRAS 10035$+$2740 & 13.6 & 13.0 & 0.6 & $<$0.8 $+$ $<$0.9 \\
IRAS 10190$+$1322 E & 12.3 & 11.6 & 0.6 & 0.8 \\
IRAS 10190$+$1322 W & 12.6 & 12.3 & 0.3 & 0.4 \\
IRAS 11506$+$1331 & 12.9 & 11.5 & 1.4 (AGN) & 1.3 (AGN) \\
IRAS 12072$-$0444 & 12.0 & 10.7 & 1.3 (AGN) & 2.1 $+$ 1.4 (AGN) \\
IRAS 12112$+$0305 & 12.3 & 11.5 & 0.8 (AGN) & 1.1 $+$ 1.1 (AGN?)\\
IRAS 13539$+$2920 & 12.9 & 12.1 & 0.8 & 1.6 (AGN) \\
IRAS 14202$+$2615 & 12.0 & 10.9 & 1.1 (AGN) & 2.0 (AGN) \\
IRAS 14394$+$5332 E & 12.4 & 11.0 & 1.4 (AGN) & 1.5 (AGN) \\
IRAS 15206$+$3342 & 12.5 & 11.3 & 1.2 (AGN) & 2.0 $+$ $<$0.9 (AGN) \\
IRAS 20414$-$1651 & 13.0 & 12.2 & 0.8 & 0.6 \\
IRAS 21219$-$1757 & 9.9 & 8.9 & 1.0 (AGN) & 2.0 (AGN) \\
IRAS 22491$-$1808 & 12.7 & 12.2 & 0.5 & $<$1.3 $+$ $<$1.0 \\ 
\hline
\enddata

\tablecomments{
Col.(1): Object name.
Col.(2): WISE $W1$ (3.4 $\mu$m) magnitude.
Col.(3): WISE $W2$ (4.6 $\mu$m) magnitude. 
Col.(4): WISE $W1-W2$ color in mag.
When the $W1-W2$ color exceeds the WISE AGN selection criterion of 
$>$0.8 mag \citep{ste12}, a note of ``(AGN)'' is added.
For reference, the WISE $W1-W2$ color of IRAS 12112$+$0305 is slightly 
above 0.8 mag, but those of IRAS 13539$+$2920 and IRAS 20414$-$1651 
are slightly below 0.8 mag.
Col.(5): $K'-L'$ color in mag based on our 0$\farcs$5 radius 
aperture photometry of the Subaru AO data.
For some fraction of sources, our high-spatial-resolution AO data 
resolve the WISE-detected component into multiple galaxy nuclei.  
When the primary galaxy nucleus is more than 1 mag brighter than 
other fainter galaxy nuclei, 
we assign the color of the primary galaxy nucleus, 
because it should dominate the observed WISE color.
When the difference in the $K'$-band magnitude (0$\farcs$5 radius 
aperture photometry) between the primary and secondary galaxy 
nucleus is less than 1 mag, we give the $K'-L'$ colors of both nuclei 
for reference.
When the $K'-L'$ color exceeds our AGN selection criterion of 
$>$1.0 mag, a note of ``(AGN)'' is added.
Given the potential $\lesssim$0.2 mag uncertainty in the $K'-L'$ 
color, a source with $K'-L'$ $=$ 1.1 mag is noted as ``(AGN?)''.
}

\end{deluxetable}

\begin{deluxetable}{lcrccc}
\tabletypesize{\scriptsize}
\tablecaption{Luminosity Ratio and Nuclear Separation in Spatially 
Resolved Multiple Nuclei ULIRGs \label{tbl-5}}
\tablewidth{0pt}
\tablehead{
\colhead{Object} & \colhead{$K'$(stellar) ratio}  & \colhead{$L'$ ratio} &
\colhead{$L' $(AGN) ratio} & \colhead{Separation} &
\colhead{Separation} \\ 
\colhead{} & \colhead{} & \colhead{} & \colhead{} & \colhead{(arcsec)} &
\colhead{(kpc)} \\
\colhead{(1)} & \colhead{(2)} & \colhead{(3)} & \colhead{(4)} 
& \colhead{(5)} & \colhead{(6)} 
}
\startdata
IRAS 00456$-$2904 SW, NE & 26 (14.0, 17.5) & $>$1.5 & $>$0.3 & 11 & 23 \\
IRAS 09116$+$0334 W, SE & 18 (13.9, 17.1) & $>$2.7 & $>$0.5 & 6.8 & 17 \\
IRAS 10190$+$1322 E, W & 1.4 (13.6, 14.0) & 2.4 & $\infty$ & 4.1 & 5.7 \\
IRAS 12072$-$0444 N, S (2019 Apr) & 2.3 (14.4, 15.3) $\tablenotemark{A}$ 
& 4.3 & 5.7 & 0.96 & 2.2 \\
IRAS 12112$+$0305 SW, NE & 1.2 (14.3, 14.5) & 2.0 & 2.0 & 3.0 & 4.1 \\
IRAS 13539$+$2920 NW, SE & 14 (14.3, 17.1) & $>$1.6 & $>$1.3 & 3.8 & 7.6 \\
IRAS 14202$+$2615 SE, NW & 6.2 (14.5, 16.4) & $>$3.0 & $>$3.0 & 6.0 & 16 \\
IRAS 14394$+$5332 SW, NE & 2.2 (13.9, 14.8) & $>$10 & $\infty$ & 1.3 & 2.5 \\
IRAS 15206$+$3342 NE, SW & 1.9 (14.7, 15.4) $\tablenotemark{A}$ & $>$5.3 
& $>$12 & 0.71 & 1.6 \\
\enddata

\tablenotetext{A}{We compare our 0$\farcs$5 radius aperture $K'$-band 
photometry of two galaxy nuclei, because they are not spatially resolved 
in a seeing-limited image \citep{kim02}.}

\tablecomments{
Col.(1): Object name.
Col.(2): $K'$-band (2.1 $\mu$m) flux ratio between two galaxy nuclei, 
measured with central 4 kpc diameter aperture photometry 
by \citet{kim02}.
The 4 kpc diameter aperture $K'$-band magnitudes of primary and 
secondary galaxy nuclei are shown in the first and second columns, 
respectively, in parentheses.
Col.(3): Nuclear $L'$-band (3.8 $\mu$m) luminosity ratio based on our
0$\farcs$5 radius aperture photometry (Table 3, column 3).
Col.(4): Nuclear AGN-origin $L'$-band (3.8 $\mu$m) luminosity ratio 
after the subtraction of possible stellar emission component 
(Table 3, column 8),  
based on our 0$\farcs$5 radius aperture photometry.
See $\S$5.1.
In columns (2)--(4), the luminosity at the primary galaxy 
nucleus listed first in column 1 is divided by that at the secondary 
galaxy nucleus listed second in column 1.
Col.(5): Apparent nuclear separation (in arcsec) calculated from our 
Subaru AO $K'$-band images. 
Col.(6): Apparent nuclear physical separation (in kpc) calculated with 
our adopted cosmological parameters in $\S$1.}

\end{deluxetable}

\clearpage

\begin{figure}
\begin{center}
\includegraphics[angle=0,scale=.45]{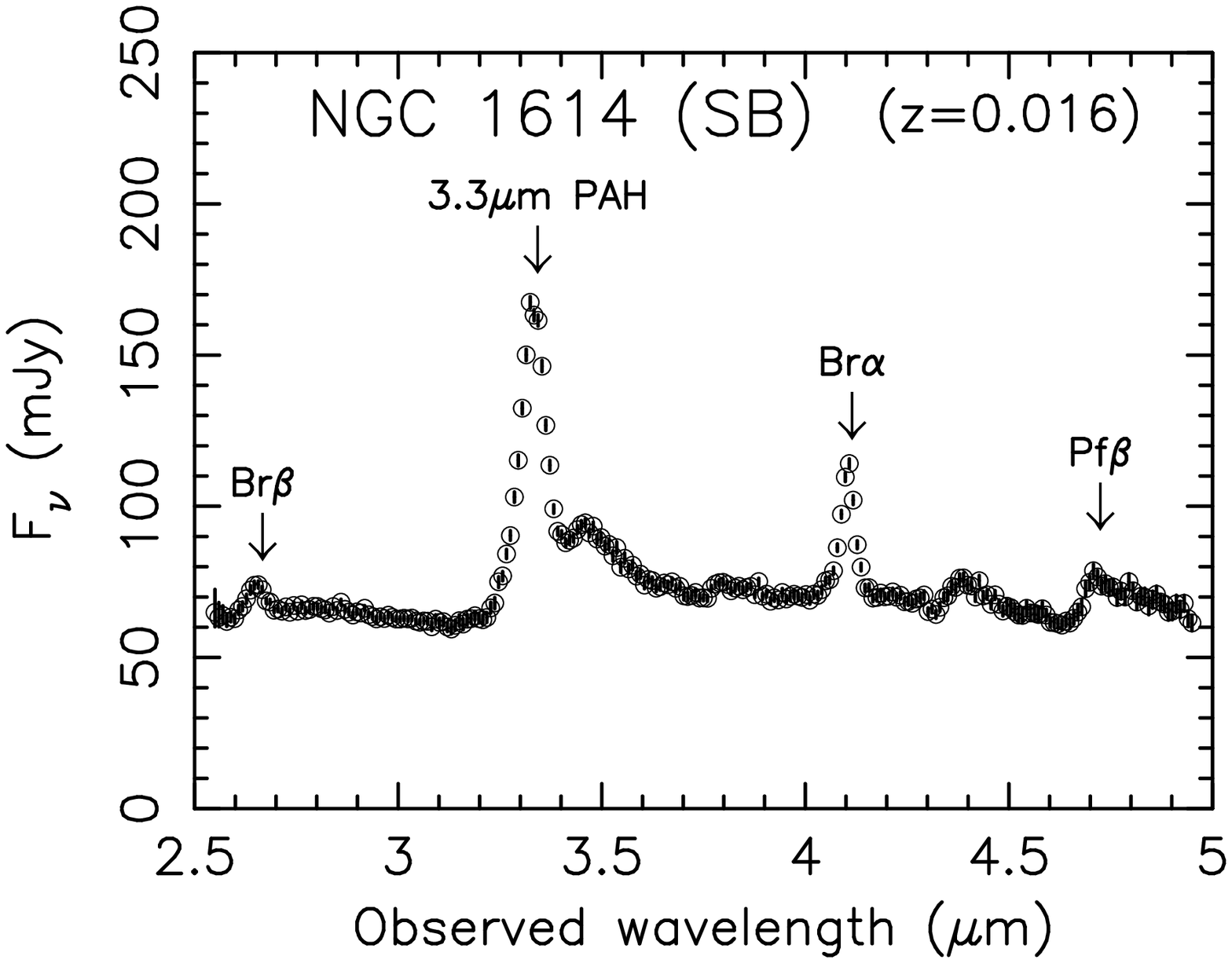} 
\includegraphics[angle=0,scale=.45]{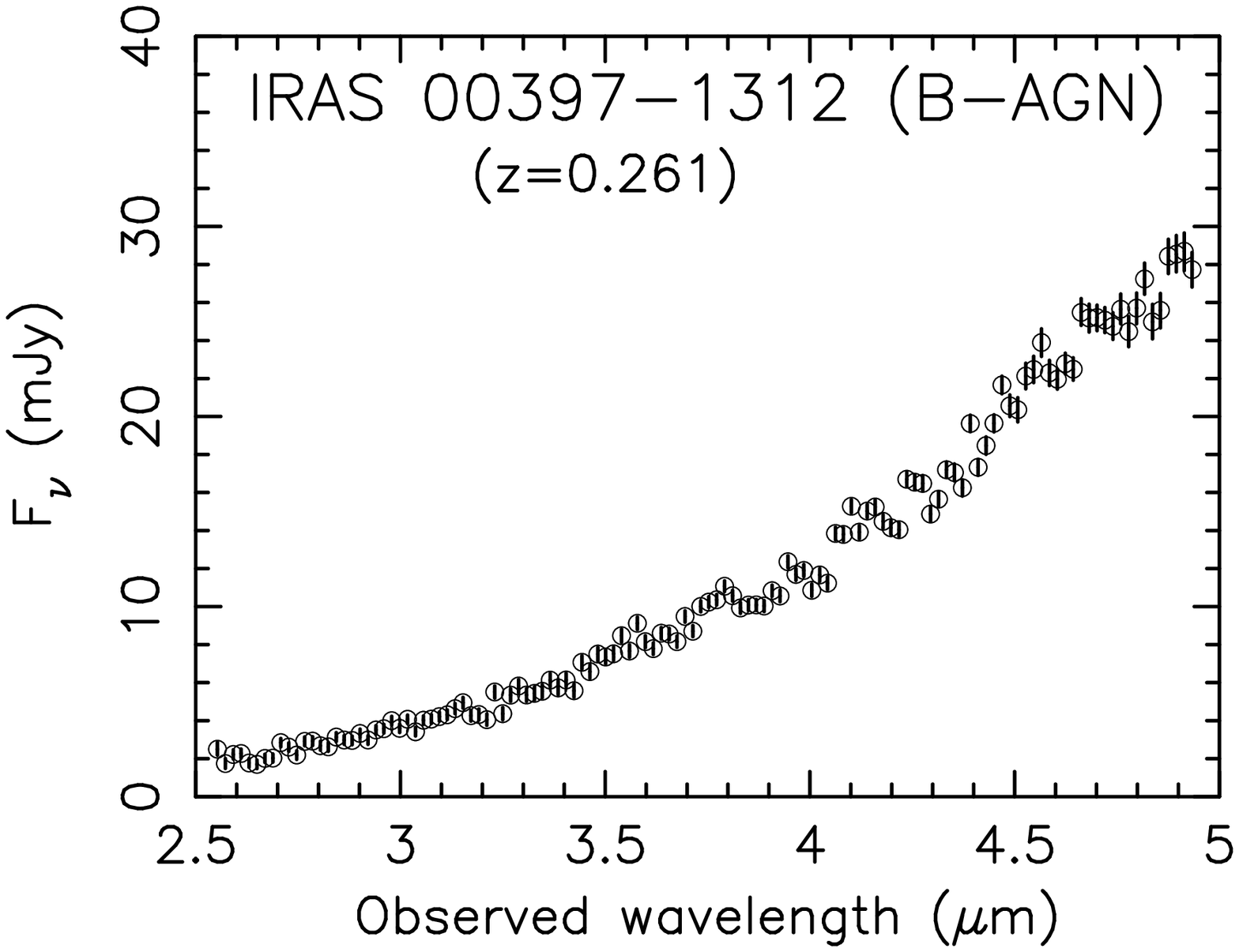} 
\end{center}
\caption{
Examples of infrared 2.5--5 $\mu$m spectra of a starburst-dominated 
infrared luminous galaxy NGC 1614 ($z =$ 0.016) ({\it Left}) 
and a buried-AGN-dominated infrared luminous galaxy 
IRAS 00397$-$1312 ($z =$ 0.261) 
({\it Right}), taken with the AKARI IRC instrument \citep{ima10b}.
The abscissa is the observed wavelength in $\mu$m, and the ordinate 
is flux (F$\nu$) in mJy.
A luminous buried AGN shows a much more steeply rising continuum 
from 2.5 $\mu$m to 5 $\mu$m, and thereby much redder 
$K'$(2.1$\mu$m)$-$$L'$(3.8$\mu$m) color, than a starburst-dominated galaxy.
``SB'' ({\it Left}) and ``B-AGN'' ({\it Right}) mean a starburst and 
a buried AGN, respectively. 
In the left panel, the 3.3 $\mu$m polycyclic aromatic hydrocarbons (PAH) 
emission feature at rest wavelength $\lambda_{\rm rest}$ = 3.29 $\mu$m, 
and hydrogen recombination lines 
(Br$\beta$ at $\lambda_{\rm rest}$ = 2.63 $\mu$m, 
Br$\alpha$ at $\lambda_{\rm rest}$ = 4.05 $\mu$m, 
and Pf$\beta$ at $\lambda_{\rm rest}$ = 4.65 $\mu$m) are indicated.
}
\end{figure}

\begin{figure}
\begin{center}
\includegraphics[angle=0,scale=.424]{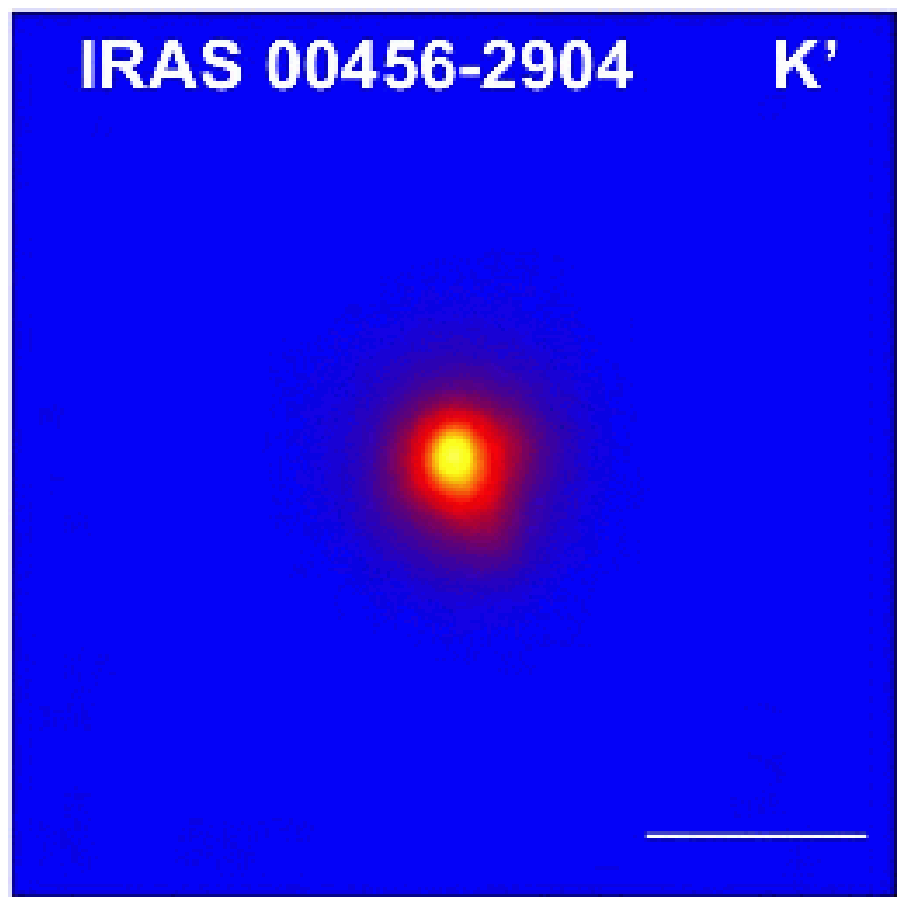} 
\includegraphics[angle=0,scale=.424]{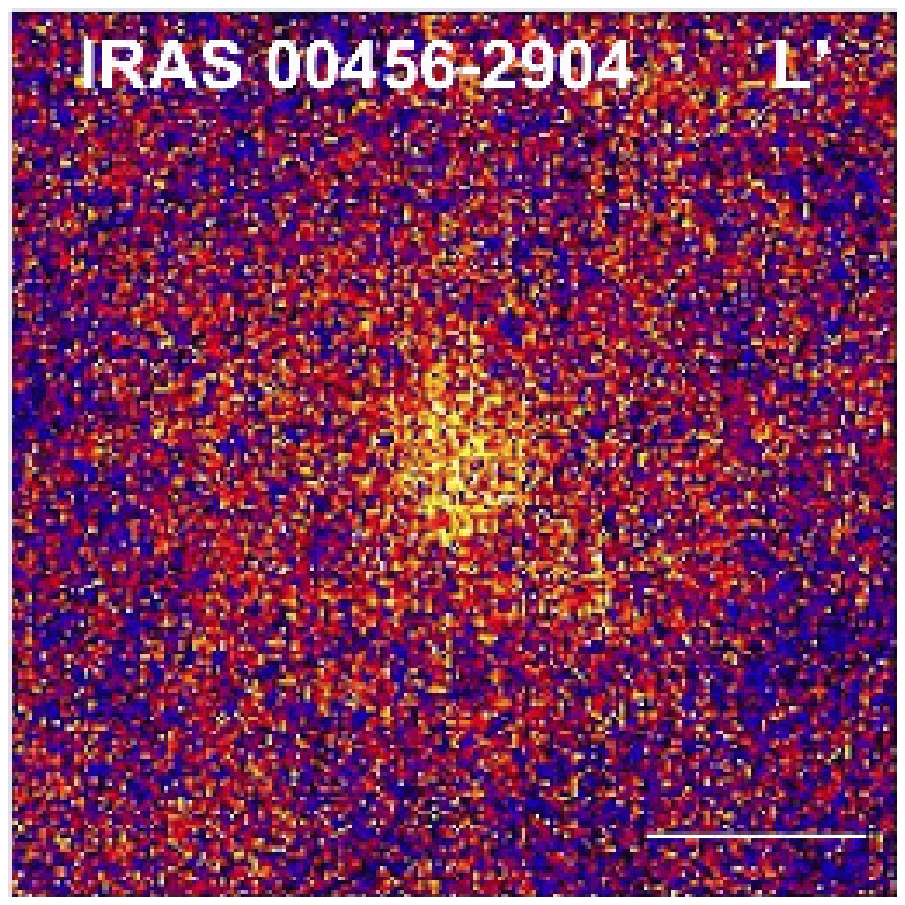} 
\includegraphics[angle=0,scale=.424]{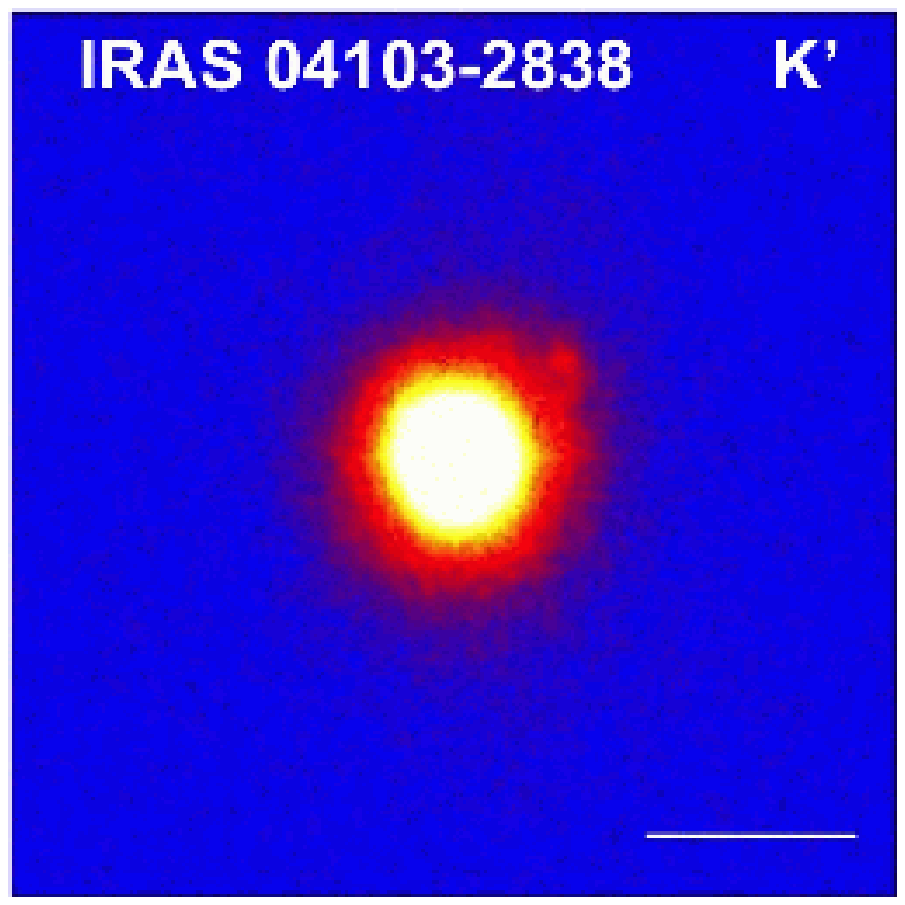} 
\includegraphics[angle=0,scale=.424]{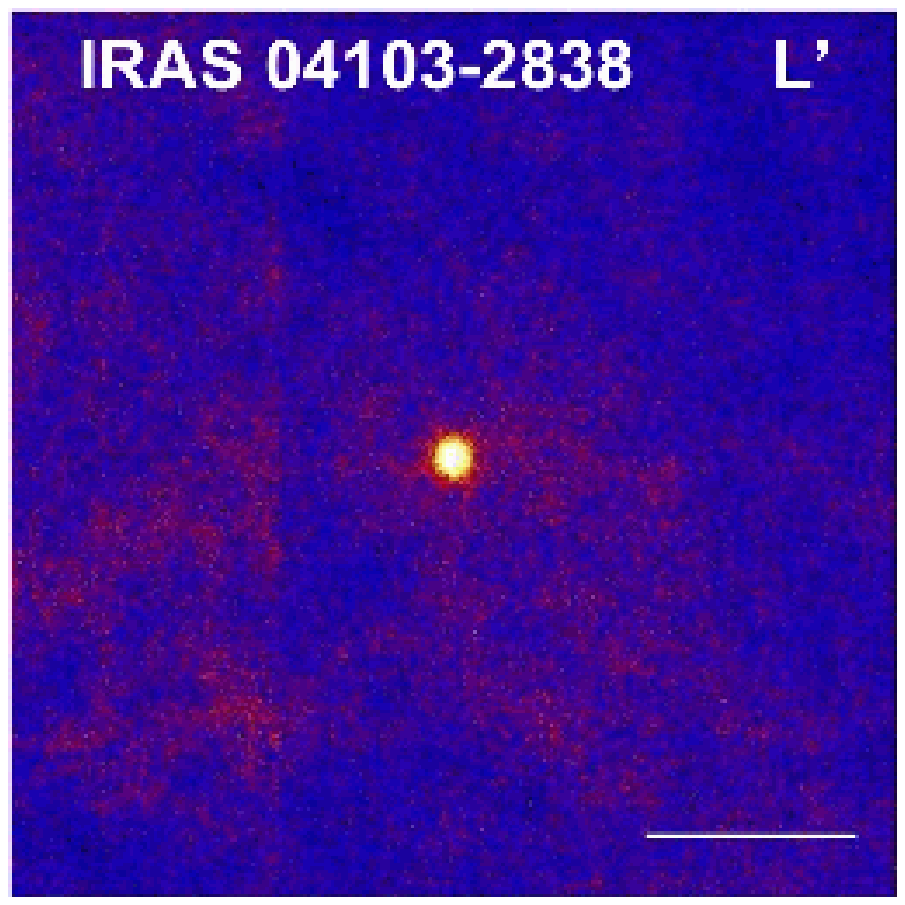} \\ 
\includegraphics[angle=0,scale=.424]{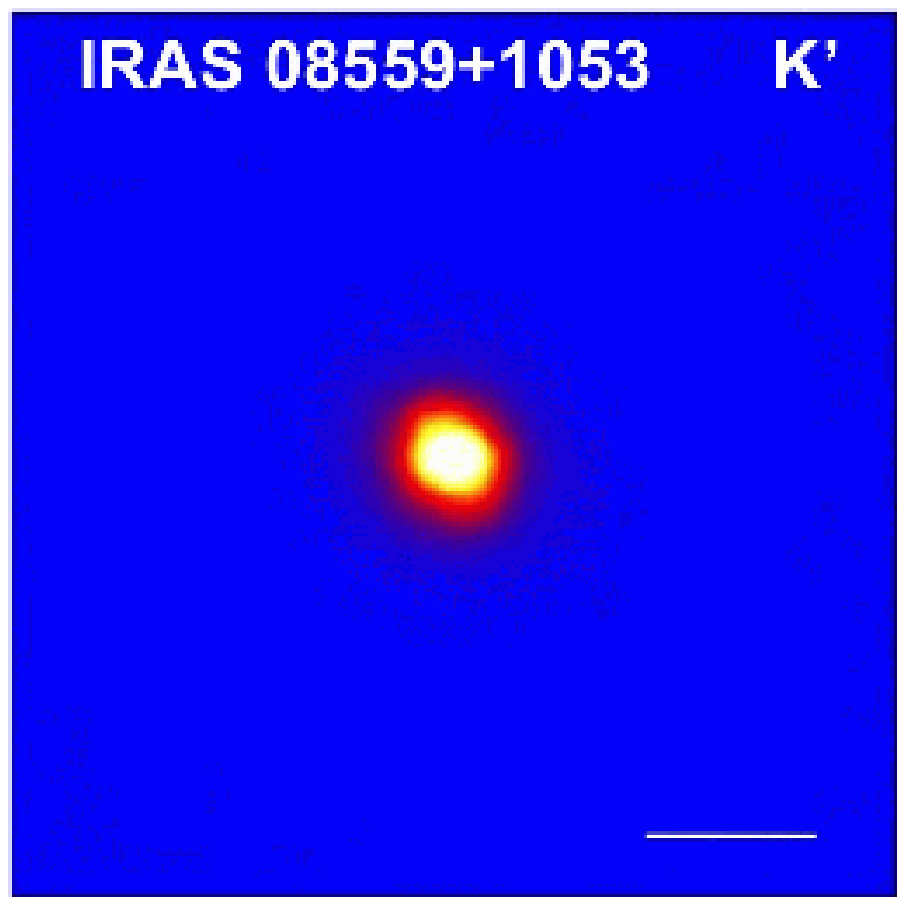} 
\includegraphics[angle=0,scale=.424]{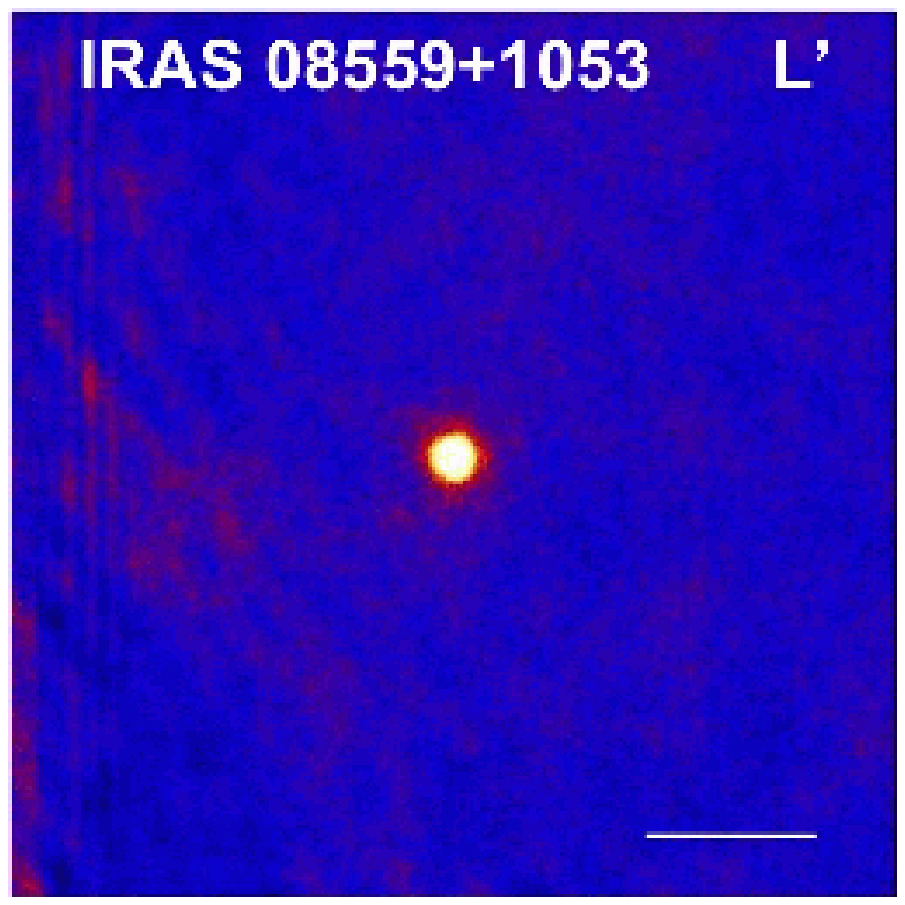} 
\includegraphics[angle=0,scale=.424]{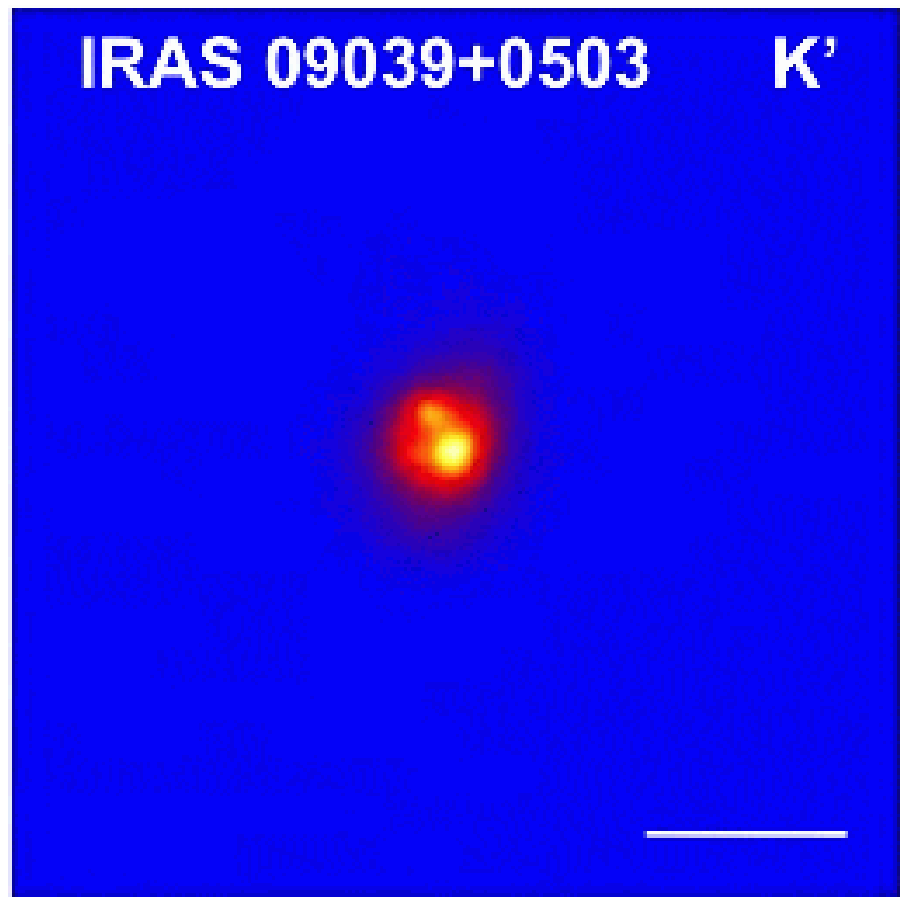} 
\includegraphics[angle=0,scale=.424]{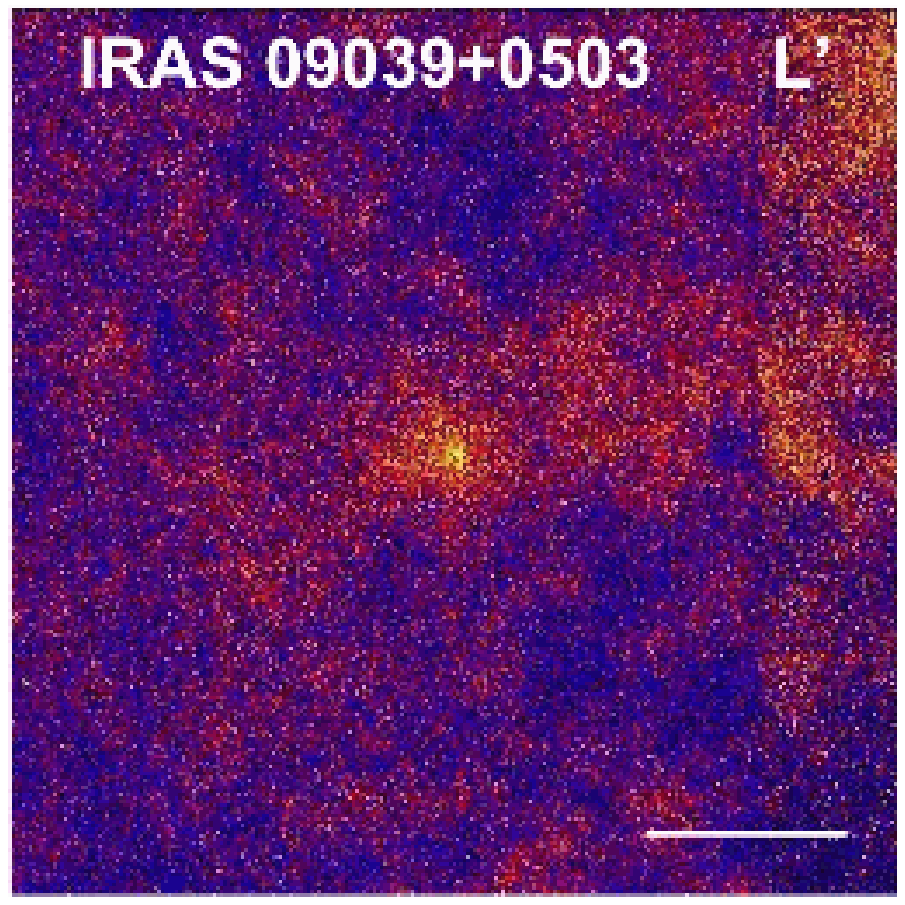} \\
\includegraphics[angle=0,scale=.424]{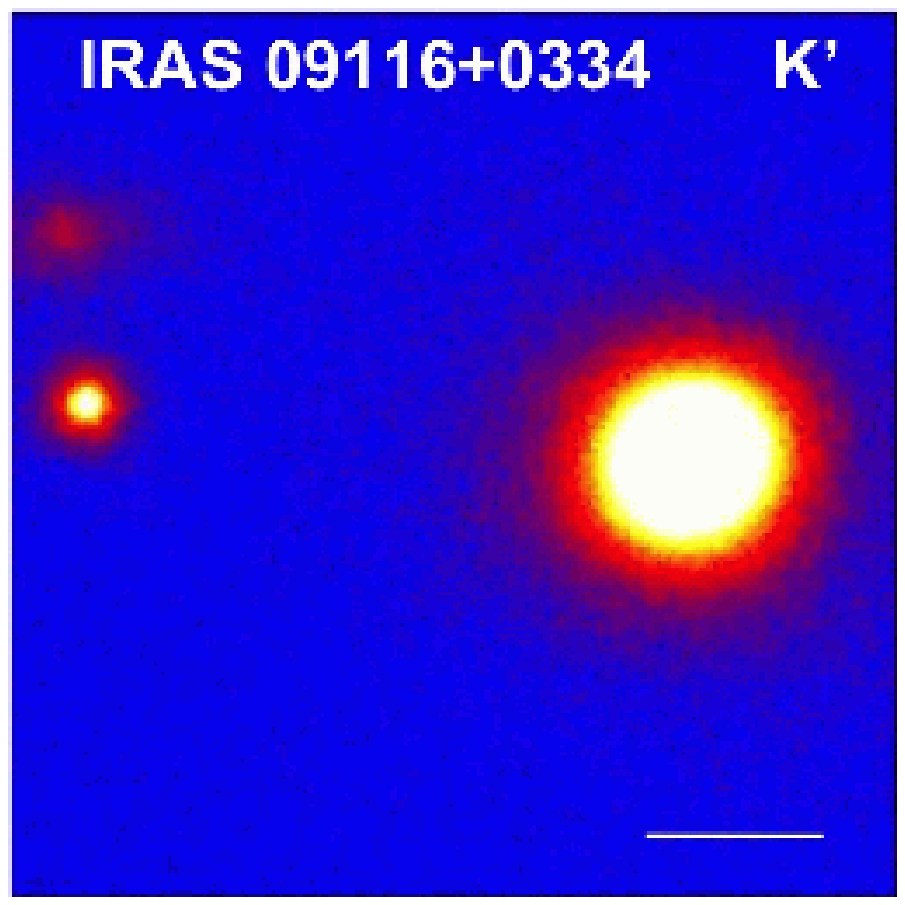} 
\includegraphics[angle=0,scale=.424]{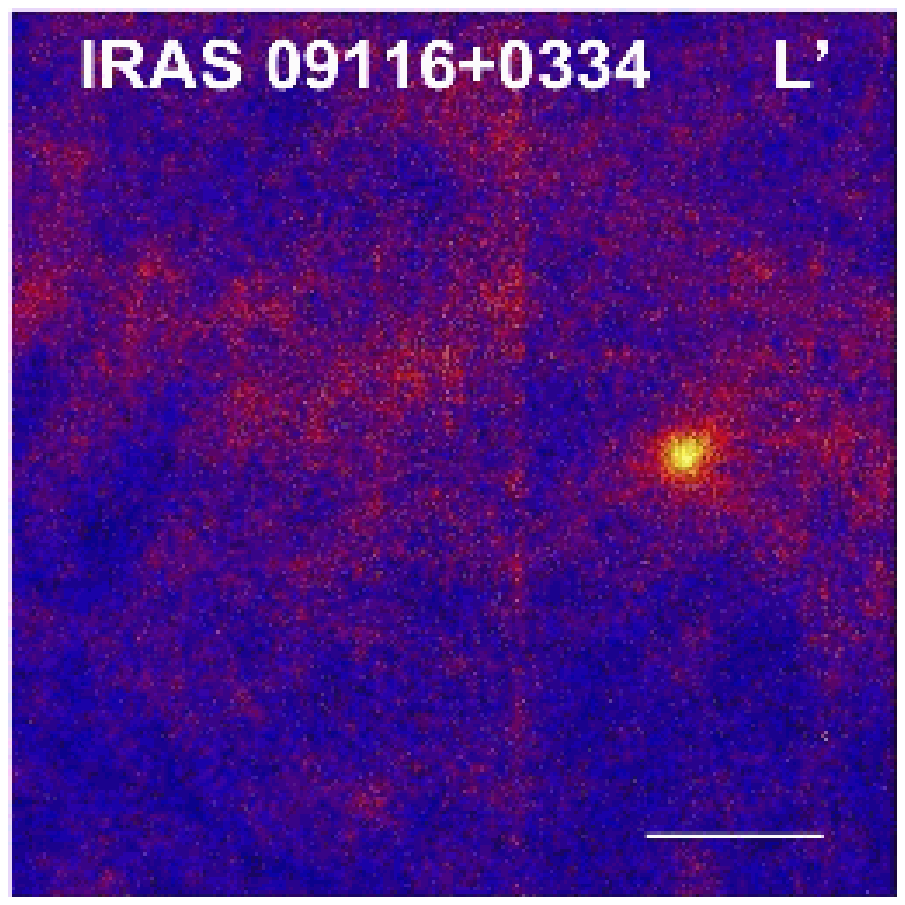} 
\includegraphics[angle=0,scale=.424]{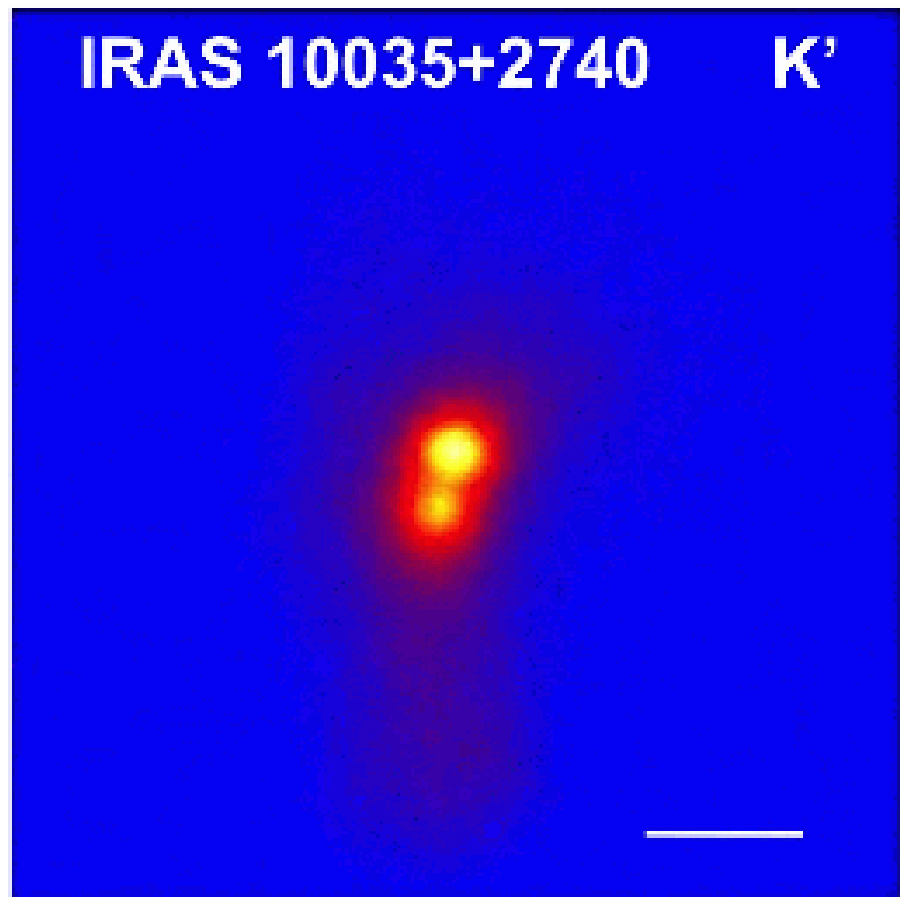} 
\includegraphics[angle=0,scale=.424]{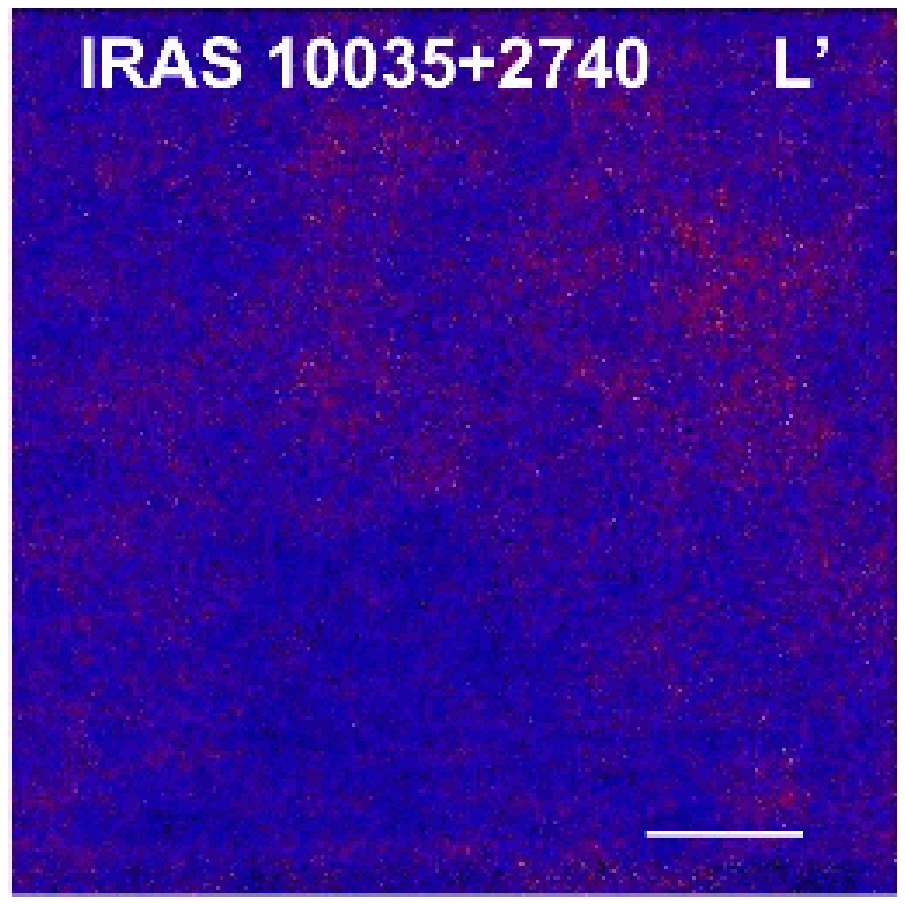} \\
\includegraphics[angle=0,scale=.424]{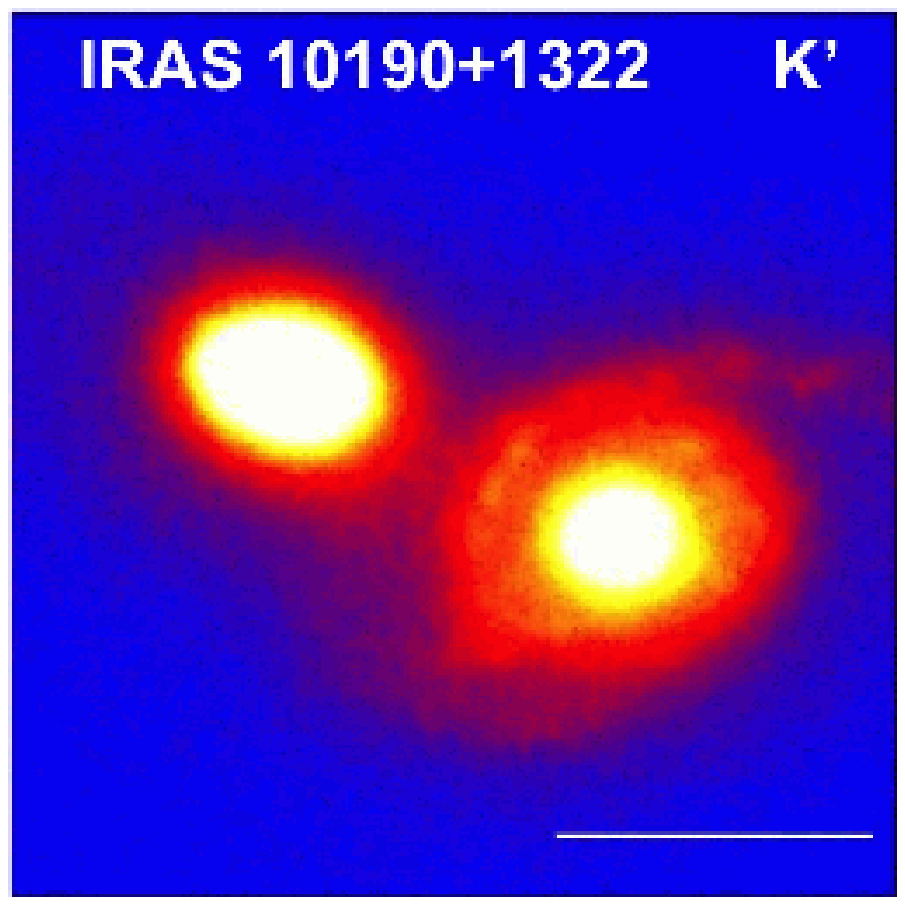} 
\includegraphics[angle=0,scale=.424]{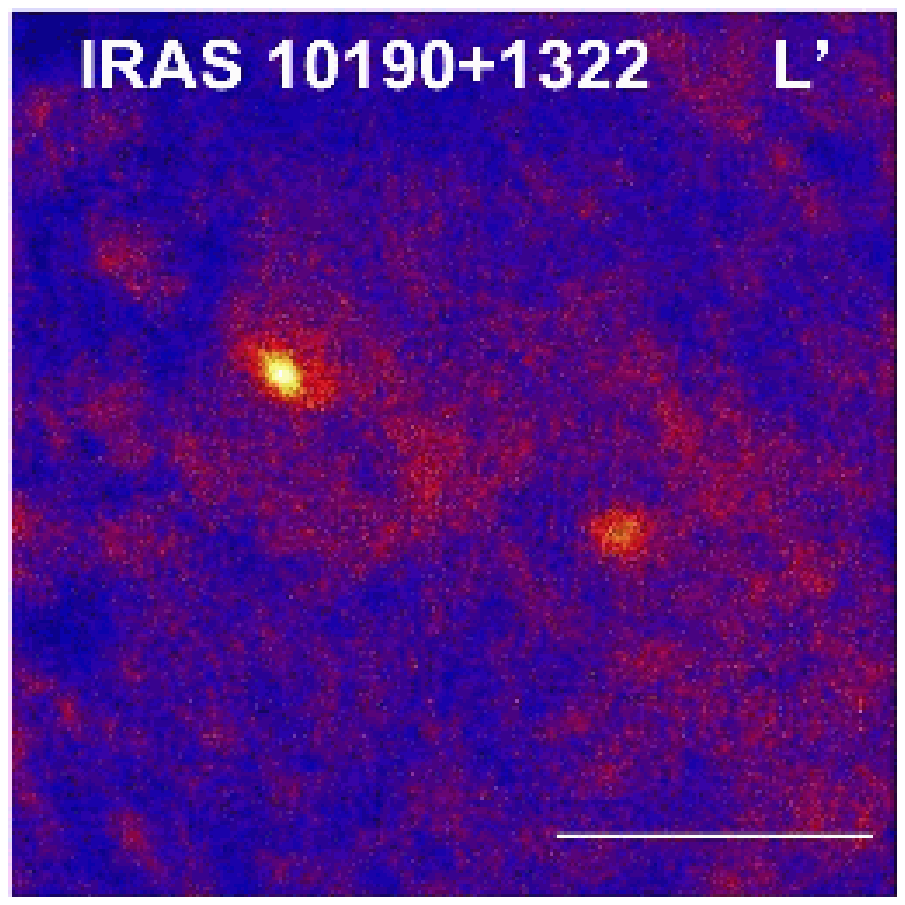} 
\includegraphics[angle=0,scale=.424]{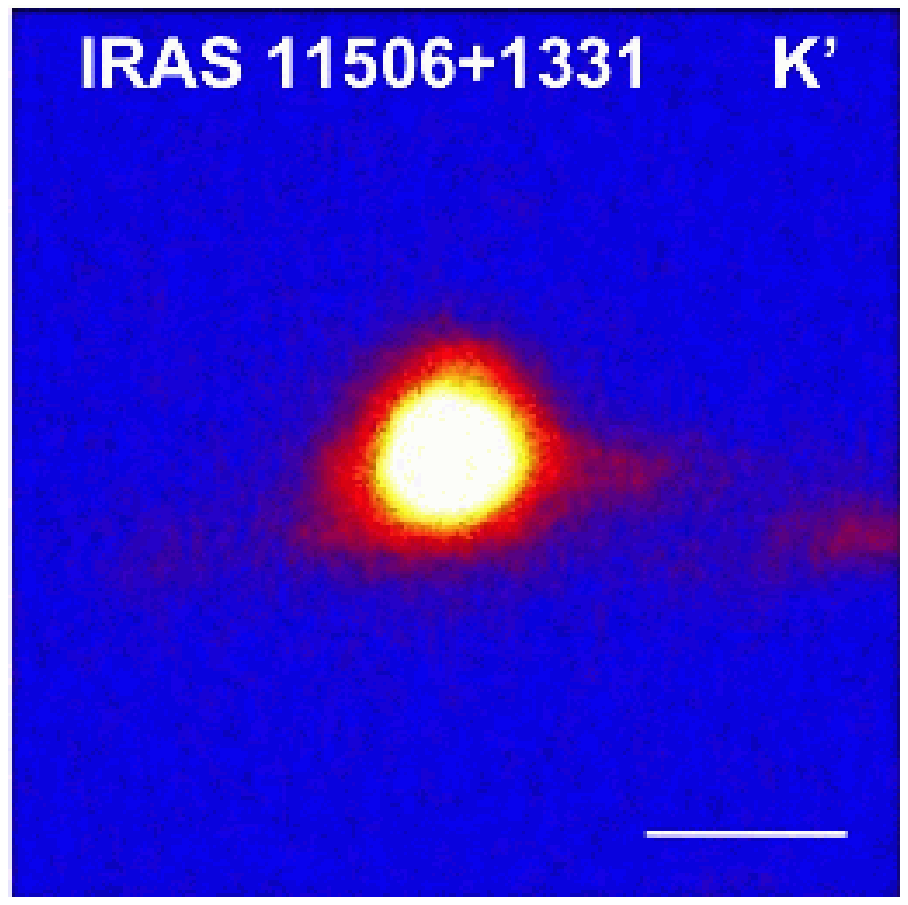} 
\includegraphics[angle=0,scale=.424]{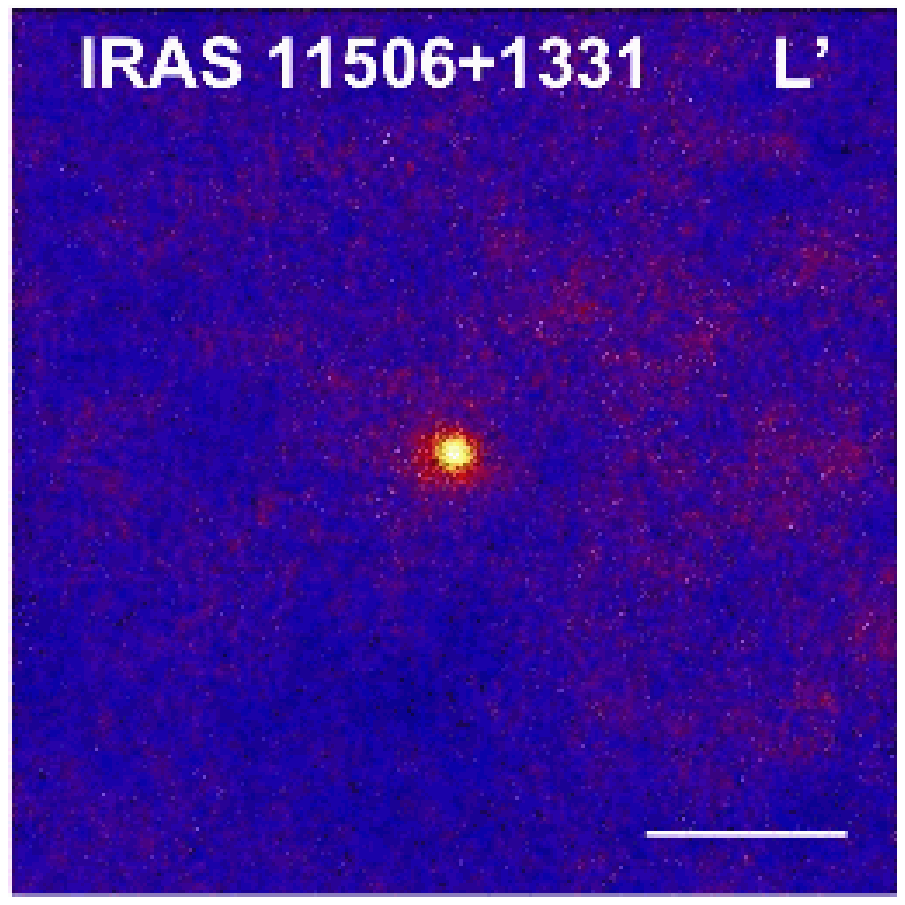} \\
\includegraphics[angle=0,scale=.424]{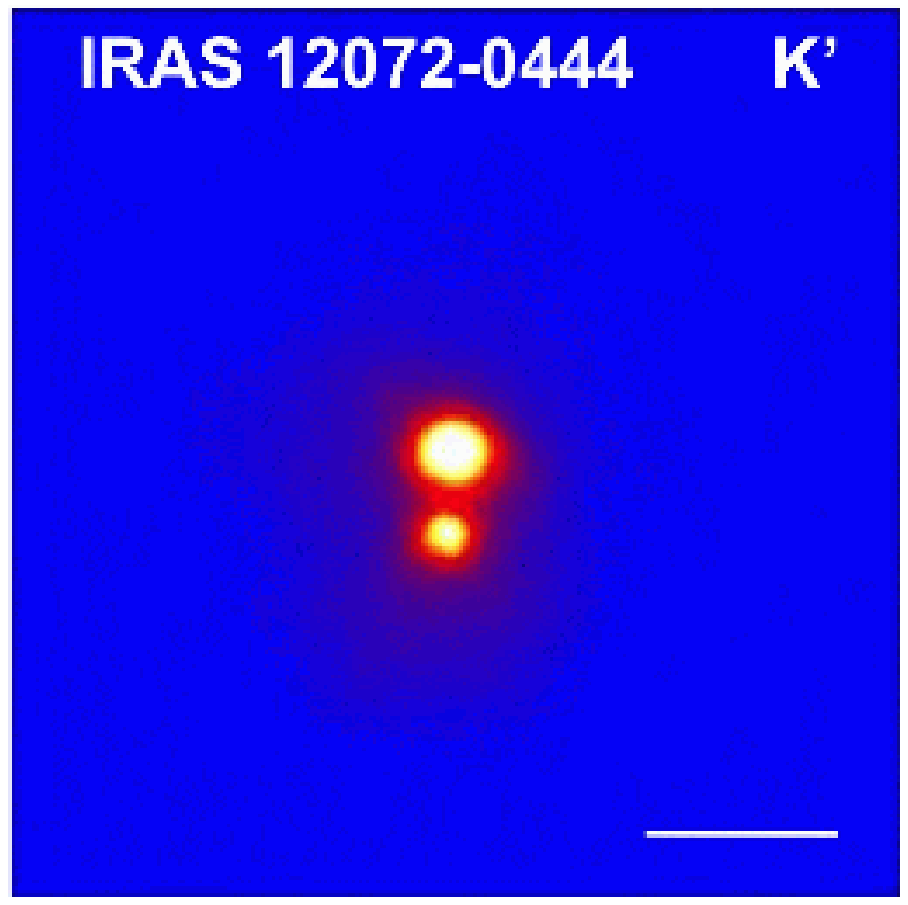} 
\includegraphics[angle=0,scale=.424]{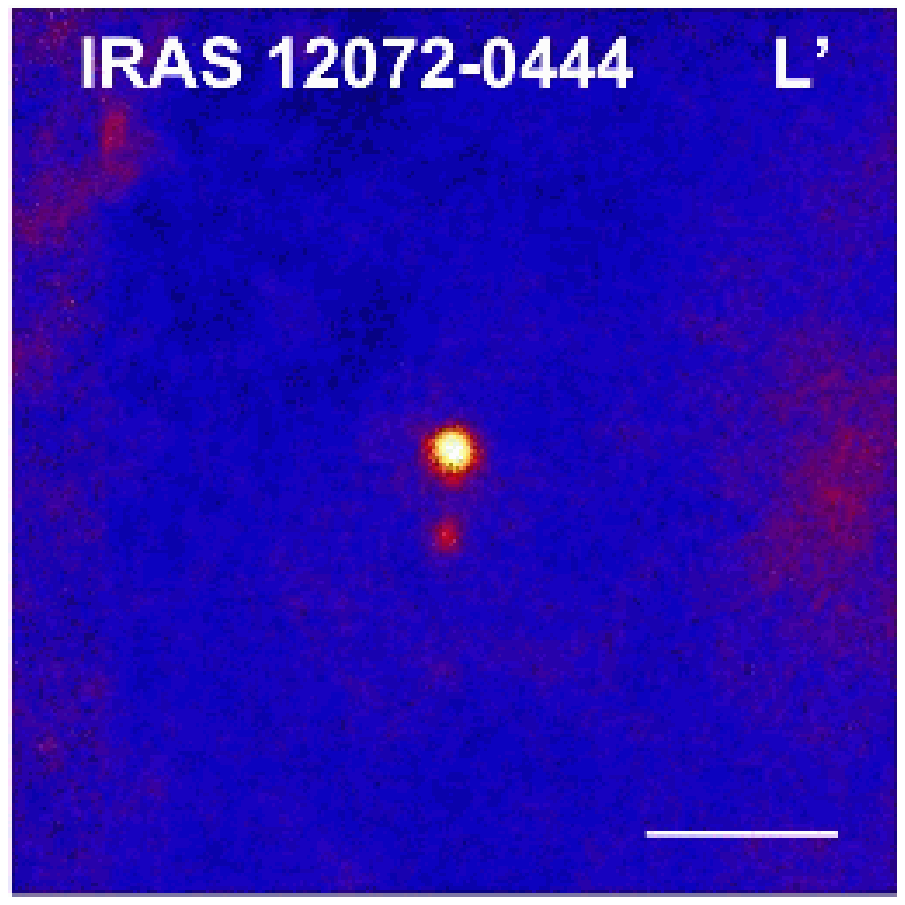} 
\includegraphics[angle=0,scale=.424]{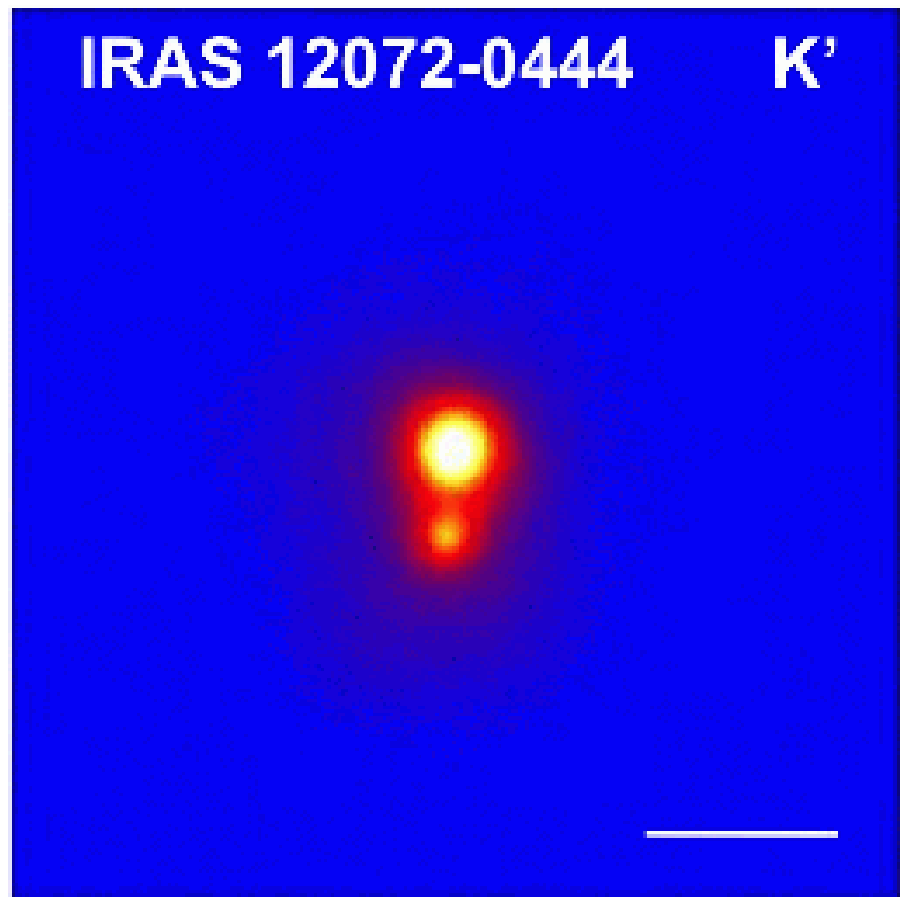} 
\includegraphics[angle=0,scale=.424]{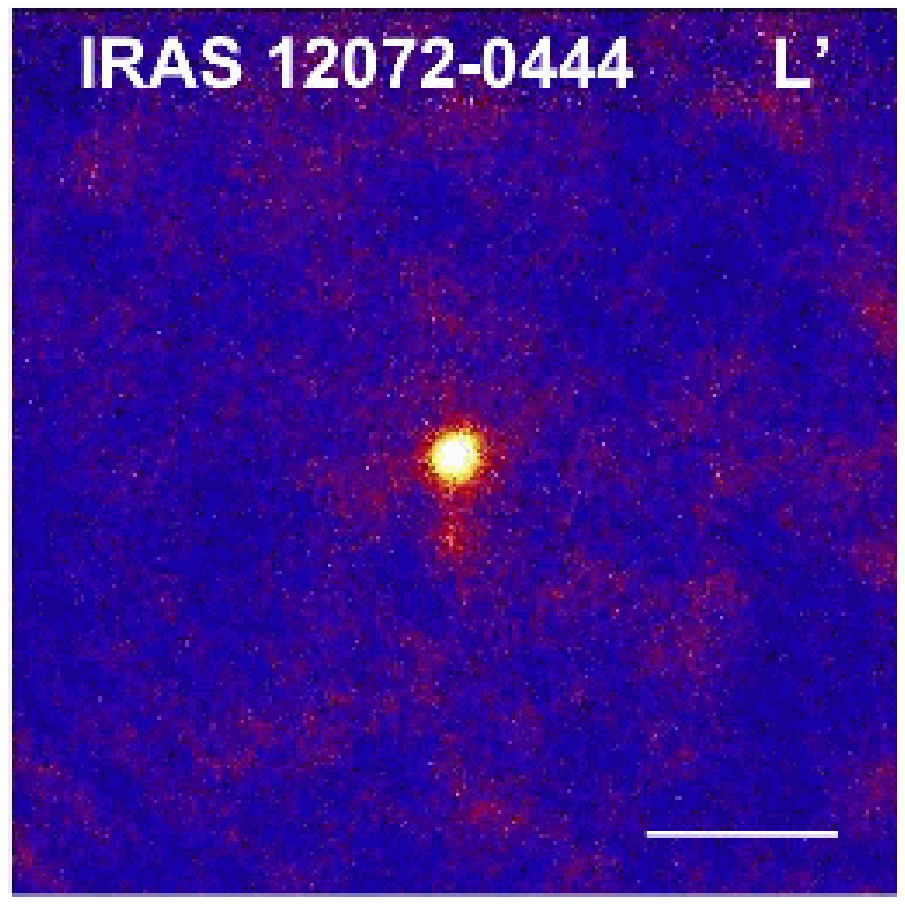} \\
\includegraphics[angle=0,scale=.424]{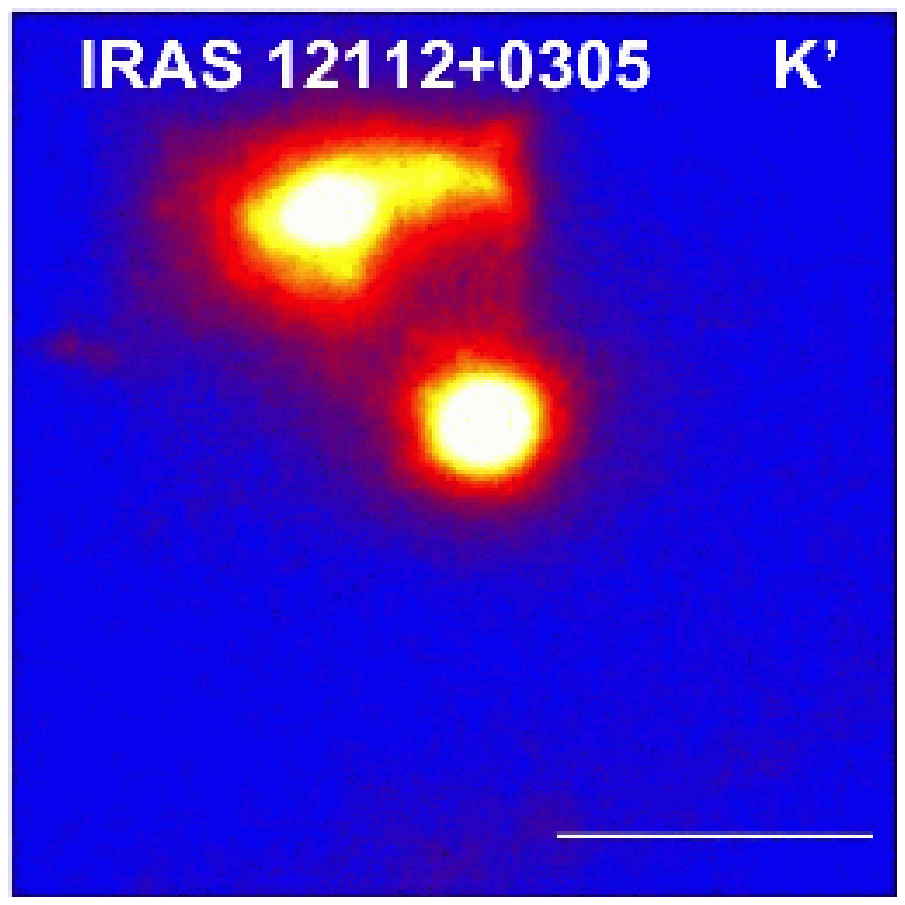} 
\includegraphics[angle=0,scale=.424]{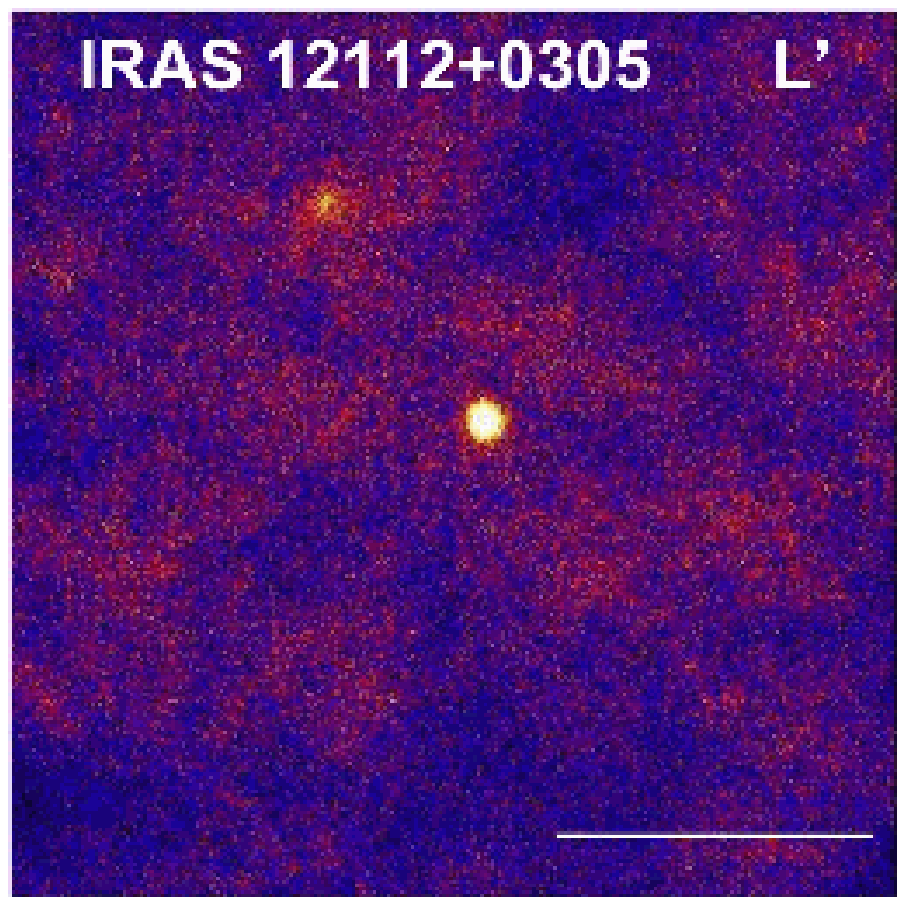} 
\includegraphics[angle=0,scale=.424]{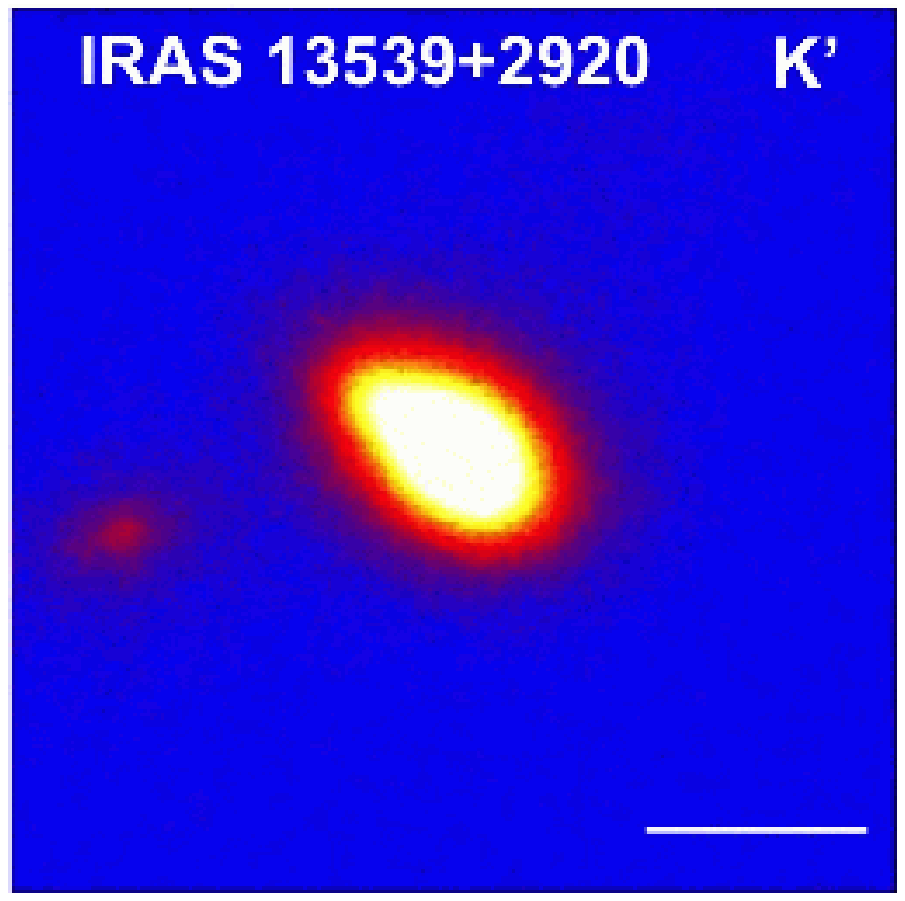} 
\includegraphics[angle=0,scale=.424]{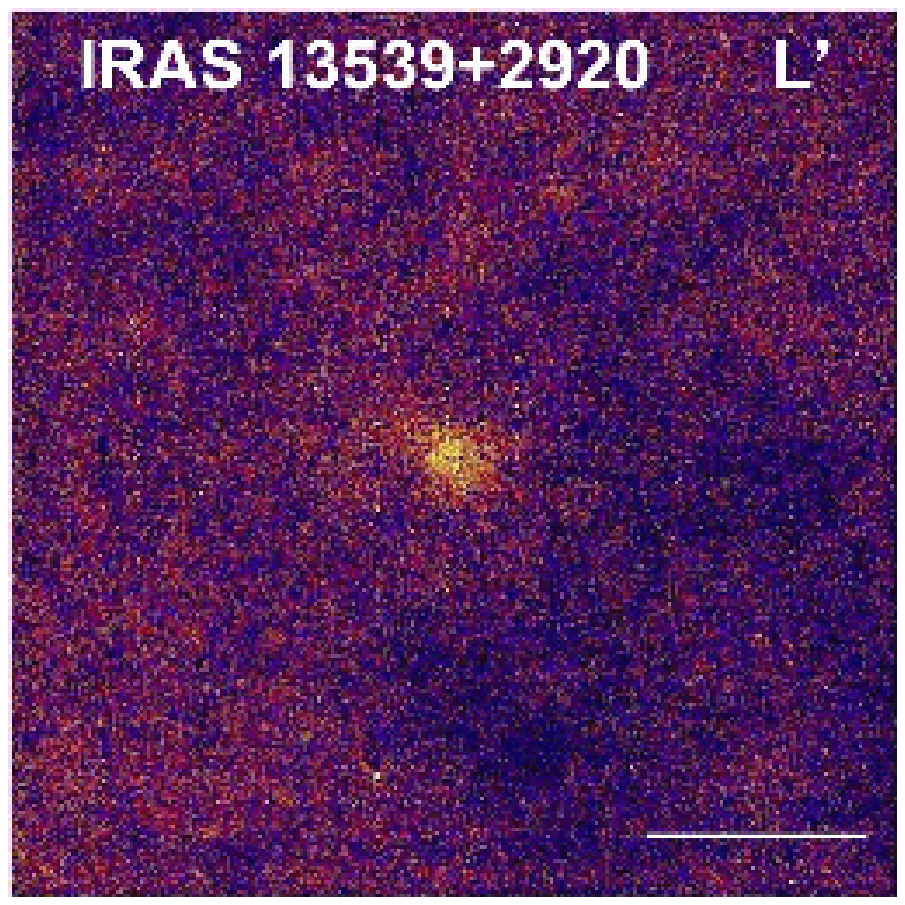} 
\end{center}
\end{figure}

\clearpage

\begin{figure}
\begin{center}
\includegraphics[angle=0,scale=.424]{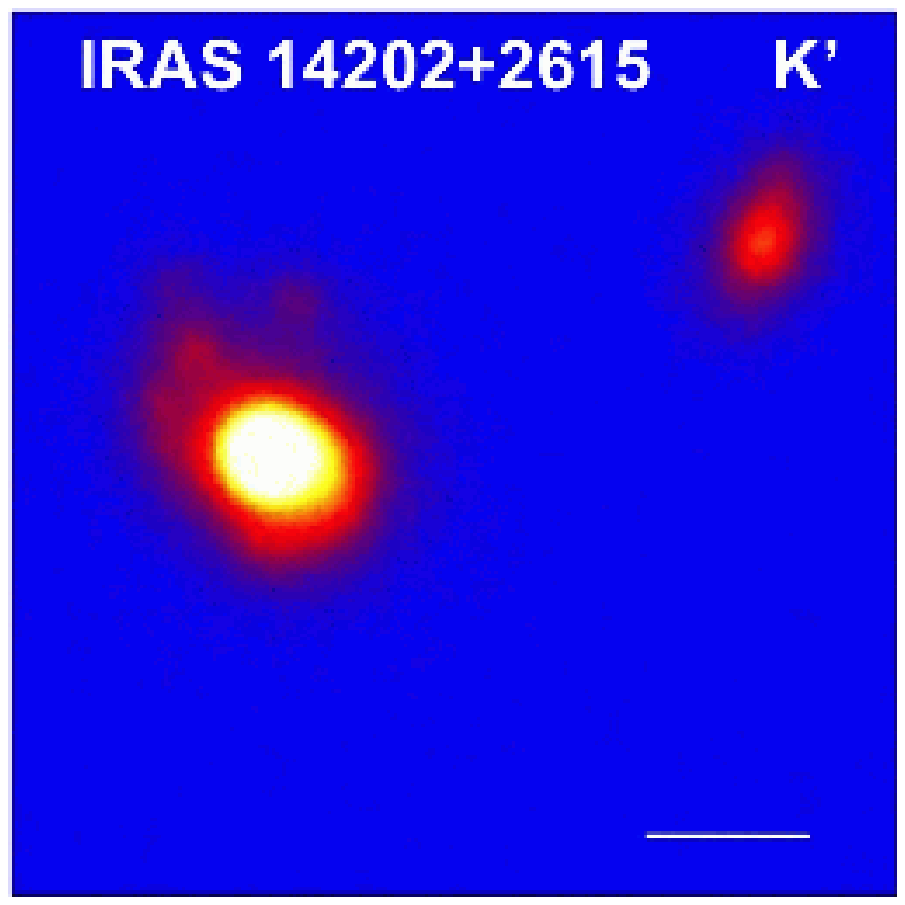} 
\includegraphics[angle=0,scale=.424]{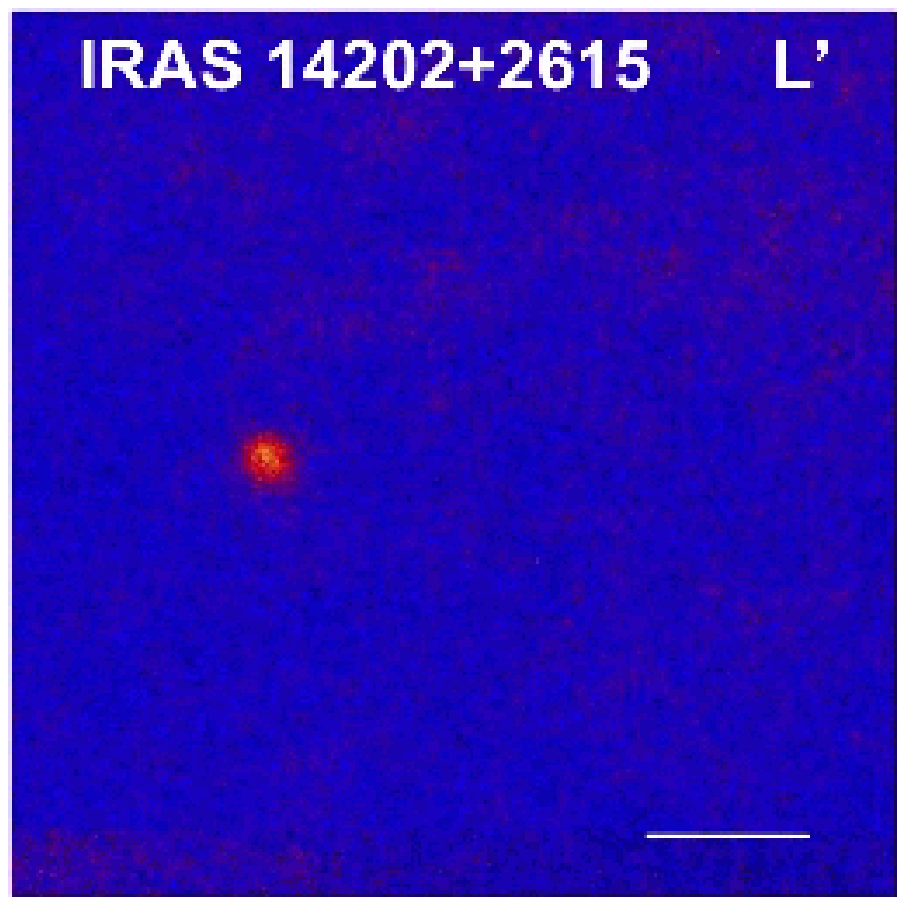} 
\includegraphics[angle=0,scale=.424]{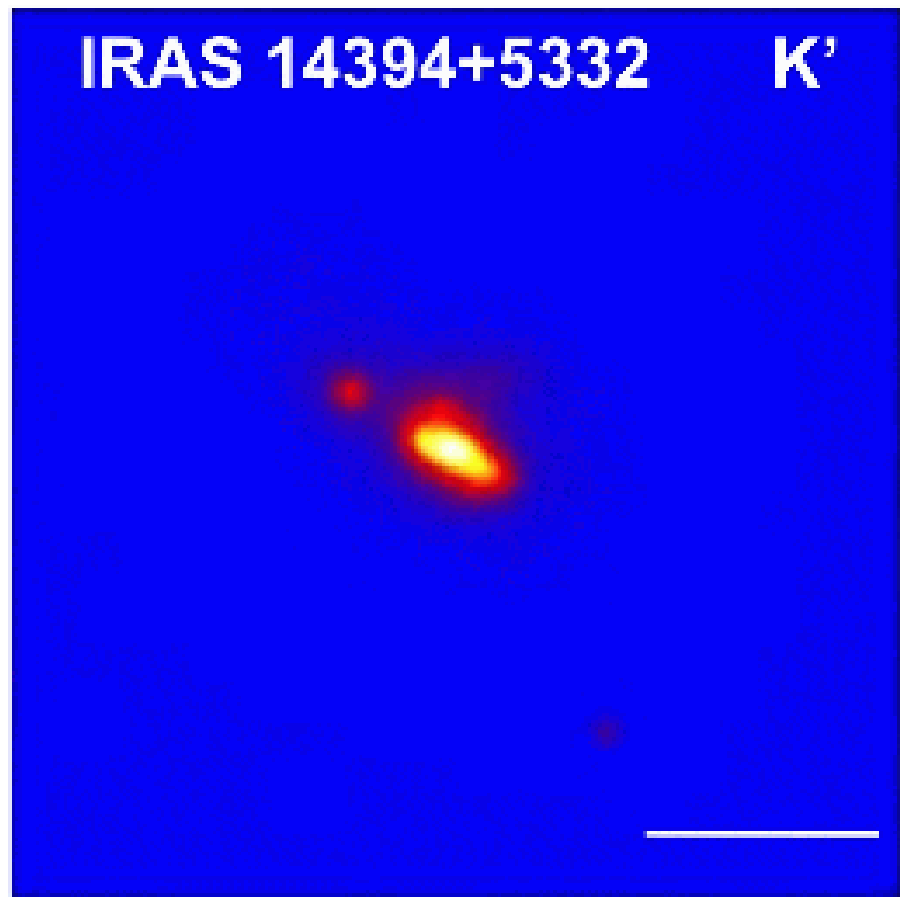} 
\includegraphics[angle=0,scale=.424]{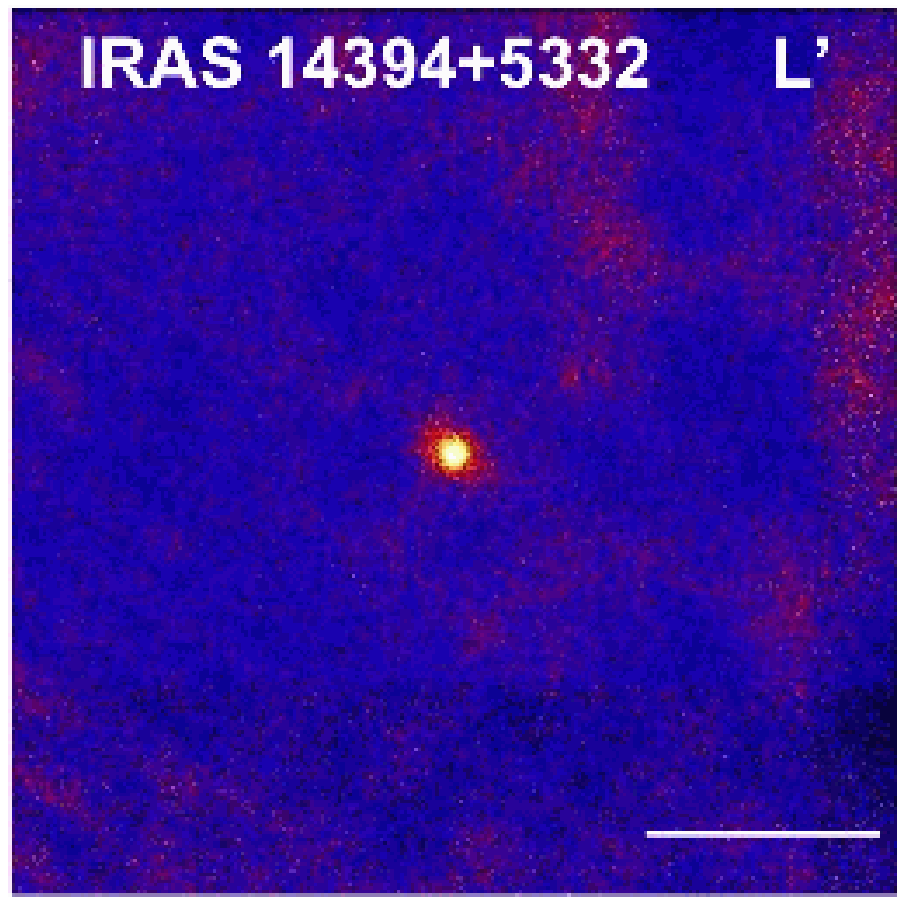}\\ 
\includegraphics[angle=0,scale=.424]{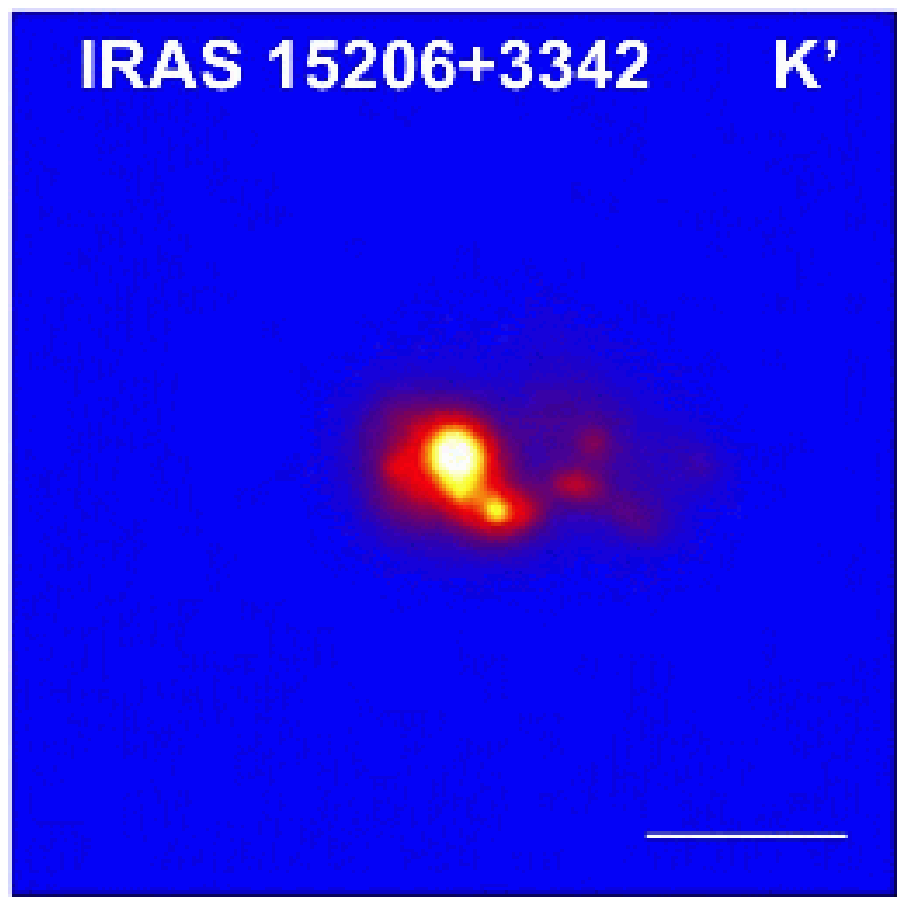} 
\includegraphics[angle=0,scale=.424]{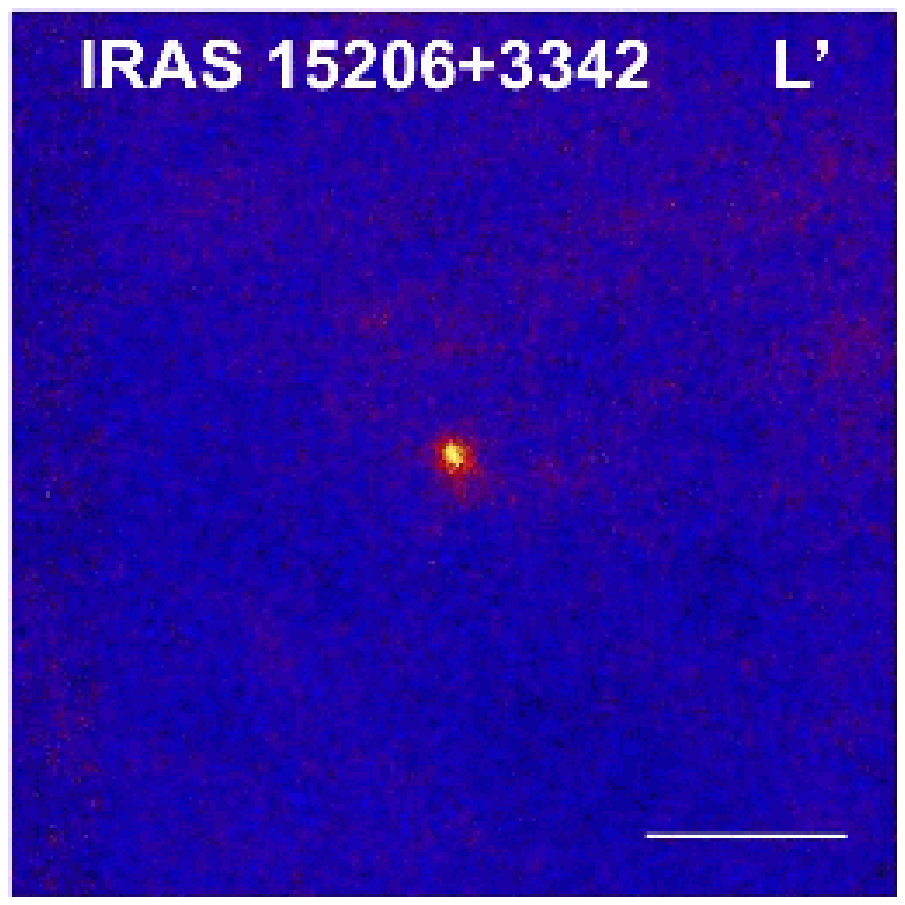} 
\includegraphics[angle=0,scale=.424]{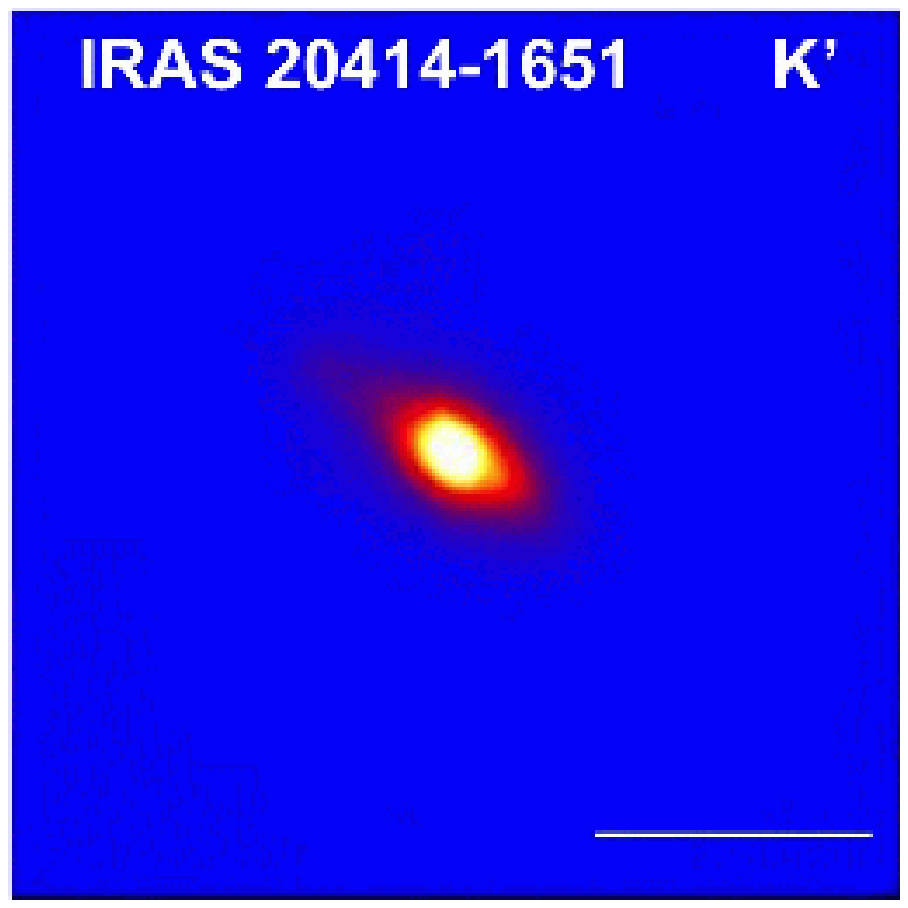} 
\includegraphics[angle=0,scale=.424]{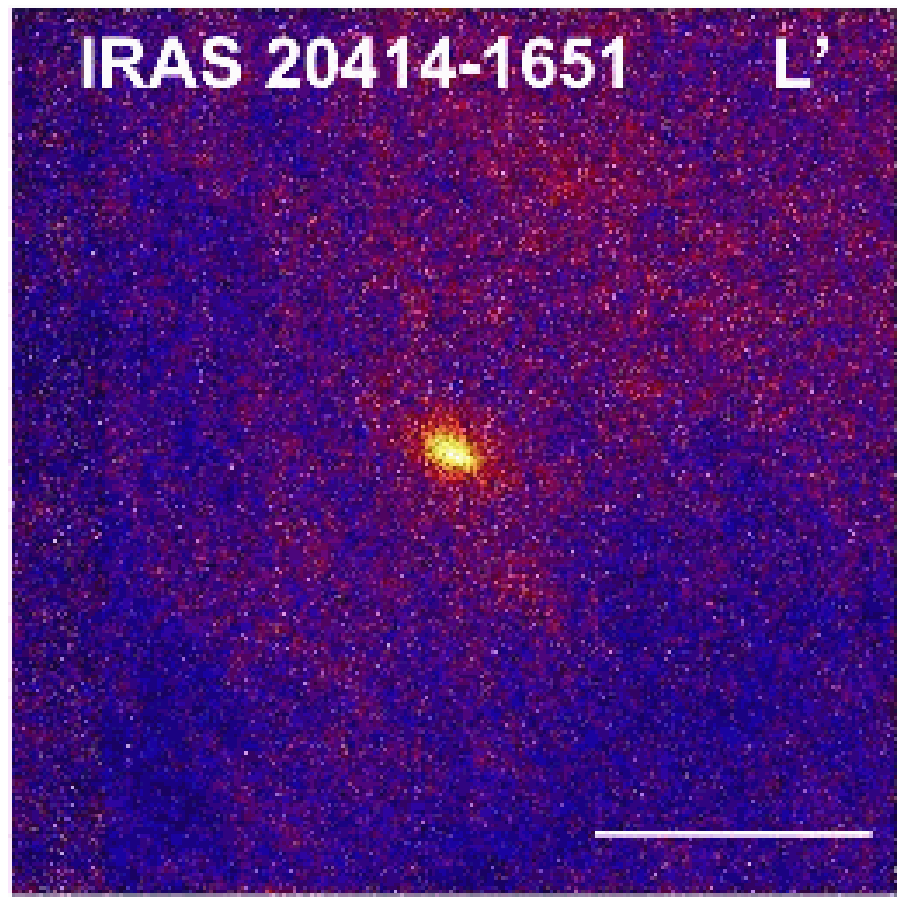} \\
\includegraphics[angle=0,scale=.424]{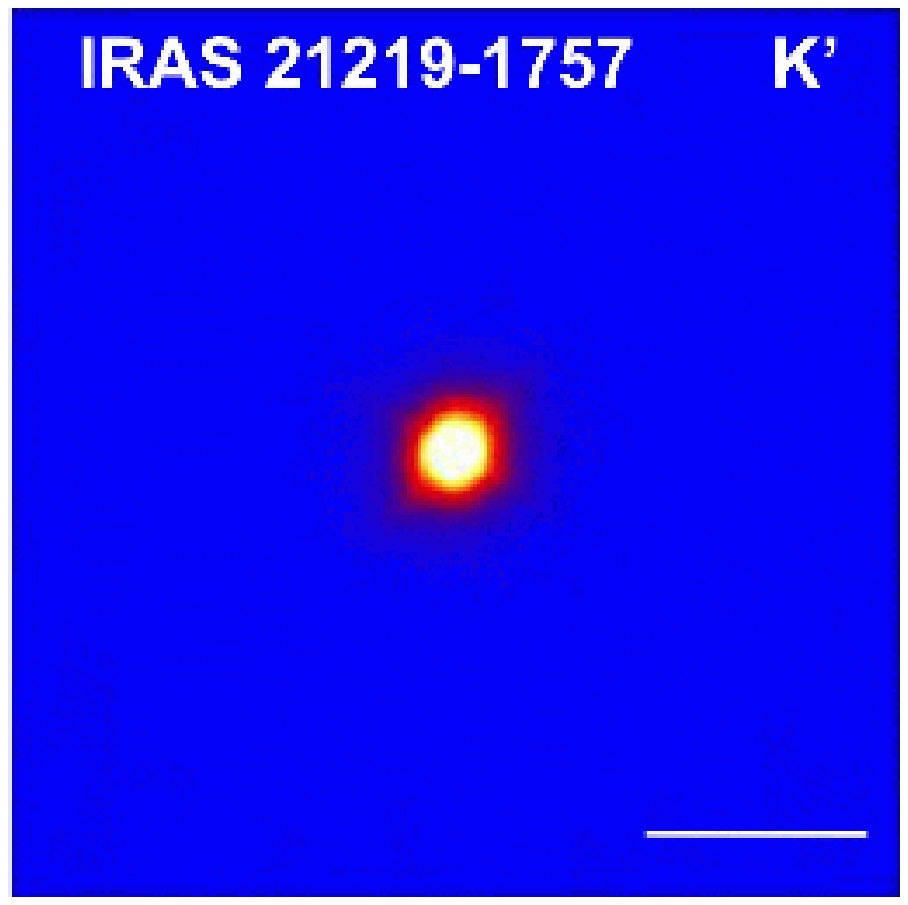} 
\includegraphics[angle=0,scale=.424]{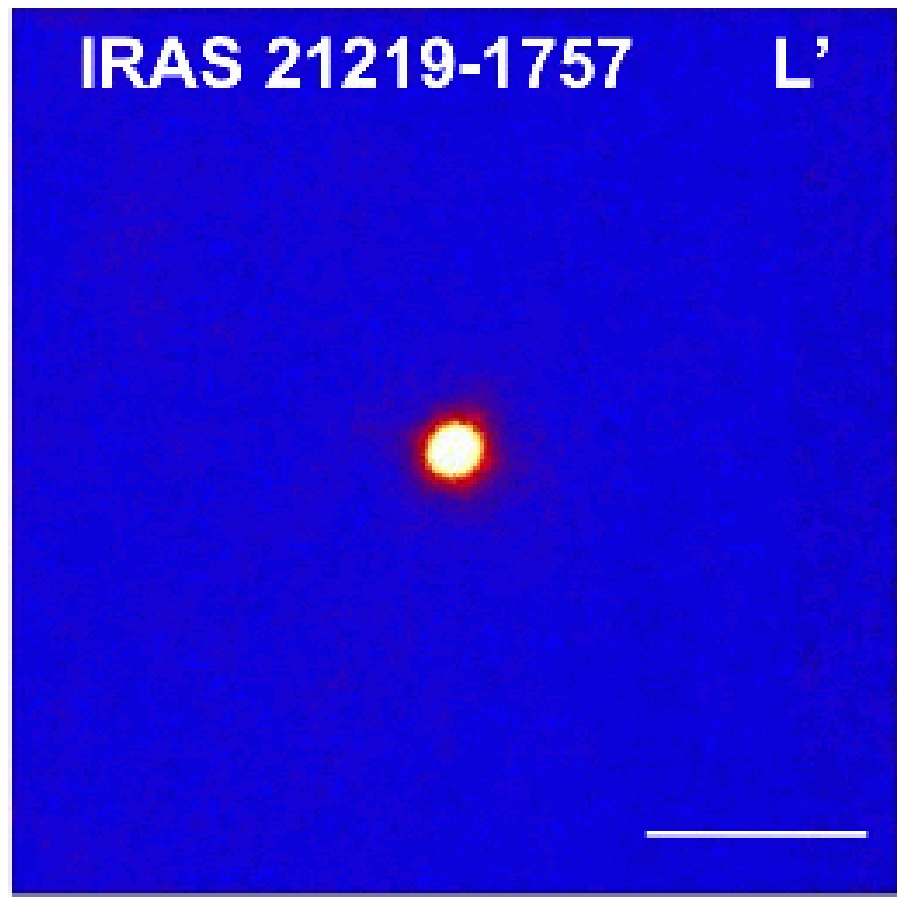} 
\includegraphics[angle=0,scale=.424]{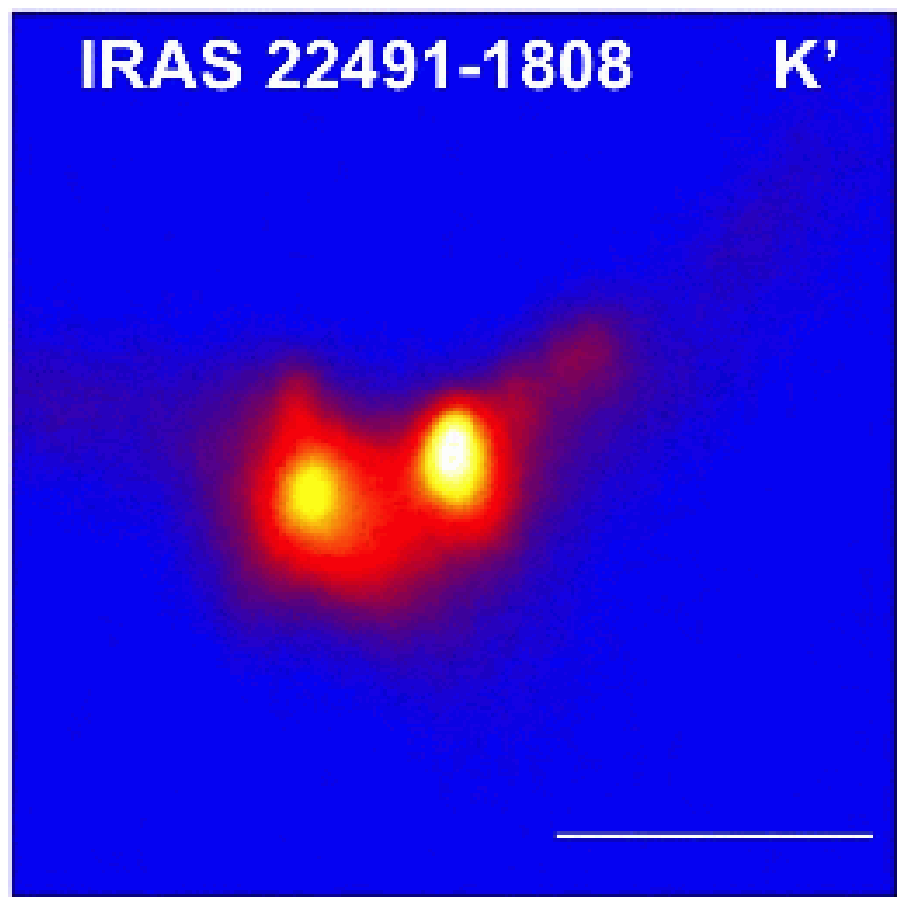} 
\includegraphics[angle=0,scale=.424]{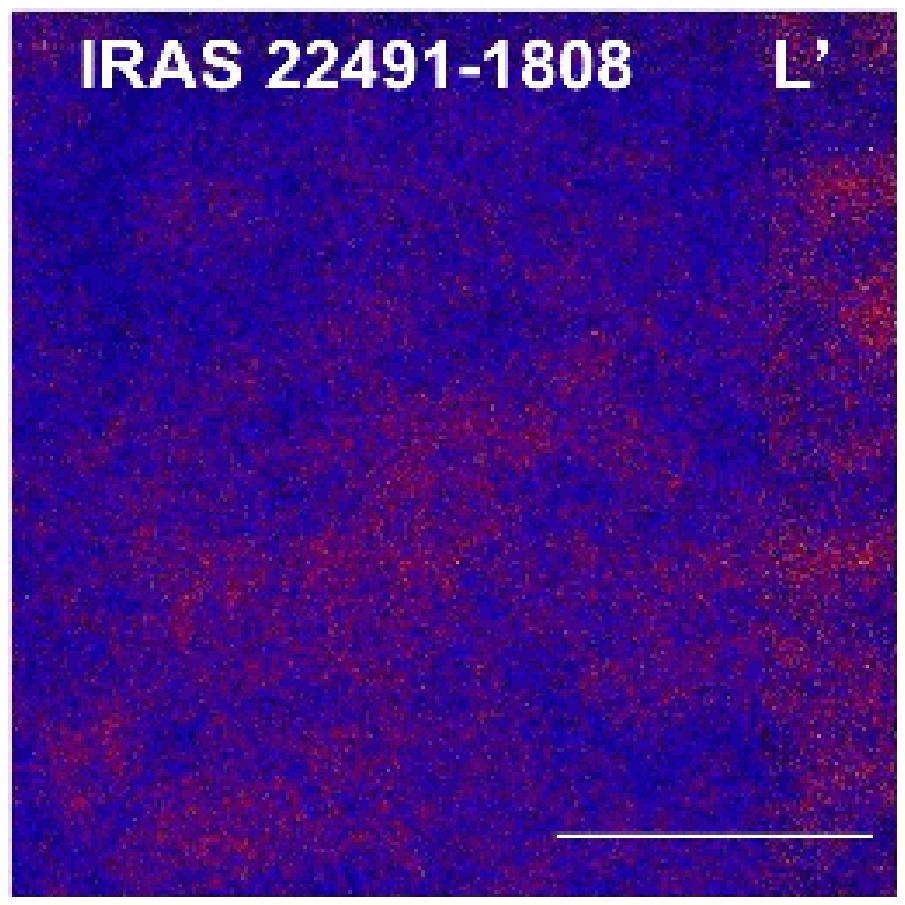} 
\end{center}
\caption{
Infrared $K'$- (2.1 $\mu$m) and $L'$-band (3.8 $\mu$m) AO images of 
observed ULIRGs.  
North is up, and east is to the left. 
The field of view (FOV) is 10$''$ $\times$ 10$''$. 
The length of the white horizontal bar at the lower right side of each 
image corresponds to 5 kpc at the distance of each ULIRG.
For IRAS 12072$-$0444 (fifth row), the left and right two panels are data 
taken in 2019 April (LGS-AO) and 2018 May (NGS-AO), respectively.
The maximum and minimum signal display scale of each image is adjusted 
to make interesting features clearly visible.
For IRAS 14394$+$5332, we observed the E and EE nuclei in \citet{kim02}.
For IRAS 00456$-$2904, the fainter NE nucleus $\sim$11$\farcs$3 
separated from the brighter SW nucleus is outside the field of view and 
is not displayed. 
This fainter NE nucleus was detected by \citet{kim02} in a seeing-limited 
infrared image at $\sim$2.2 $\mu$m and is not newly 
resolved into multiple emission components in our AO-assisted 
higher-spatial-resolution $K'$-band image.
IRAS 04103$-$2838, IRAS 08559$+$1053, IRAS 11506$+$1331, 
IRAS 20414$-$1651, and IRAS 21219$-$1757 are classified as single 
nucleus ULIRGs in our $K'$-band images.
For $L'$-band undetected ULIRGs, we have no way to accurately identify 
the nuclear peak position using actual signals. 
The nuclear position may be slightly displaced from the center of 
each image, but we see no significant emission even in the surrounding 
area.
}
\end{figure}

\clearpage

\begin{figure}
\begin{center}
\includegraphics[angle=0,scale=.45]{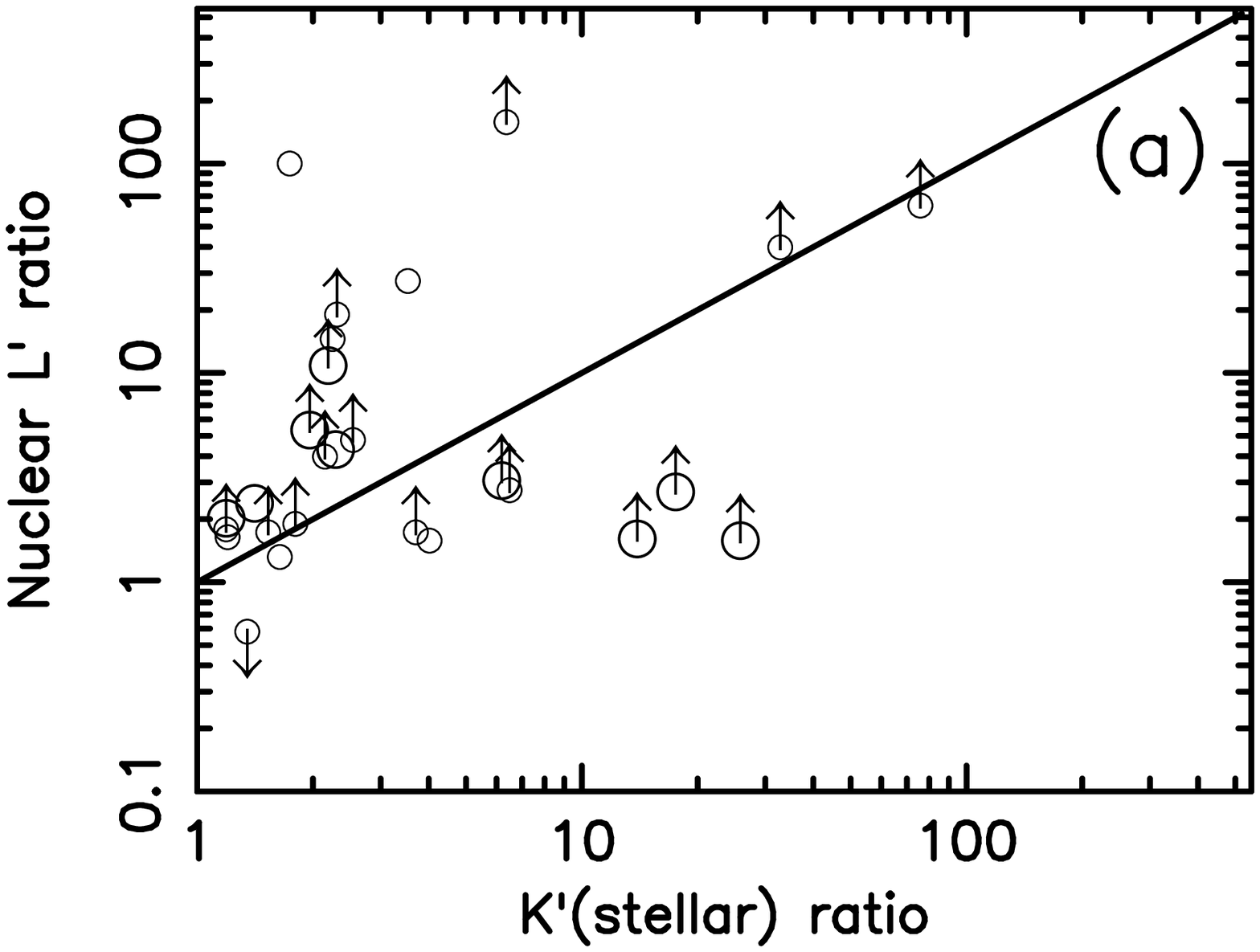} 
\includegraphics[angle=0,scale=.45]{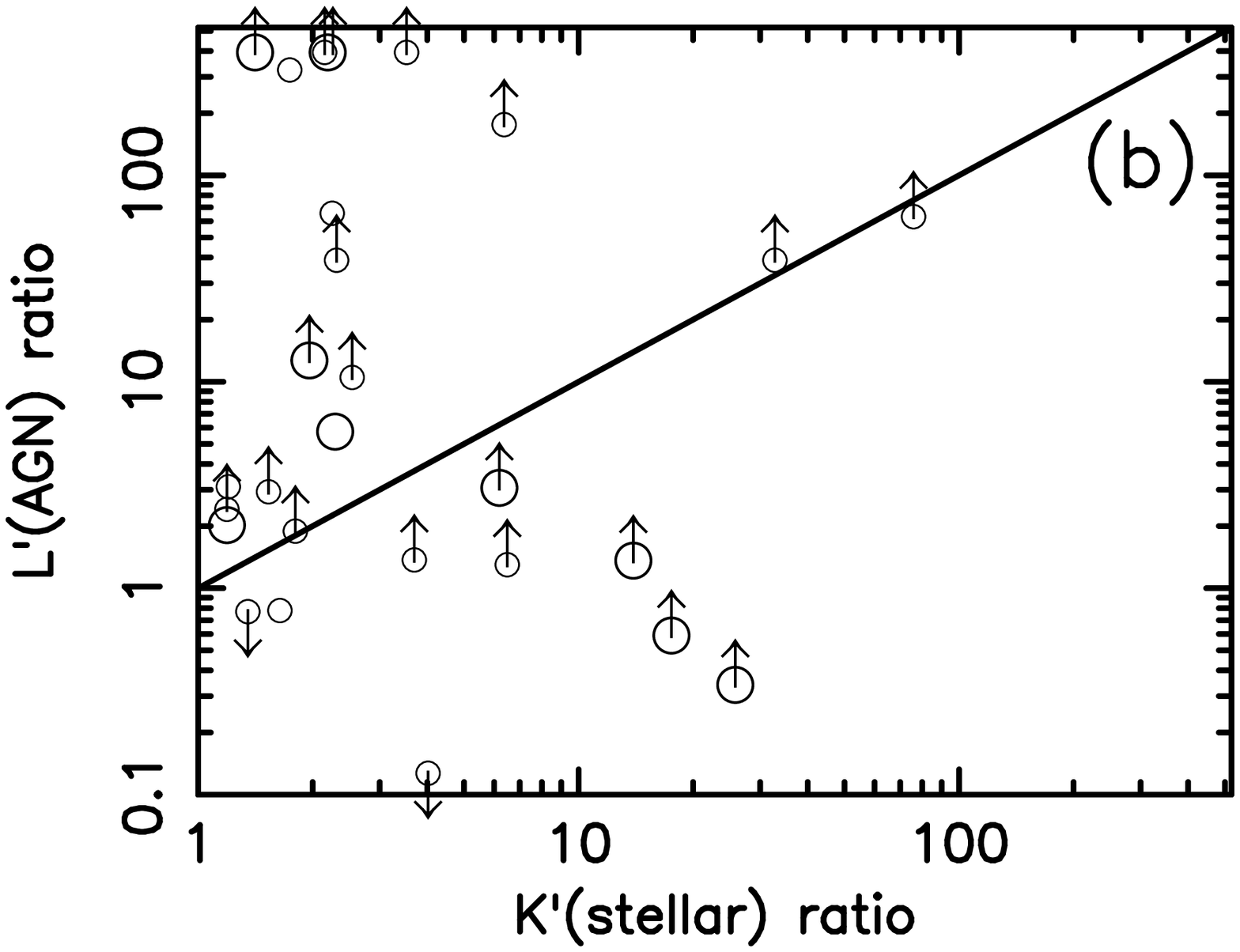} 
\caption{
{\it (a)}: Luminosity ratio between the $K'$-band brighter primary and 
fainter secondary galaxy nucleus. 
The luminosity at the primary nucleus is divided by that at the 
secondary nucleus.
The abscissa is the ratio of $K'$-band emission (central 
4 kpc diameter aperture photometry) \citep{kim02}, which is regarded 
as a stellar luminosity ratio or approximately a central SMBH mass 
ratio between the primary and secondary galaxy nuclei (see $\S$5.2).
The ordinate is the ratio of nuclear $L'$-band emission (0$\farcs$5 
radius aperture photometry), largely coming from a luminous AGN, 
and so is taken as the AGN luminosity ratio between the primary and 
secondary galaxy nuclei. 
The solid straight line indicates the same ratio between the abscissa 
and ordinate. 
If the primary galaxy nucleus has a more active SMBH with higher 
SMBH-mass-normalized AGN luminosity than the secondary galaxy nucleus, 
this source is located above the solid straight line.
{\it (b)}: Same as (a), but the ordinate is the AGN-origin $L'$-band 
luminosity ratio, after removing possible stellar contaminations 
to the observed nuclear $L'$-band emission in each galaxy nucleus 
(Table 3, column 8).
In both (a) and (b), larger circles indicate ULIRGs observed in the 
current study, and smaller circles are sources presented in \citet{ima14}.
}
\end{center}
\end{figure}

\begin{figure}
\begin{center}
\includegraphics[angle=0,scale=.45]{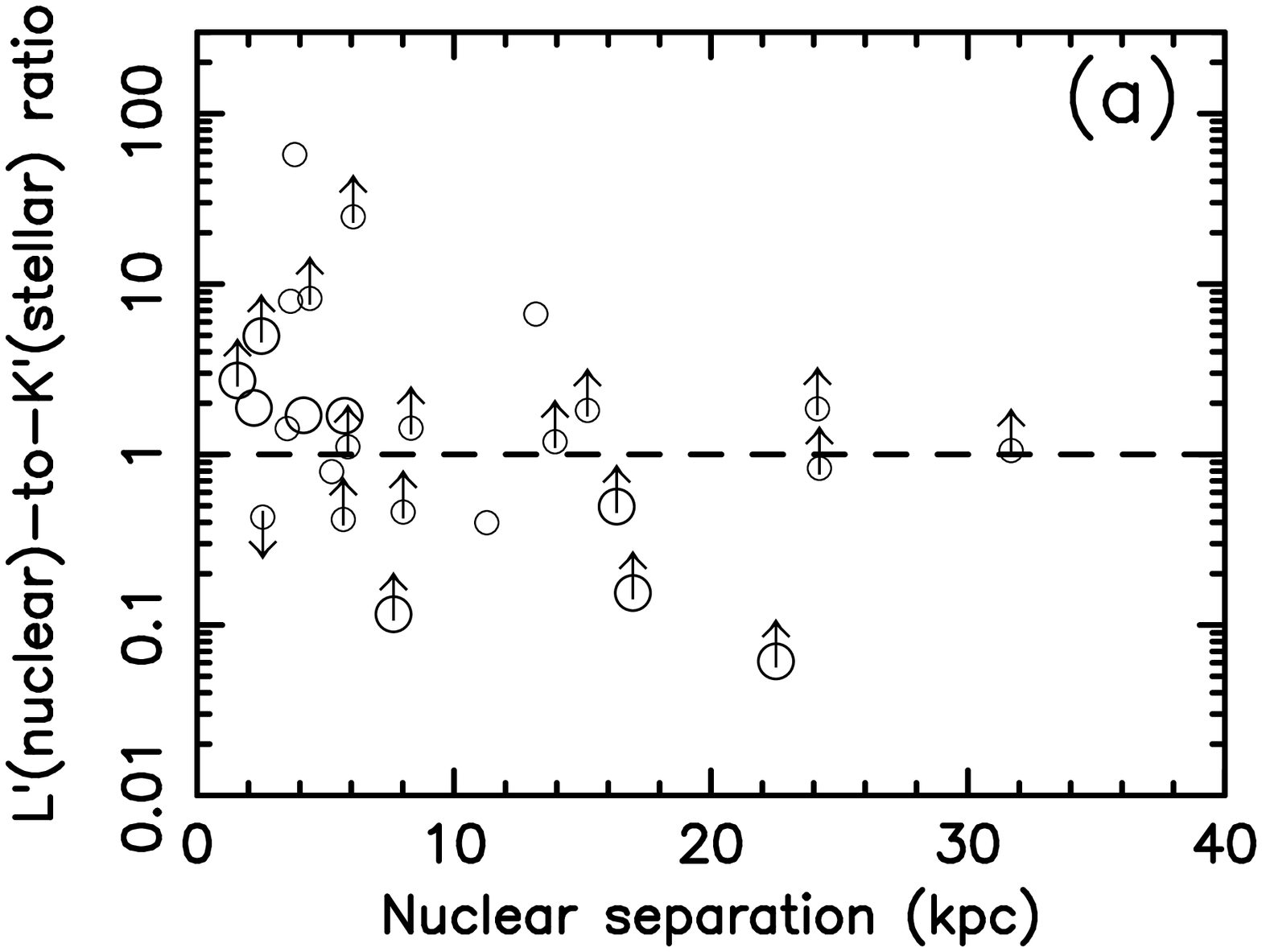} 
\includegraphics[angle=0,scale=.45]{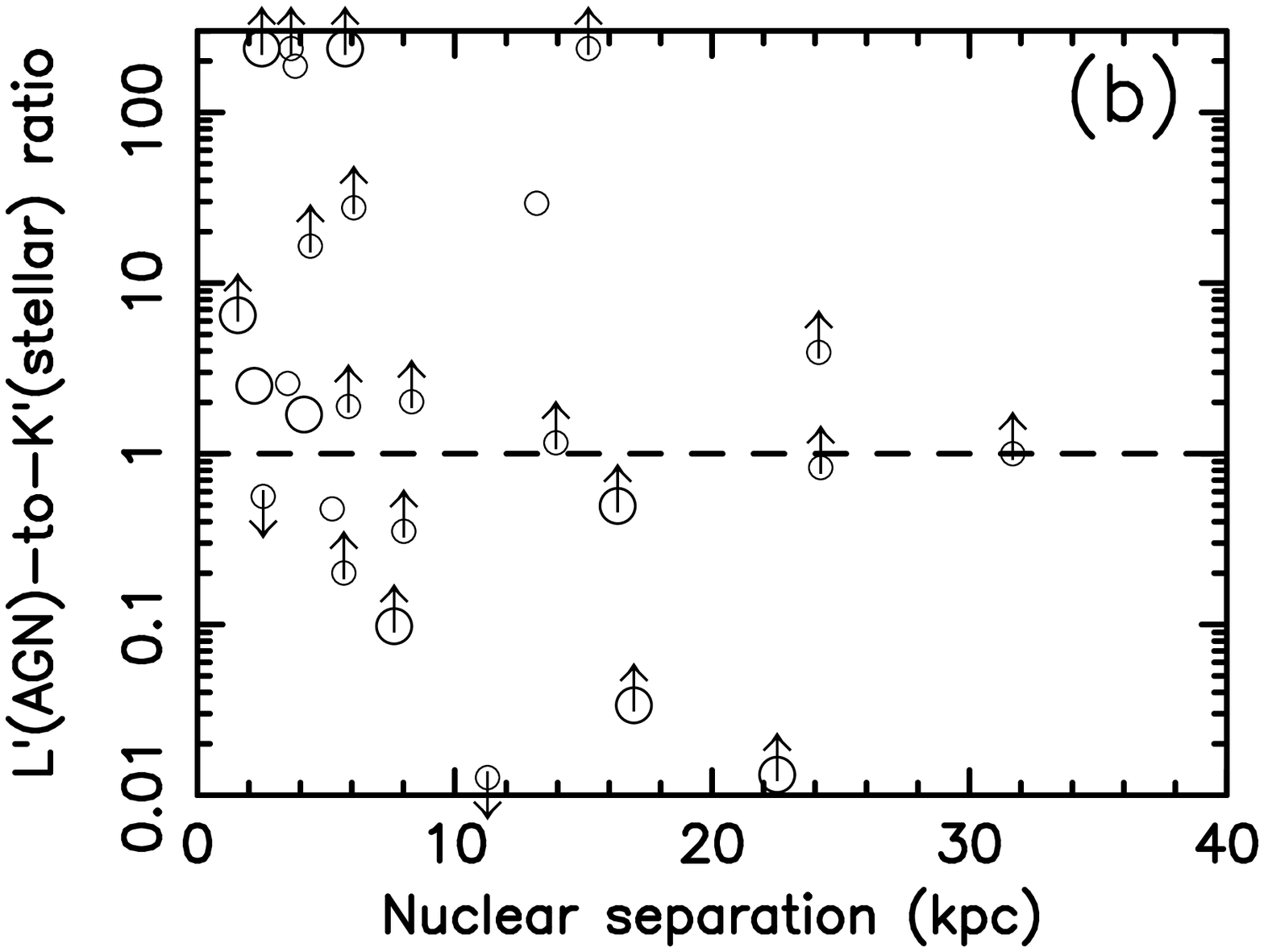} 
\caption{
{\it (a)}: Ratio of SMBH-mass-normalized AGN luminosity between 
primary and secondary galaxy nuclei (see Figure 3) as a function of 
nuclear separation.
The abscissa is projected nuclear separation in kiloparsecs.
The ordinate is the ratio of ``the primary to secondary nuclear $L'$-band 
luminosity ratio'' to ``the primary to secondary nuclear 4 kpc diameter 
$K'$-band luminosity ratio''.
The horizontal dashed line indicates the ratio of unity. 
Sources to the upper left (lower right) of the solid straight line 
in Figure 3(a) are now located at the upper (lower) side of the horizontal 
dashed straight line and indicate that the primary galaxy nucleus 
has a more active (less active) SMBH with higher (lower) 
SMBH-mass-normalized AGN luminosity than the secondary galaxy nucleus. 
{\it (b)}: Same as (a), but AGN-origin nuclear $L'$-band luminosity, 
after removing possible stellar contributions to the nuclear 
$L'$-band emission (Table 3, column 8), is used in the ordinate.  
In both (a) and (b), larger and smaller circles indicate ULIRGs observed 
in this study and in \citet{ima14}, respectively.
}
\end{center}
\end{figure}

\clearpage

\begin{figure}
\begin{center}
\includegraphics[angle=0,scale=.4]{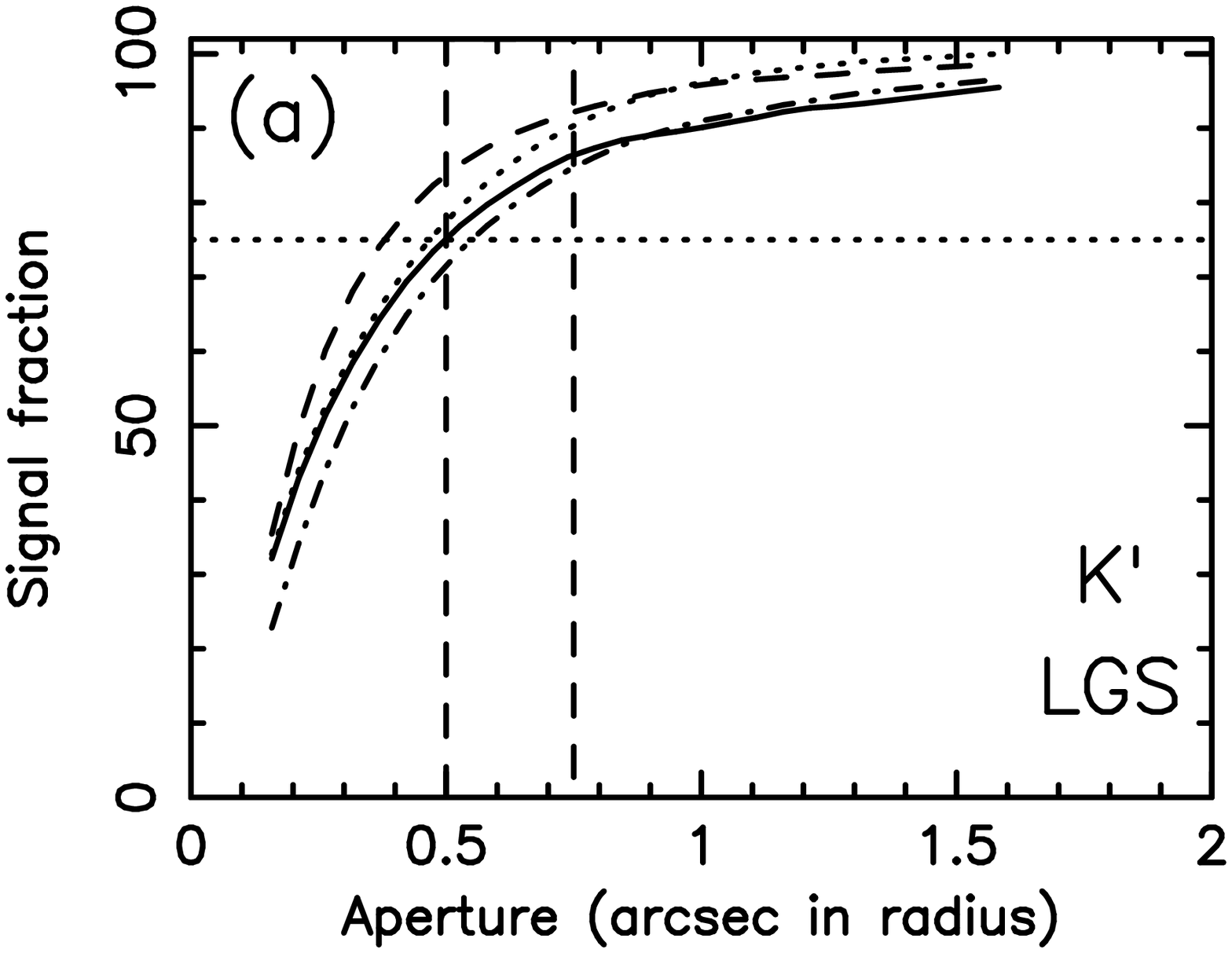} 
\includegraphics[angle=0,scale=.4]{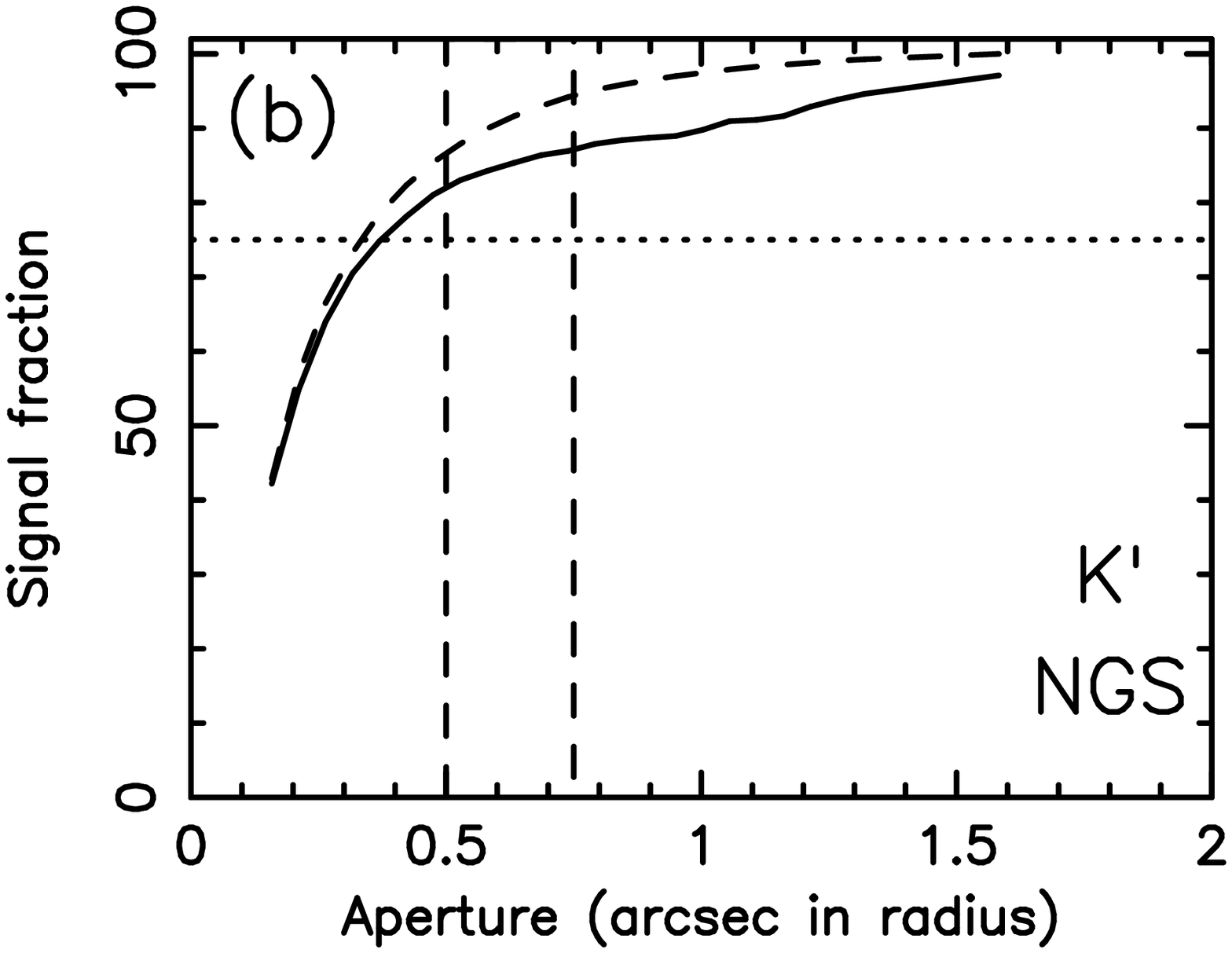} \\
\includegraphics[angle=0,scale=.4]{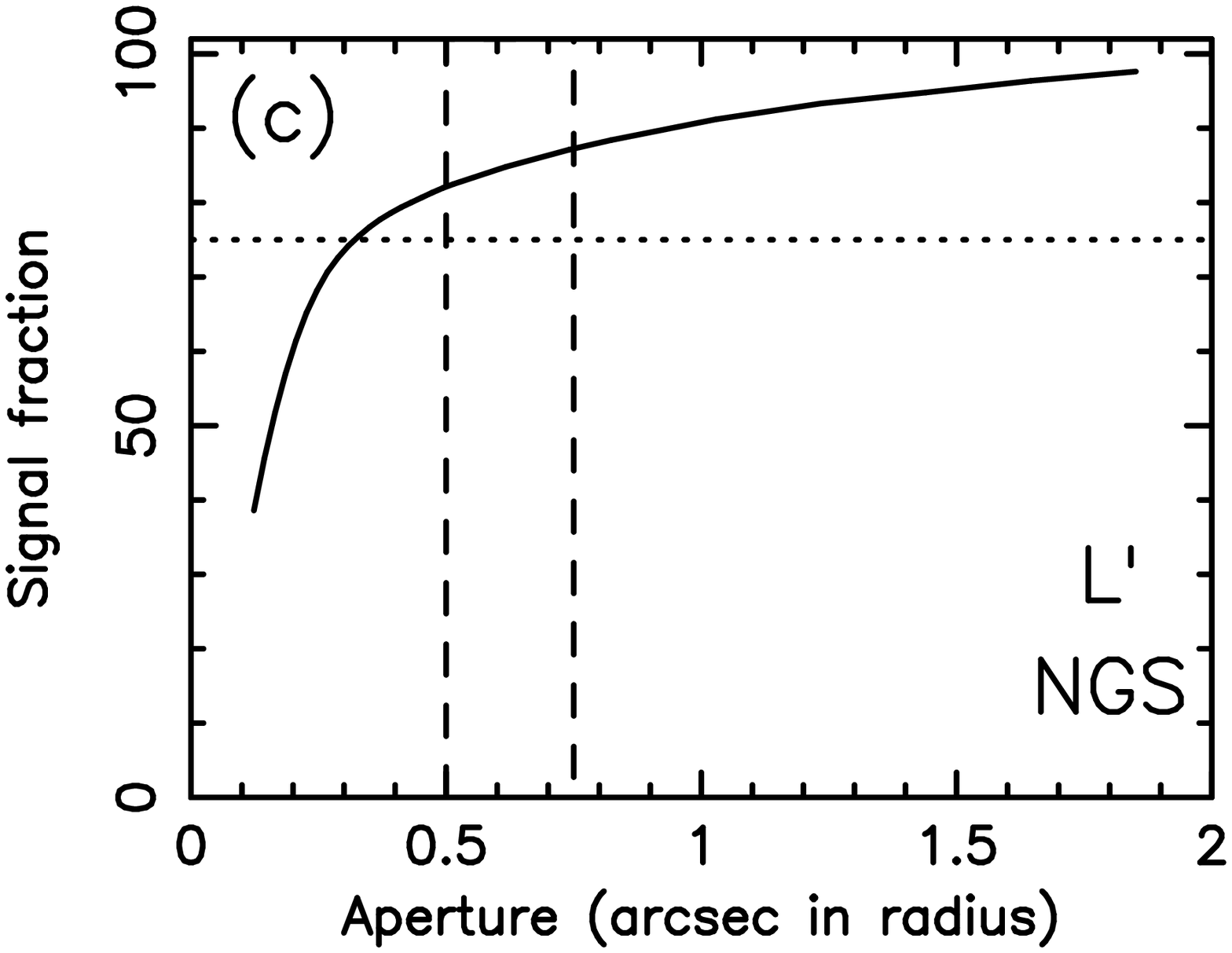} 
\end{center}
\caption{Signal growth curve of compact objects (whenever available) 
within the field of view of ULIRG images.
(a): $K'$-band data observed with LGS-AO. 
A compact object at $\sim$11$''$ north and $\sim$21$''$ west of 
IRAS 09039$+$0503 (2019 April) (solid line), 
that at $\sim$18$''$ south and $\sim$5$''$ west of 
IRAS 14394$+$5332 (2019 April) (dashed line), 
that at $\sim$11$''$ north and $\sim$2$''$ east of 
IRAS 15206$+$3342 (2016 April) (dash-dotted line), and 
that at $\sim$19$''$ south and $\sim$0$\farcs$4 west of 
IRAS 10190$+$0334 (2015 February) (dotted line).
(b): $K'$-band data observed with NGS-AO.
A compact object at $\sim$14$''$ south and $\sim$11$''$ east of 
IRAS 20414$-$1651 (2015 September) (solid line) and 
that at $\sim$23$''$ north and $\sim$20$''$ west of 
IRAS 21219$-$1757 (2015 September) (dashed line).
(c): $L'$-band data observed with NGS-AO.
IRAS 21219$-$1757 itself (compact ULIRG) (2015 September).
The vertical dashed lines indicate 0$\farcs$5 and 0$\farcs$75 
radius apertures.
The horizontal dotted line indicates 75\% signal fraction relative 
to a 2$\farcs$5 radius aperture measurement.
}
\end{figure}

\begin{figure}
\begin{center}
\includegraphics[angle=0,scale=.4]{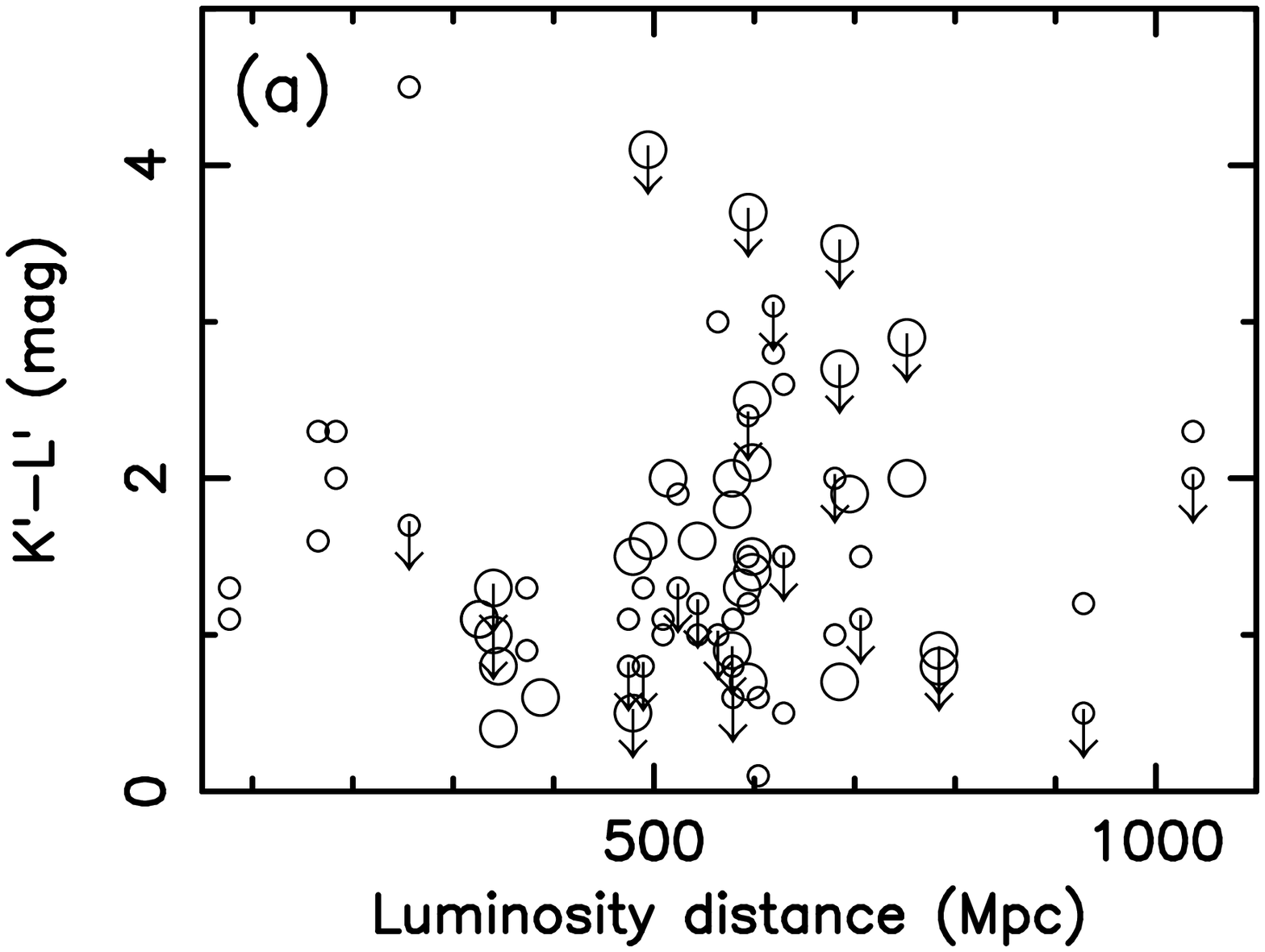} 
\includegraphics[angle=0,scale=.4]{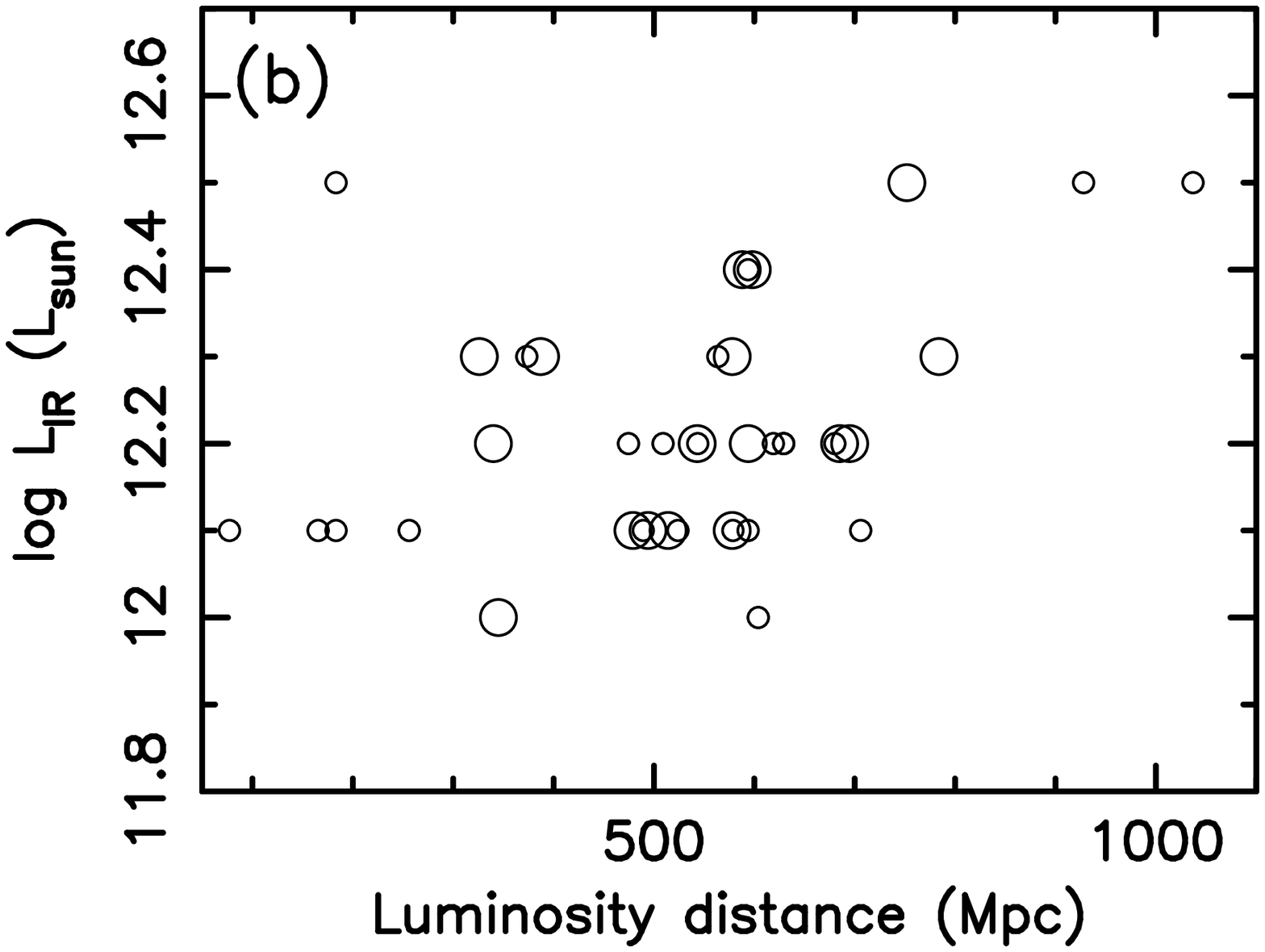} \\
\end{center}
\caption{
{\it (a)}: Observed $K'-L'$ color in mag of compact emission at ULIRG 
nuclei, measured with a 0$\farcs$5 radius aperture (ordinate), 
as a function of luminosity distance in Mpc (abscissa).
{\it (b)}: Decimal logarithm of infrared luminosity in L$_{\odot}$ 
(ordinate) as a function 
of luminosity distance in Mpc (abscissa).
In both (a) and (b), larger and smaller open circles mean ULIRGs observed 
in this study and \citet{ima14}, respectively.
}
\end{figure}

\end{document}